\documentclass[nofootinbib,showpacs,showkeys]{revtex4-2}
\usepackage{graphicx,color}
\usepackage{amssymb}
\usepackage{amsmath}
\usepackage{bm}
\usepackage{tensor} 
\usepackage{lipsum}
\usepackage{latexsym}
\usepackage{dcolumn}
\usepackage{float}
\usepackage{array}
\usepackage{graphicx}
\usepackage{dcolumn}
\usepackage{bm}
\usepackage{tensor} 
\usepackage{natbib}
\usepackage{subcaption}
\graphicspath{{Images/}}
\usepackage[hyperindex,colorlinks]{hyperref}
\usepackage{cleveref}

\usepackage{acro}

\DeclareAcronym{kds}{
  short=KdS,
  long=Kerr-de Sitter,
}
\DeclareAcronym{qnm}{
  short=QNMs,
  long=Quasinormal Modes,
}
\DeclareAcronym{rkds}{
  short=RKdS,
  long=Kerr-de Sitter Revisited,
}
\DeclareAcronym{sds}{
  short=SdS,
  long=Schwarzschild-de Sitter,
}
\DeclareAcronym{egb}{
  short=EGB,
  long=Einstein-Gauss-Bonnet,
}
\DeclareAcronym{egbds}{
  short=EGBdS,
  long=Einstein-Gauss-Bonnet de Sitter,
}
\DeclareAcronym{gr}{
  short=GR,
  long=General Relativity,
}

\DeclareAcronym{vlbi}{
  short=VLBI,
  long=Very
Long Baseline Interferometry,
}
\DeclareAcronym{eht}{
  short=EHT,
  long=Event Horizon Telescope,
}
\DeclareAcronym{sgrA}{
  short=Sgr$A^{*}$,
  long=Sagittarius $A^{*}$,
}
\DeclareAcronym{zamo}{
  short=ZAMO,
  long=Zero Angular Momentum Orbit,
}

\begin{document}

\title{Black hole lensing in Kerr-de Sitter spacetimes}


\author{Eunice Omwoyo}%
\email{eunice.m.omwoyo@aims-senegal.org}
\affiliation{%
PPGCosmo, CCE, Universidade Federal do Esp\'irito Santo (UFES)\\
Av. Fernando Ferrari, 540, CEP 29.075-910, Vit\'oria, ES, Brazil.}%

\author{Humberto Belich}%
\email{humberto.belich@ufes.br}
\affiliation{%
N\'ucleo Cosmo-ufes \& Departamento de F\'isica,  Universidade Federal do Esp\'irito Santo (UFES)\\
Av. Fernando Ferrari, 540, CEP 29.075-910, Vit\'oria, ES, Brazil.}

\author{J\'ulio C. Fabris}
\email{julio.fabris@cosmo-ufes.org}%
\affiliation{%
N\'ucleo Cosmo-ufes \& Departamento de F\'isica,  Universidade Federal do Esp\'irito Santo (UFES)\\
Av. Fernando Ferrari, 540, CEP 29.075-910, Vit\'oria, ES, Brazil.}%
\affiliation{%
National Research Nuclear University MEPhI, Kashirskoe sh. 31, Moscow 115409, Russia}%

\author{Hermano Velten}%
\email{§ hermano.velten@ufop.edu.br
} %
\affiliation{%
Departamento de F\'isica, Universidade Federal de Ouro Preto (UFOP), Campus Universit\'ario Morro do Cruzeiro, 35.400-000, Ouro Preto, Brazil}%
\date{\today}

\begin{abstract}
We have derived analytical solutions using Jacobi elliptic functions for bound and nearly bound photon orbits in \ac{kds} and \ac{rkds} spacetimes. Leveraging our obtained solutions, we have  conducted an analytic ray-tracing in both spacetimes. We have obtained direct images, lensing rings and photon rings for equatorial disks considering inclined locally static observers. Images corresponding to $n=(2,3)$ exhibit a significantly closer resemblance to the critical curve as compared to the $n=1$ image. This highlights the remarkable potential of these higher-order images as robust testing grounds for \ac{gr}. Furthermore in both spacetimes, we have obtained analytical solutions for the critical parameters governing the structure of the photon ring and analyzed these parameters in details. 
\end{abstract}

\maketitle

\section{Introduction}

It is well-established that in the vicinity of rotating black holes there exists a photon region, also known as a photon sphere for static black holes, which consists of orbits where photons move at a constant radial distance, referred to as bound photon orbits. When these orbits are unstable to radial perturbations, photons on nearby orbits that are not precisely in the photon region/sphere may undergo numerous half orbits close to the bound orbit before either plunging into the black hole or escaping. These orbits are known as nearly bound photon orbits. In the image plane of an observer, the photons on nearly bound orbits manifest as an annular feature that is located in close proximity to the critical curve, and this feature is known as the photon ring [\cite{gralla2020lensing}, \cite{gralla2020shape},\cite{gralla2019black}]. The term critical curve describes the locus on the image plane that arises from the photons moving along bound orbits and its shape follows from  General Relativity (GR). Recently, the \ac{eht} collaboration has provided interferometric observations of the black holes M87 \cite{collaboration2019first} and \ac{sgrA} \cite{akiyama2022first} at the center of the galaxies Messier 87 and Milky Way, respectively, thus providing valuable insights into the emission structure on horizon scales. Calculations within the framework of \ac{gr}, \cite{gralla2019black}, show that the photon ring is not a single continual entity. Rather, it is made up of an infinite sequence of self-similar subrings converging to the critical curve. Each subring is uniquely identified by the number of half orbits, denoted as $n=0,1,2,..\infty$, which nearly bound photons execute before escaping or plunging into the black hole.
The $n=0$ image, also known as the primary or direct image, is formed by photons on weakly lensed orbits and its characteristics strongly depend on the astrophysical source profile around the black hole. On the other hand, the $n=1$ image is referred to as the lensed image \cite{gralla2020lensing}. The photon ring consists of subrings with $n\geq 2$ resulting from photons on orbits that have undergone extreme lensing due to the strong gravity of the black hole, causing them to execute multiple half orbits around the black hole. Probing the photon ring is thus a gravity probe because the presence of it is due to strong lensing and is insensitive of the astrophysical source profile's details. 

As a photon on a nearly bound orbit escapes or falls into a black hole, the distance between it and a bound orbit grows exponentially. The Lyapunov exponent of the respective bound orbit measures the rate of this exponential deviation. Furthermore, when these photons are projected onto the celestial sphere of an observer, they generate subrings whose thickness is controlled by the Lyapunov exponent \cite{gralla2020lensing}. With each additional half orbit that photons on nearly bound orbits execute, there is a corresponding change in the azimuthal angle, which causes subsequent subrings to rotate. Moreover, the detection of subsequent subrings by an observer occurs at different times due to the extra half orbits executed by the photons, thereby causing a delay in their detection. Thus the photon ring is controlled by  three critical parameters, the change in azimuthal angle $\delta$, Lyapunov exponent $\gamma$, and time delay $\zeta$ \cite{gralla2020lensing}. The time delay parameter controls the arrival time of subsequent subrings by determining how much time photons take to complete half orbits. Meanwhile, the change in azimuthal angle parameter controls the rotation of the subrings by specifying the amount of change in the azimuthal angle per half orbit. Finally, the Lyapunov exponent parameter measures the exponential demagnification of subsequent subrings. These critical parameters are defined on the critical radii, which is the radial coordinate of the bound photon orbits and are independent of the emitting matter around the black hole. Thus, to decipher the photon ring alongside these critical parameters, a comprehensive understanding of bound photon orbits is vital.

In this work, we begin with performing an analysis of bound and nearly bound photon orbits in Kerr-de Sitter (KdS) \cite{novikov1973houches} and Kerr-de Sitter Revisited (RKdS) \cite{ovalle2021kerr} spacetimes. 
The \ac{rkds} is a newly proposed solution that has been obtained through gravitational decoupling for axially symmetric systems (\cite{contreras2021gravitational},\cite{ovalle2017decoupling},\cite{ovalle2019decoupling}). Unlike the \ac{kds} solution, the \ac{rkds} solution is neither a $\Lambda$ Vacuum solution - it does not belong to the Plebanski-Demianski class of metrics - nor is it a constant curvature solution. The salient feature of the \ac{rkds} solution is the emergence of a warped curvature induced by rotational effects which is absent in \ac{kds} solution. This deformation is likely to be closely related to the warped vacuum energy in the direct proximity of extreme gravitational sources \cite{ovalle2022warped}. It is possible that the deformation of the vacuum energy in the vicinity of the \ac{rkds} black hole could be of relevant astrophysical significance and thus extensive study of this solution is necessary.

In Ref.\cite{hackmann2010analytical}, analytic solutions to geodesics in \ac{kds} spacetime have been obtained using Weierstrass' elliptic $\wp$ function and Weierstrass' $\zeta$ and $\sigma$ functions. The Weierstrass' elliptic $\wp$ function and $\sigma$ function are multivalued while the Weierstrass' $\zeta$ function is single-valued. A multivalued function, also known as a set-valued function, is a mapping that assigns a set of output values to each input value in its domain while a single-valued function can be defined mathematically as a mapping that assigns a unique output value to each input value in its domain. Nearly bound orbits involve photons winding around the black hole multiple times before escaping or falling into the black hole, thus, there can be multiple values of the angular coordinate at certain radial positions. To capture this winding behaviour explicitly, it is important to obtain solutions that are solely in terms of multivalued functions. As a result, the first goal of this work is to obtain solutions to null geodesic equations of \ac{kds} and \ac{rkds} spacetime in terms of Jacobi elliptic functions, which are all multivalued in nature. Furthermore, solutions obtained in terms of Jacobi elliptic functions do not require gluing and instead extend through various regions, providing explicit complete trajectories. The Jacobi elliptic functions approach has been considered to fully characterize null geodesics in the exterior of a Kerr black hole \cite{gralla2020null}, so we refer the reader to this reference in the case of vanishing cosmological constant. Using our solutions, we perform an analytic ray-tracing to obtain direct images, lensing rings and photon rings in \ac{kds} and \ac{rkds} spacetime. Besides, we also obtain analytic forms of the critical parameters in both spacetimes.

This work is structured as follows: In \cref{sectionII}, a comprehensive analysis of \ac{kds} and \ac{rkds} spacetimes is presented, accompanied by the derivation of the solutions to their respective null geodesic equations utilizing Jacobi elliptic functions. This section establishes the groundwork upon which the subsequent sections are built. Moreover, the classification and investigation of the general structure of photon orbits, integral to this work, are examined in this section. \Cref{lensingandphotonrings} capitalizes on the derived solutions to perform an analytical ray-tracing in \ac{kds} and \ac{rkds} spacetimes. By considering inclined locally static observers, we investigate images of equatorial disks around \ac{kds} and \ac{rkds} black holes, encompassing direct images, lensing rings and photon rings. Building upon the properties of the aforementioned images, \cref{sectionIII} delves into the critical parameters that govern the structure of the photon ring. Analytical expressions for these parameters are derived and their detailed investigation is provided. Lastly but not least, in \cref{sectionIV} we give a conclusion of this work. The common derivation steps involved in this work are elucidated in \cref{ferari}, (\ref{apB}), (\ref{apI}), (\ref{apII}) and (\ref{apIII}).

\section{Solutions of the Equations of Motion} \label{sectionII}
In this section, we commence by providing an overview of \ac{kds} and \ac{rkds} spacetimes. Leveraging the general forms derived for the radial and angular integrals, as outlined in Appendix \ref{apI}, we proceed to derive solutions for the $r$, $\theta$, $\phi$, and $t$ components encompassing the trajectory of the photons. The primary objective of this section is to unravel the general structure characterizing bound and nearly bound photon orbits thus we keep the solutions simple. In so doing, we make the assumption that a turning point has yet to be reached \cite{gralla2020null}. The periodic nature of Jacobi elliptic functions allows for this assumption. Hence, in spite of this assumption, the solutions will still provide trajectories that go through the turning points, and the physical effects on the turning points, such as gravitational lensing, will still hold. Throughout our analysis, the terms "source" and "observer" are frequently employed. Herein, the "source" refers to the location of photon emission, while the "observer" denotes the location of the photon detector. It is imperative to emphasize that our considerations are confined to observers situated within the domain of outer communication.

\subsection{Kerr-de Sitter spacetime }
\ac{kds} metric is a solution of Einstein's field equations that describes a rotating black hole in a universe with a positive cosmological constant. In the Boyer-Lindquist coordinate system the metric takes the form
\cite{novikov1973houches},
\begin{multline}
\text{ds}^{2}= \left(\frac{a^2 \Delta_{\theta } \sin ^2(\theta )}{L^2 \Sigma }-\frac{\text{$\Delta_{r} $}}{L^2 \Sigma }\right) \text{dt}^2 -\frac{2 a \sin ^2(\theta ) \left(\Delta_{\theta }  \left(a^2+r^2\right)-\text{$\Delta_{r}$}\right)}{L^2 \Sigma } \text{dt} \text{d$\phi $} \\+\frac{ \sin ^2(\theta ) \left(\Delta_{\theta }  \left(a^2+r^2\right)^2-a^2 \text{$\Delta_{r} $} \sin ^2(\theta )\right)}{L^2 \Sigma } \text{d$\phi $}^2+\frac{ \Sigma }{\Delta_{\theta }  } \text{d$\theta $}^2+\frac{ \Sigma }{\text{$\Delta_{r} $}} \text{dr}^2. \label{kds1}
\end{multline} 
The parameters $\Delta_{r}$, $\Delta_{\theta}$,$L$ and $\Sigma$ are defined as,
\begin{align}
    \Delta_{r}=\left( 1-\dfrac{\Lambda r^{2}}{3}\right) (r^{2}+a^{2})-2Mr,
\Delta_{\theta}=\left( 1+\dfrac{a^{2}\Lambda \cos^{2}\theta}{3}\right), 
L=\left(1+\dfrac{a^{2}\Lambda}{3} \right), 
\Sigma=r^{2}+a^{2}\cos^{2}\theta. \label{26}
\end{align}
Furthermore, the coordinates $t$ and $r$ range over all $\mathbb{R}$ while $\theta \in [0,\pi]$ and $\phi \in [0,2\pi)$. The coefficients of metric \cref{kds1} are independent of $t$ and $\phi$ thus $\partial_{t}$ and $\partial_{\phi}$ are Killing vector fields. The roots of $\Delta_{r}=0$ and $\Sigma=0$ are singularities of (\ref{kds1}). Additionally, one can immediately see that when $a=0$, \cref{kds1} reduces to the \ac{sds} metric and when $\Lambda=0$, it reduces to the Kerr metric with the Schwarzschild metric being the special case when both $\Lambda$ and $a$ are set to zero. 

Equations motion of photons are determined by four fundamental quantities: the Lagrangian, the angular momentum in the $z$ axis $L_z$, the energy $E$, and Carter's constant $K$. Specifically, $L_z$ and $E$ are associated to the axial symmetry and stationarity, while $K=K^{\mu \nu } p_{\nu } p_{\nu }$, the Carter's constant \cite{Carter:1968rr}, is as a result of a hidden symmetry in \ac{kds} spacetime. The discussion in this work will be based on the modified form of the Carter's constant, $Q=K-L(a E-L_z)^2$ \cite{hackmann2010analytical},\cite{charbulak2017photon}. For photons, the quantities $L_z$, $K$ and $E$ can be reparametrized as,
\begin{align}
    \lambda = \frac{L_{z}}{E}, 
    \eta = \frac{Q}{E^2}.\label{28}
\end{align}
Null geodesics in \ac{kds} spacetime follow Carter's equations given by
\cite{novikov1973houches},
	\begin{align}
\dfrac{\Sigma}{E}p^{r}=\pm_{r}\sqrt{R(r)}, 
\dfrac{\Sigma}{E}p^{\theta}=\pm_{\theta} \sqrt{\Theta(\theta)},\label{22}\\
\dfrac{\Sigma}{E}p^{\phi}=\dfrac{aL^{2}}{\Delta_{r}}(a(a-\lambda)+r^{2})-\dfrac{L^{2}}{\Delta_{\theta}\sin^{2}\theta}(a\sin^{2}\theta-\lambda), 
\dfrac{\Sigma}{E}p^{t}=\dfrac{L^{2}}{\Delta_{r}}((r^{2}+a^{2})^{2}-a\lambda(a^{2}+r^{2}))-\dfrac{aL^{2}}{\Delta_{\theta}}(a\sin^{2}\theta-\lambda). \label{23}
\end{align}
$p^{\mu}=\frac{d x^\mu}{d\sigma}$ is the photon's four momentum while $\pm_{r}$ and $ \pm_{\theta}$ are the signs of $p^r$ and $p^\theta$ respectively.
We have as well defined the radial and angular potentials as,
\begin{align}
    R(r)=L^{2}(r^{2}+a^{2}-a\lambda)^{2}-\Delta_{r}(\eta+L^{2}(\lambda-a)^{2}), \label{24}
\end{align}
\begin{align}
\Theta(\theta)=a^2 \Delta_{ \theta}  L^2+a^2 L^2 \cos ^2(\theta )-a^2 L^2-2 a \Delta_{ \theta } \lambda  L^2+2 a \lambda  L^2+\Delta_{ \theta}  \eta +\Delta_ {\theta}  \lambda ^2 L^2-\lambda ^2 L^2 \cot ^2(\theta )-\lambda ^2 L^2. \label{25} 
\end{align}

The physically allowed ranges of $r$ and $\theta$ are determined by the conditions $R(r) \geq 0$ and $\Theta(\theta) \geq 0$ respectively.\\
To obtain analytic solutions to \cref{22} and \cref{23}, one can introduce the Mino parameter $\tau$, which helps to decouple \cref{22} and is defined as \cite{mino2003perturbative},
\begin{align}
    \frac{\Sigma}{E}p^{\mu}=\frac{\text{dx}^{\mu}}{d\tau}. \label{29}
\end{align}
The integral forms for \cref{22} and (\ref{23}) are shown in \cref{a30}-(\ref{32}).
\subsubsection{Latitudinal Motion}
In this part we examine the angular potential and its roots. The equation for the photon's trajectory in the $\theta$ direction will then be deduced.

Substituting the variable $u=\cos^{2}\theta$ in the angular potential gives,
\begin{align}
\Theta(u)= \frac{1}{1-u}\left(\eta + (\frac{1}{3} a^2 \Lambda  \left(L^2 (a-\lambda )^2+\eta \right)+L^2 (a-\lambda ) (a+\lambda )-\eta)u+(-\frac{1}{3} a^2 \left(L^2 \left(\Lambda  (a-\lambda )^2+3\right)+\eta  \Lambda \right))u^{2} \right).\label{33}
\end{align}
 \Cref{33} is a second-order polynomial and, as a consequence, it possesses two roots, $u_{\pm}$,
\begin{align}
u_{\pm}=\chi \pm \sqrt{\frac{3 \eta }{a^2 \left(L^2 \left(\Lambda  (a-\lambda )^2+3\right)+\eta  \Lambda \right)}+\chi ^2},\quad\chi =\frac{1}{2}-\frac{3 \left(\eta +\lambda ^2 L^2\right)}{2 a^2 \left(L^2 \left(\Lambda  (a-\lambda )^2+3\right)+\eta  \Lambda \right)} .  \label{34}
\end{align}
The inversion of $u=\cos^{2}\theta$ leads to the expression $\theta=\arccos(\pm\sqrt{u})$. Consequently, the roots of \ref{25} can be represented in terms of $u_{\pm}$,
\begin{align}
\theta_{1}=\arccos(\sqrt{u_{+}}), \label{36}\\
\theta_{2}=\arccos(\sqrt{u_{-}}),\label{37}\\
\theta_{3}=\arccos(-\sqrt{u_{-}}),\label{38}\\
\theta_{4}=\arccos(-\sqrt{u_{+}}). \label{39}
\end{align}
The ranges of $u$ and the nature of the roots in equations (\ref{36})-(\ref{39}) are determined by the sign of $\eta$. This can be investigated by the relation,
\begin{align}
 u_{+}u_{-}=-\frac{3 \eta }{a^2 \left(L^2 \left(\Lambda  (a-\lambda )^2+3\right)+\eta  \Lambda \right)}. \label{40}
\end{align}
Examining the expression in \eqref{40} for different $\eta$, namely $\eta > 0$, $\eta < 0$, and $\eta = 0$, results in the following insights,
\begin{enumerate}
    \item For $\eta>0$, it can be inferred that $u_{-}u_{+}<0$. Given that $u_{+}>0$, then $u_{-}<0$. Therefore, the roots in \cref{37} and (\ref{38}) will be complex conjugates, while those in \cref{36} and (\ref{39}) will be real. Specifically, $\theta_{1}$ will lie in the first quadrant and $\theta_{4}$ will be in the second quadrant, indicating that $\theta_{1}<\pi/2<\theta_{4}$. Whereby, the term quadrant refers to the region of the two-dimensional Cartesian coordinate system that is delimited by the intersection of its horizontal and vertical axes hence demarcating four distinct sections (the first quadrant, the second quadrant and the third quadrant). The photon will oscillate between $\theta_{1}$ and $\theta_{4}$, crossing the equatorial plane each time. This type of orbit is referred to as an ordinary geodesic.
    \item For the case where $\eta<0$, we can observe that $u_{-}u_{+}>0$. In order for the angular potential to be non-negative, the condition
    \begin{align}
        a^2 \left(L^2 \left(\Lambda  (a-\lambda )^2+3\right)+\eta  \Lambda \right)-3 \left(\eta +\lambda ^2 L^2\right)>0, \label{41}
    \end{align}
        has to be satisfied. As both $u_{+}$ and $u_{-}$ are positive, it follows that all four roots given by \cref{36} to \cref{39} will be real. Thus, $\theta_{1}$ and $\theta_{2}$ lie in the first quadrant, while $\theta_{3}$ and $\theta_{4}$ are in the second quadrant. This means that $\theta_{1}<\theta_{2}<\pi/2<\theta_{3}<\theta_{4}$. As a result, the photon will oscillate between $\theta_{1}$ and $\theta_{2}$ or between $\theta_{3}$ and $\theta_{4}$. The photon will be confined within either the northern or southern hemisphere, and cannot cross the equatorial plane. Such orbits are referred to as vortical geodesics.
    \item When $\eta=0$, we have $u_{+}u_{-}=0$. This implies two scenario : either ($u_{+}=0$ $u_{-}=2\chi$) or ($u_{-}=0$ $u_{+}=2\chi$). In our analysis, we find that the possible scenario is $u_{+}=0$ and $u_{-}=2\chi$ such that $2\chi<0$. Therefore, $\theta_{1}$ and $\theta_{4}$ both equal $\pi/2$, while $\theta_{2}$ and $\theta_{3}$ are complex conjugates. This implies that the orbit is confined to the equatorial plane. 
 \end{enumerate}
We then derive the $\theta$ component of the trajectory by considering the path of a photon emitted from a source point $s$ and detected by an observer $o$. We will consider the case for which $\eta \geq 0$ and will show in the next subsection that this is because for bound photon orbits, $\eta<0$ is not physically relevant. The source is located at a latitude $\theta_s$, while the observer is located at $\theta_o$. 

By utilizing \cref{58}, we obtain,
\begin{align}
    \tau=I_{\theta}=\dfrac{1}{\sqrt{-u_{-}\Xi}} \left[\nu_{\theta} F(x_{s}|k)-\nu_{\theta} F(x_{o}|k) \right] \label{75}.
\end{align}
We invert \cref{75} using the relations,
\begin{align}
       am(F(x,k),k)=x, \label{Finvert}\\
        \sin (am(x\mid |k))=sn(x\mid |k), \label{87b}
\end{align}
which results in,
\begin{align}
     \frac{\cos \theta_{o}}{\sqrt{u_{+}}}=-\nu_{\theta} sn\left[\left(\tau \sqrt{-u_{-}\Xi}-\nu_{\theta} F\left(\varphi_{s} | k\right) \right) |k\right]. \label{82}
\end{align}
Using equation \eqref{82}, we can easily obtain $\theta_{o}$, which represents the latitude of the observer,
\begin{align}
\theta_{o}(\tau)=\arccos\left(-\sqrt{u_{+}}\nu_{\theta} sn\left[\left(\tau \sqrt{-u_{-}\Xi}-\nu_{\theta} F\left(\varphi_{s} | k\right) \right) |k\right] \right). \label{83} 
\end{align}
Accordingly, this equation encapsulates the motion of the photon in the latitudinal direction in terms of Mino parameter from a point $\theta_{s}$ to $\theta_{o}$. Regardless of the fact that it is assumed that a turning point has not yet been reached, \cref{83} is in terms of the Jacobi elliptic sine function which oscillates between $-1$ and $+1$ with period $4 K(k)$. As a result, the solution in \cref{83} will indeed oscillate between the turning points, giving rise to full trajectories.
\subsubsection{Radial Motion}
In this subsection, we will investigate the conditions for bound orbits and obtain the constants of motion characterizing them as well as the radii at which they occur in the exterior of a \ac{kds} black hole. Furthermore, for photons on unbound orbits, we will derive the radial component of their trajectory by taking into account the nature of the roots of the radial potential.

Bound photon orbits are characterized by the condition,
\begin{align}
    R(r)=0,\quad \frac{dR}{dr}=0 .\label{42}
\end{align}
Simultaneously solving \cref{42} for $\eta$ and $\lambda$ results in two cases of equations,
\begin{enumerate}
    \item \begin{align}
        \eta=-\frac{L^2 r^4}{a^2},\quad
        \lambda=\frac{a^2+r^2}{a}. \label{43}
    \end{align}
    \item \begin{align}
        \eta = -\frac{L^2 r^3 \left(6 a^2 \left(\Lambda  r^2 (3 M+r)-6 M\right)+a^4 \Lambda ^2 r^3+9 r (r-3 M)^2\right)}{a^2 \left(r \left(a^2 \Lambda +2 \Lambda  r^2-3\right)+3 M\right)^2} ,\label{44} \\
\lambda = \frac{r \left(a^2 \left(6-\Lambda  r^2\right)+3 r (r-3 M)\right)}{a \left(r \left(a^2 \Lambda +2 \Lambda  r^2-3\right)+3 M\right)}+a .\label{45}
    \end{align}
\end{enumerate}
$\eta$ is clearly negative for the first case. Substituting the parameters of the first case into \cref{41}  results in : $a^2 \left(L^2 \left(\Lambda  (a-\lambda )^2+3\right)+\eta  \Lambda \right)-3 \left(\eta +\lambda ^2 L^2\right)=-6 L^2 r^2<0$.  For bound photon orbits, this case is then not physical since it violates the condition in \cref{41} . From this analysis, it follows that vortical motion is not possible for \ac{kds} bound photon orbits. Bound photon orbits in Kerr spacetime are also found to lack the presence of vortical motion, as shown by Teo \cite{teo2003spherical}. It should be emphasized that the absence of vortical motion is specific to bound orbits, unbound photon orbits may still exhibit such behavior. However, we shall refrain from delving into the intricacies of unbound vortical orbits in this work.\\ The second case is only relevant for $\eta>0$ and when $\eta$ vanishes the photon orbit will be confined to the equatorial plane. Solving for $r$ in $\eta=0$ gives the radial coordinate of the equatorial circular photon orbits \cite{omwoyo2022remarks},
\begin{align}
r_{ph+}=-\frac{2 M (\mathcal{Y}-1)}{(\mathcal{Y}+1)^2}+2 \sqrt{\frac{M^2 ((\mathcal{Y}-14) \mathcal{Y}+1)}{(\mathcal{Y}+1)^4}}  \cos \left(\frac{\kappa }{3}+\frac{4 \pi }{3}\right), \label{3}\\
r_{ph-}=-\frac{2 M (\mathcal{Y}-1)}{(\mathcal{Y}+1)^2}+2 \sqrt{\frac{M^2 ((\mathcal{Y}-14) \mathcal{Y}+1)}{(\mathcal{Y}+1)^4}}  \cos \left(\frac{\kappa }{3}\right). \label{4}
\end{align}
We have also defined 
\begin{align}
\kappa=\arccos\left( \frac{M \left(2 a^2 (\mathcal{Y}+1)^4+M^2 (\mathcal{Y}-1) (\mathcal{Y} (\mathcal{Y}+34)+1)\right)}{(\mathcal{Y}+1)^6 \left(\frac{M^2 ((\mathcal{Y}-14) \mathcal{Y}+1)}{(\mathcal{Y}+1)^4}\right)^{3/2}} \right), \quad \mathcal{Y}=\frac{a^{2}\Lambda}{3}. \label{kappa}
\end{align}
Thus $r_{ph+} \leq r \leq r_{ph-}$ delineates the radii at which bound photon orbits exist in the exterior of a \ac{kds} black hole, hence the photon region.

$\eta$ increases from $r_{ph+}$ and attains a maximum at $ r_{\eta max}=-\frac{9 M}{a^2 \Lambda -3}$ where it begins to decrease towards $r_{ph-}$. 

$\lambda$ on the other hand is monotonic and it decreases from positive values at $r_{ph+}$ towards negative values at $r_{ph-}$. This parameter vanishes at,   
\begin{align}
r_{zamo} =\frac{3 M}{a^2 \Lambda +3} +2 \sqrt{\frac{9 M^2}{\left(a^2 \Lambda +3\right)^2}-\frac{a^2}{3}} \cos\left(\dfrac{1}{3}\arccos\left[ -\frac{9 \sqrt{3} M \left(a^2 \left(a^2 \Lambda +3\right)^2-9 M^2\right)}{\left(a^2 \Lambda +3\right)^3 \left(\frac{27 M^2}{\left(a^2 \Lambda +3\right)^2}-a^2\right)^{3/2}}\right]  \right) . \label{50}
\end{align}
$\lambda$ is related to the angular momentum of the photons about the $z$-axis. Thus, the case where $\lambda=0$ corresponds to photons that possess no angular momentum in the $z$ direction, and can therefore traverse the entire range of the angular coordinate, $\theta \in [0,\pi]$, allowing them to cross the symmetry axis of the black hole.

A discussion of the roots of the radial potential has been given in \cref{ferari} with the roots in \cref{34ra}-(\ref{34r}). In this work we will consider the case of double roots whose physical interpretation implies that the photon orbits are bound at a constant radial coordinate $r_{3}=r_{4}$. In addition, we will briefly examine solutions with $(r_{1}<r_{2}<r_{3}<r_{4}<r_s)$, which represents orbits that go through a turning point at $r_{4}$ and get detected by an observer. In some instances all these roots can be less than the Cauchy horizon hence the orbits do not have a turning point in the exterior of a black hole. Therefore, the configuration of four distinct real roots, $(r_{1}<r_{2}<r_{3}<r_{4}<r_s)$,  can represent orbits either escaping or falling into the black hole. We will also consider orbits with $r_{1}<r_{2}<r_{s}$ and $r_{3}=\Bar{r}_{4}$, which represent photon trajectories that always plunge into the black hole as they never encounter a turning point in the exterior region. Our main objective for considering these two scenarios is to investigate the structure of photon orbits that are nearly bound since they play a vital role in the formation of the photon ring.\\
Nearly bound photon orbits refer to those orbits that, with a slight change in initial conditions from bound photon orbits, result in trajectories that remain close to the bound trajectories for a certain period before exponentially diverging and escaping or plunging into the black hole. We will solve the radial integrals for the two aforementioned cases of escape and plunge orbits.
\begin{itemize}
    \item \underline{Escape orbits:}\\ 
Making use of \cref{Ixg} results in,
\begin{align}
   \tau= I_{r}=\nu_{r}g_{E}(F(\arcsin \sqrt{\frac{(r_{3}-r_{1})(r_{o}-r_{4})}{(r_{4}-r_{1})(r_{o}-r_{3})}} |k_{E})-F(\varphi_{E,s} |k_{E})). \label{161}
\end{align}
Inverting for $r_{o}$ gives,
\begin{align}
     r_{o}^{E}[\tau]=\frac{r_3 \left(r_1-r_4\right) \text{sn}^2\left(\left.\frac{\tau }{g_{E}}+\nu_{r} F(\varphi_{E,s}|k_{E})\right|k_{E}\right)+\left(r_3-r_1\right) r_4}{(r_1-r_4) \text{sn}^2\left(\left.\frac{\tau }{g_{E}}+\nu_{r} F(\varphi_{E,s}|k_{E})\right|k_{E}\right)-r_{1}+r_{3}}.\label{162}
\end{align}
\Cref{162} describes the radial component of a photon's trajectory that is emitted as a point $r_s$ and has a turning point at $r=r_{4}$ where it escapes towards an observer at located at $r_o$. It is important to note that this case can also represent a plunge orbit in the case that $r_{1}<r_{2}<r_{3}<r_{4}<r_-< r_+<r_s$, where $r_-$ and $r_+$ are the Cauchy and event horizons respectively.
\item \underline{Plunge orbits:}\\ 
Making use of \cref{Irgp} results in,
	\begin{align}
	    \tau=I_{r}=\nu_{r}g_{P}\left(F\left(\arccos\left[ \frac{(A-B)r_{o}+r_{2} B-r_{1} A}{(A+B)r_{o}-r_{2} B-r_{1}A} \right]|k_{P}\right)-F(\varphi_{P,s}|k_{P})\right). \label{162b}
	\end{align}
 Inverting for $r_{o}$ gives,
	\begin{align}
	    r_{o}^{P}[\tau]=\frac{(r_{2} B+A r_{1}) \text{cn}\left(\left.\nu_{r} F(\varphi_{P,s}|k_{P}) +\frac{\tau }{g_{P}}\right|k_{P}\right)+r_{2} B-A r_{1}}{(A+B) \text{cn}\left(\left.\nu_{r} F(\varphi_{P,s}|k_{P}) +\frac{\tau }{g_{P}}\right|k_{P}\right)-A+B}. \label{kdp}
	\end{align}
\Cref{kdp} represents the radial component of a photon's trajectory which is emitted at the radial coordinate $r_s$ and eventually plunges into the black hole as it has no turning point in the exterior region of the black hole
\end{itemize}
\subsubsection{Azimuthal Motion}
The equation of motion in the $\phi$ direction, \cref{156}, incorporates both radial and angular integrals. This subsection discusses the solution to these integrals.

Making use of the general solutions presented in \cref{61} and \cref{66}, we obtain the angular integrals in the azimuthal direction as follows:
\begin{align}
    I_{\phi}=\dfrac{1}{\sqrt{-u_{-}\Xi}}\left[\nu_{\theta} \Pi(-u_+ \mathcal{Y},\varphi_{s},k)-\nu_{\theta}\Pi(-u_+ \mathcal{Y},\varphi_{\tau},k) \right], \label{88} \\
    \Bar{I}_{\phi}=\dfrac{1}{\sqrt{-u_{-}\Xi}} \dfrac{1}{1+\mathcal{Y}}\left[\nu_{\theta}(\Pi(u_{+},\varphi_{s},k)+\mathcal{Y}\Pi(-u_+ \mathcal{Y},\varphi_{s},k))-\nu_{\theta}(\Pi(u_{+},\varphi_{\tau}|k)+\mathcal{Y}\Pi(-u_+ \mathcal{Y},\varphi_{\tau}|k))\right], \label{89}
\end{align}
where $\varphi_{\tau}=\varphi_{o}$ can also be expressed in terms of only the Mino parameter and the initial conditions through ,
\begin{align}
  \varphi_{\tau}=  -\nu_{\theta} am \left[ \left(\tau \sqrt{-u_{-}\Xi}-\nu_{\theta} F\left(\varphi_{s} | k\right) \right) |k \right]. \label{87a}
\end{align}
Equations (\ref{88}) and (\ref{89}) are the solutions to the angular integrals in the azimuthal direction.

For the radial integral $I_x$, which we have shown in \cref{159*}, we proceed to solve for both escape and plunge orbits.
\begin{itemize}
    \item \underline{Escape orbits:}\\
   By utilizing the solution provided in \cref{Irg}, we obtain the expression for $I_x$ as,
\begin{align}
    I_{x}=-\frac{2  (r_{4}-r_{3})(\Pi_{o}-\Pi_{s})}{(r_{x}-r_{4}) (r_{x}-r_{3}) \sqrt{(r_{4}-r_{2}) (r_{3}-r_{1})}}-\frac{2 (F_{o}-F_{s})}{(r_{x}-r_{3}). \sqrt{(r_{4}-r_{2}) (r_{3}-r_{1})}}.\label{163}
\end{align}
$\Pi_i$ and $F_i$ have been defined in \cref{Ixsym} where $i$ represents $s$ or $o$.
Furthermore, we re-define $\varphi_{E,o}$ as,
\begin{align}
   \varphi_{E,o}=\arcsin \sqrt{\frac{(r_{3}-r_{1})(r_{o}-r_{4})}{(r_{4}-r_{1})(r_{o}-r_{3})}}=am(\frac{ \tau+\nu_{r}g_{E}F(\varphi_{E,s} |k_{E})}{g_{E}} |k_{E}). \label{Ixsymnew}
\end{align}
The second equality has been obtained by inverting for $\arcsin \sqrt{\frac{(r_{3}-r_{1})(r_{o}-r_{4})}{(r_{4}-r_{1})(r_{o}-r_{3})}}$ in \cref{161} using \cref{Finvert}. 

Equations (\ref{88}), (\ref{89}) and (\ref{163}) will then be substituted into \cref{156} to obtain the azimuthal motion of photons on unbound geodesics that escape or plunge into the black hole in the case that $r_{1}<r_{2}<r_{3}<r_{4}<r_-< r_+<r_s$.
\item \underline{Plunge orbits:}\\
Utilizing \cref{Izgp}, we obtain $I_x$ as,
\begin{align}
I_{x,P}  =  \frac{(g_{P} (B-A)) \left((\alpha -\alpha_{2}) (R_{1,o}-\nu_{r}R_{1,s})+\alpha_{2}( F(\varphi_{P,o} \mid k_{P})-\nu_{r}F(\varphi_{P,s} \mid k_{P}))\right)}{r_{2} B+r_{1}A -A r_{x}-B r_{x}}. \label{164b}
\end{align}
The various parameters in \cref{164b} have been defined in (\ref{A1})-(\ref{A3}). We also define $\varphi_{P,o}$ as,
\begin{align}
   \varphi_{P,o}=\arccos\left[ \frac{(A-B)r_{o}+r_{2} B-r_{1} A}{(A+B)r_{o}-r_{2} B-r_{1}A} \right]=am\left[ \frac{\tau+\nu_{r}g_{P}F(\varphi_{P,s} \mid k_{P})}{g_{P}} \right]. \label{164c}
\end{align}
Specifically, we obtained the second equality in (\ref{164c})
by inverting for $\arccos\left[ \frac{(A-B)r_{o}+r_{2} B-r_{1} A}{(A+B)r_{o}-r_{2} B-r_{1}A} \right]$ in \cref{162b} using (\ref{Finvert}).
\end{itemize}
Similarly, equations (\ref{88}), (\ref{89}) and (\ref{164b}) will then be substituted into \cref{156} to obtain the $\phi$ component of a photon's trajectory that plunges into the black hole.
\subsubsection{Motion in $t$ direction}
Radial and angular integrals are additionally incorporated in the equation of $t$ direction, \cref{157}. The radial integral is still $I_x$, and we have solutions for both escape and plunge orbits in \cref{163} and \cref{164b}, respectively.
The solution of the angular integral is obtained from \cref{69} as,
\begin{align}
     I_{t}=\dfrac{u_{+}}{\sqrt{-u_{-}\Xi}}\left[\nu_{\theta}J(-u_+ \mathcal{Y},\varphi_{s} |k)-\nu_{\theta}J(-u_+ \mathcal{Y},\varphi_{\tau} |k)\right],\label{90}
\end{align}
where $\varphi_\tau=\varphi_o$ is defined in \cref{87a}.
These solutions will then be substituted into \cref{157} to obtain the $t$ component of the photon's escape and plunge trajectory.

\begin{figure}[ht]
  \centering
  \begin{tabular}{ccccc}
    \begin{subfigure}[b]{0.2\textwidth}
      \includegraphics[width=\textwidth]{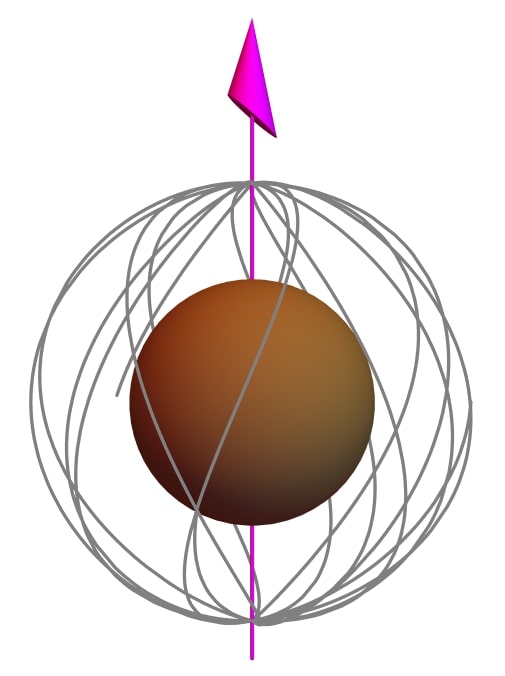}
      \caption{A bound photon orbit at $r=r_{zamo,KdS}=2.56M$ with $\lambda=0$ and $\eta = 23.3576792$.}
      \label{fig:zamokds}
    \end{subfigure} & \quad \quad \quad \quad \quad
    \begin{subfigure}[b]{0.2\textwidth}
      \includegraphics[width=\textwidth]{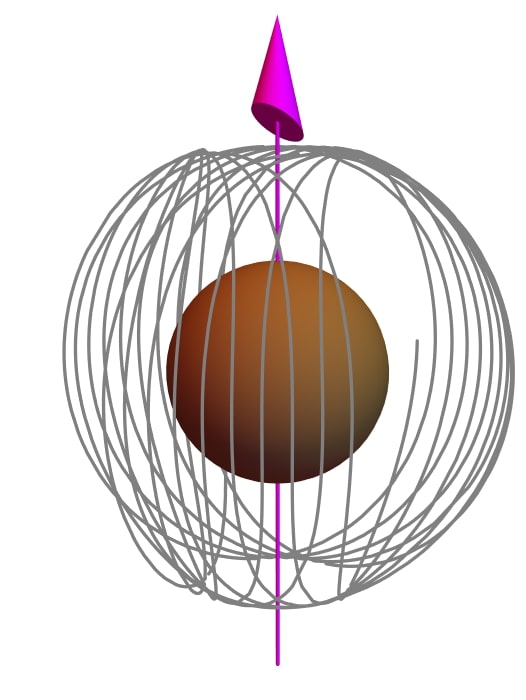}
      \caption{A bound photon orbit at $r = 3.00M$ with $\lambda=-1.8$ and $\eta=27$. }
      \label{fig:etamaxkds}
    \end{subfigure}
  \end{tabular}
   \begin{tabular}{ccc}
    \begin{subfigure}[b]{0.4\textwidth}
      \includegraphics[width=\textwidth]{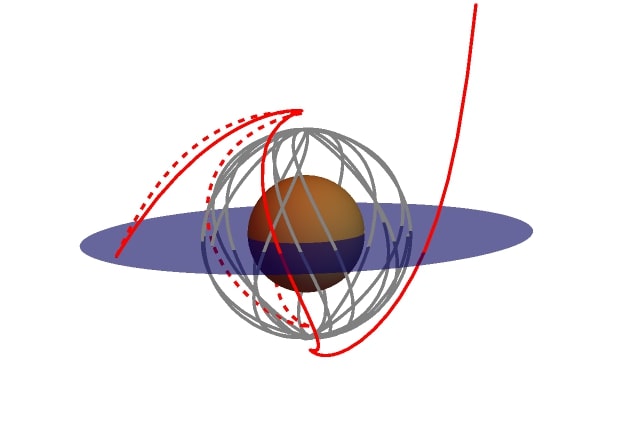}
      \caption{The escape orbit has $r_{4} = 2.749 M$, $\lambda=0$ and $\eta= 23.591256$ whereas the plunge orbit has $\lambda=0$ and $\eta = 23.124102$.}
      \label{fig:nbc1p1}
    \end{subfigure} & \quad \quad 
    \begin{subfigure}[b]{0.4\textwidth}
      \includegraphics[width=\textwidth]{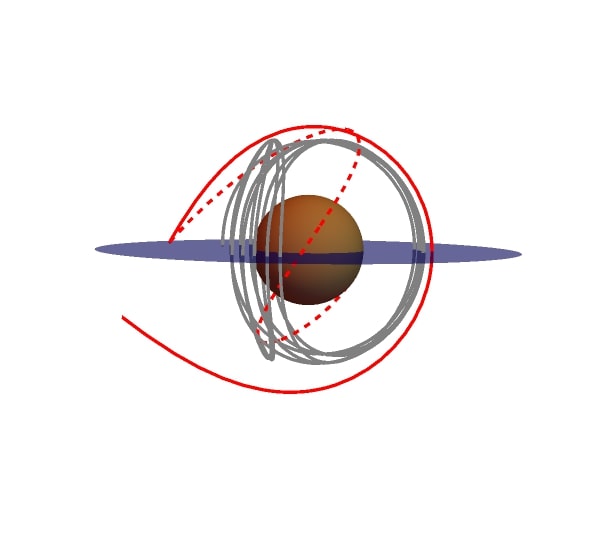}
      \caption{The escape orbit is such that $r_{4}=3.2206 M$,$\lambda=-1.818$ and $\eta=27.27$ while the plunge orbit has $\lambda=-1.782$ and $\eta=26.73$.}
      \label{fig:nbc2p2}
    \end{subfigure} 
  \end{tabular}
   \begin{tabular}{ccc}
    \begin{subfigure}[b]{0.4\textwidth}
      \includegraphics[width=\textwidth]{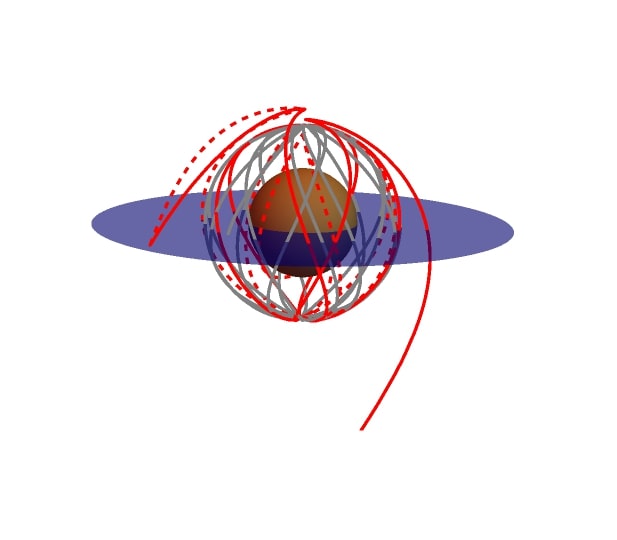}
      \caption{The escape orbit is such that $r_{4}=2.560052765 M$, $\lambda=0$, $\eta = 23.3576794$ while the plunge orbit has $\lambda=0$ and $\eta =23.3576789$.}
      \label{fig:nbc1p8}
    \end{subfigure} & \quad \quad 
    \begin{subfigure}[b]{0.4\textwidth}
      \includegraphics[width=\textwidth]{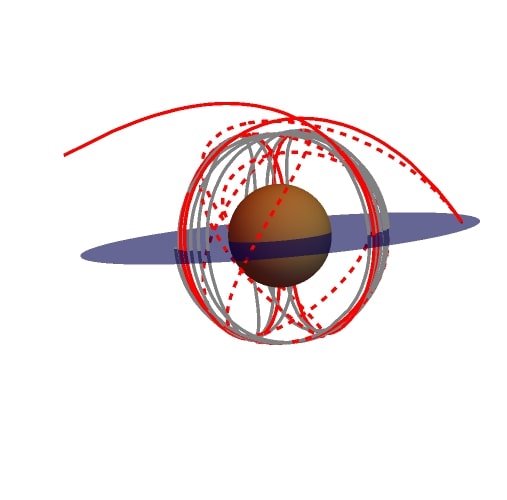}
      \caption{The escape orbit is such that $r_{4}=3.0000647 M$,$\lambda=-1.800000018$ and $\eta=27.00000027$. The plunge orbit is such that $\lambda=-1.799999982$ and $\eta=26.99999973$.}
      \label{fig:nbc2p8}
    \end{subfigure} 
  \end{tabular}
   \caption{ Illustration of bound, escape (solid red) and plunge (dotted red) orbits in \ac{kds} spacetime with $a=0.9$ and $\Lambda=1.11 \times 10^{-52} m^{-2}$. The blue disk is an equatorial source.}
  \label{fig:kdsbillustration}
\end{figure}

Using these solutions, we have illustrated examples of bound, escape and plunge orbits of \ac{kds} spacetime in \cref{fig:kdsbillustration}. Photons on the escape and plunge geodesics in the first column are orbiting close to a bound orbit at $r=r_{zamo}=2.56M$. Photons on the escape and plunge geodesics in the second column on the other hand are orbiting close to a bound orbit at $r=3.00M$. The photons have been emitted from a source located at a radial coordinate $r_s=5M$ and latitude $\theta_s=\pi/2$. Furthermore, the observer is situated within the domain of outer communication.
\clearpage
\newpage
\clearpage
\newpage

\subsection{Kerr de Sitter Revisited}
The \ac{rkds} metric is a newly proposed solution obtained through gravitational decoupling for axially symmetric systems (\cite{contreras2021gravitational}, \cite{ovalle2017decoupling}, \cite{ovalle2019decoupling}). The metric is expressed in the Boyer-Lindquist coordinates ($t$, $r$, $\theta$, $\phi$)  \cite{ovalle2021kerr},
\begin{align}
\text{ds}^{2}=-\left( \dfrac{\hat{\Delta}-a^{2}\sin^{2}\theta}{\rho^{2}}\right) \text{dt}^2+\dfrac{\rho^{2}}{\hat{\Delta}} \text{dr}^2+\rho^{2}\text{d$\theta $}^{2}+\dfrac{((r^{2}+a^{2})^{2}-\hat{\Delta}a^{2}\sin^{2}\theta)\sin^{2}\theta}{\rho^{2}}\text{d$\phi $}^2-\dfrac{2a\sin^{2}\theta}{\rho^{2}}(r^{2}+a^{2}-\hat{\Delta})\text{dt} \text{d$\phi $}, \label{kdsrevisited}
\end{align}
with the following definitions,
\begin{align}
\hat{\Delta}=r^{2}-2 M r+a^{2}-\dfrac{\Lambda r^{4}}{3}, 
\rho^{2}=r^{2}+a^{2} \cos^{2}\theta,
\end{align}
where $a$, $M$, and $\Lambda$ represent the black hole spin, mass, and cosmological constant, respectively. The \ac{rkds} metric exhibits both stationarity and axial symmetry, and the coordinates are such that $r,t \in \mathbb{R}$ while $\theta \in [0,\pi]$ and $\phi \in [0,2\pi)$. Singularities arise in the \ac{rkds} geometry when $\hat{\Delta}=0$ and $\rho^2=0$. \ac{rkds} metric is independent of $t$ and $\phi$ thus $\partial_{t}$ and $\partial_{\phi}$ are Killing vector fields. When $a=0$, \cref{kdsrevisited} reduces to the \ac{sds} metric and when $\Lambda=0$, it reduces to the Kerr metric with the Schwarzschild metric being the special case when both $\Lambda$ and $a$ are set to zero. 

Unlike the \ac{kds} metric, the \ac{rkds} metric is not a $\Lambda$ Vacuum solution and satisfies the standard Einstein's equation, $R_{\mu \nu}-1/2 R g_{\mu \nu}= 8 \pi G T_{\mu \nu}$, where $T_{\mu \nu}$ generating this metric is given by:
\begin{align}
    T^{\mu \nu }=l^{\mu } l^{\nu } p_r+m^{\nu } m^{\nu } p_{\phi }+n^{\mu } n^{\nu } p_{\theta }+\epsilon  u^{\mu } u^{\nu }.
\end{align}
Here, the energy density and pressures satisfy
\begin{align}
    \epsilon= \frac{\Lambda r^4}{\rho^4}=-p_r, \quad p_\theta=p_\phi=\epsilon-\frac{2 \Lambda r^2}{\rho^2}.
\end{align}
Moreover, the scalar curvature of the \ac{rkds} solution is expressed as [\cite{ovalle2021kerr}, \cite{ovalle2022warped}],
\begin{align}
R= \frac{-4 \Lambda r^2}{r^2+a^2\cos^2{\theta}}.
\end{align}
Thus, besides the \ac{rkds} metric not being a $\Lambda$ vacuum solution, it is also not a constant curvature solution, but rather it exhibits a warped curvature everywhere except on the equatorial plane ($\theta=\pi/2$). Besides, this deformation is significant near the black hole i.e. $r\sim a$ and disappears away from the black hole where the scalar curvature approaches a constant value, $R\sim -4 \Lambda$.  This deformation is likely to be closely related to the warped vacuum energy in the direct proximity of extreme gravitational sources \cite{ovalle2022warped}.

Due to the symmetries of \ac{rkds} spacetime, there exists the conserved quantities,
\begin{align}
    \hat{p}_{t}=-g_{tt} \hat{p}^{t}-g_{t\phi}p^{\phi} =: \hat{E}, \hat{p}_{\phi}=g_{t\phi}p^{t}+g_{\phi \phi}p^{\phi}=:\hat{L}_z.
\end{align}

Furthermore, the Lagrangian and the Carter's constant $\hat{K}=\hat{K}^{\mu \nu } \hat{p}_{\nu } \hat{p}_{\nu }$ \cite{Carter:1968rr} are also fundamental parameters determining photon motion in \ac{rkds} spacetime. In this work, our discussion focuses on the modified Carter constant $\hat{Q}=\hat{K}-(a \hat{E}-\hat{L}_z)^2$. For photons, these parameters can be reparametrized as 
\begin{align}
    \hat{\lambda} = \frac{\hat{L}_{z}}{\hat{E}}, 
    \hat{\eta} = \frac{\hat{Q}}{\hat{E}^2}.
\end{align}

Null geodesic equations in \ac{rkds} spacetime  are given by \cite{omwoyo2022remarks},
\begin{align}
\dfrac{\rho^{2}}{E} \hat{p}^{r}=\pm_r \sqrt{\hat{R}(r)}, 
\dfrac{\rho^{2}}{E} \hat{p}^{\theta}=\pm_\theta \sqrt{\hat{\Theta}(\theta)} \label{96},\\
\dfrac{\rho^{2}}{E}\hat{p}^{\phi}=\dfrac{(a r^{2}+a^{3}-a\hat{\Delta}-a^{2}\hat{\lambda})}{\hat{\Delta}}+\dfrac{\hat{\lambda}}{\sin^{2}\theta},
\dfrac{\rho^{2}}{E}\hat{p}^{t}= \dfrac{(r^{2}+a^{2})(r^{2}+a^{2}-a \hat{\lambda})}{\hat{\Delta}}+ a\hat{\lambda}-a^{2}\sin^{2}\theta, \label{97}
\end{align}
where 
\begin{align}
\hat{R}(r)=(r^{2}+a^{2}-a^{2}\hat{\lambda})^{2}-\hat{\Delta}(\hat{\eta}+(\hat{\lambda}-a)^{2}), \\
\hat{\Theta}(\theta)=\hat{\eta}+a^{2}\cos^{2}\theta-\hat{\lambda}^{2} \cot^{2}\theta, \label{99}\\
\hat{\Delta}=r^{2}-2 M r+a^{2}-\dfrac{\Lambda r^{4}}{3}.
\end{align}
The integral forms of \cref{96} and (\ref{97}) are shown in \cref{100aa}-(\ref{101}).
To perform an analysis of bound and nearly bound photon orbits in this spacetime, we use the same approach as in \ac{kds} to generalize our calculations to \ac{rkds} in a quicker way.
\subsubsection{Latitudinal Motion}
In this subsection, we will perform an analysis of the angular potential of \ac{rkds} spacetime, analytically determining its roots. In addition, we will derive the $\theta$ component of the photon's trajectory.

Substituting $\hat{u}=\cos^{2}\theta$ to the angular potential gives,
\begin{align}
    \hat{\Theta}(\hat{u})=\frac{1}{1-u}(\hat{\eta}+(a^2-\hat{\eta} -\hat{\lambda}^{2})\hat{u}-a^{2}\hat{u}^2), \label{103}
\end{align}
with its two zeros expressed as,
\begin{align}
    \hat{u}_{\pm}= \frac{(a^2-\hat{\eta} -\hat{\lambda} ^2) \pm \sqrt{\left(-a^2+\hat{\eta} +\hat{\lambda} ^2\right)^2+4 a^2 \hat{\eta} }}{2 a^2}. \label{104}
\end{align}
From \cref{104}, we arrive at the roots of \cref{99},
\begin{align}
\theta_{1}=\arccos(\sqrt{\hat{u}_{+}}), \label{105}\\
\theta_{2}=\arccos(\sqrt{\hat{u}_{-}}),\label{106}\\
\theta_{3}=\arccos(-\sqrt{\hat{u}_{-}}),\label{107}\\
\theta_{4}=\arccos(-\sqrt{\hat{u}_{+}}) .\label{108}
\end{align}
The nature of the roots in \cref{105} to (\ref{108}) can be assessed by investigating the  relation,
\begin{align}
   \hat{u}_{+}\hat{u}_{-}=- \frac{\hat{\eta}}{a^{2}}. \label{109}
\end{align}
From \cref{109}, it is clear that when $\eta>0$, then $\hat{u}_{+}\hat{u}_{-}<0$. Since $\hat{u}_{+}>0$ it implies that $\hat{u}_{-}<0$. As a result, \cref{105} and (\ref{108}) will be real such that $\theta_{1}<\pi/2<\theta_{4}$. Thus, the photon will oscillate between $\theta_{1}$ and $\theta_{4}$ crossing the equatorial plane each time exhibiting ordinary motion.
On the other hand, when $\eta<0$, it follows that $\hat{u}_{+}\hat{u}_{-}>0$. For this case, the angular potential will be non-negative under the condition,
\begin{align}
    a^{2}-\hat{\eta}-\hat{\lambda}^{2}>0. \label{109b}
\end{align}
Consequently, \cref{105}-(\ref{108}) will be real such that, $\theta_{1}<\theta_{2}<\pi/2<\theta_{3}<\theta_{4}$. In this case, the photon oscillates between $(\theta_{1},\theta_{2})$ or $(\theta_{3},\theta_{4})$, hence  will be confined in the northern or southern hemisphere, thus, exhibiting vortical motion.\\ When $\eta=0$, $\hat{u}_{+}\hat{u}_{-}=0$, thus $\hat{u}_{+}=0$, $\hat{u}_{-}<0$. This gives $\hat{\theta}_{1}=\hat{\theta}_{4}=\pi/2$, hence the photon will be confined to the equatorial plane. 

We restrict our consideration to cases where $\eta \geq 0$. This choice will be further shown in the next subsection, where we demonstrate that bound photon orbits do not display vortical motion. It is important to note that there are instances where vortical motion can occur for unbound photon orbits. However, for the purpose of our current discussion, we will not delve into the specifics of this scenario.
  
To obtain the $\theta$ component of the trajectory, we make use of \cref{118} which results in,
\begin{align}
    \tau = \dfrac{1}{\sqrt{-u_{-}a^{2}}} \left( \nu_{\theta} F(\hat{\psi}_{s} |\hat{k})-\nu_{\theta} F(\hat{\psi}_{o} |\hat{k}) \right), \hat{\psi}_{i}= \arcsin \left( \frac{\cos \theta_{i}}{\sqrt{\hat{u}_{+}}}  \right). \label{125} 
\end{align}
By utilizing the inversion relations presented in equations \cref{Finvert} and \cref{87b} on (\ref{125}), we solve for $\theta_o$,
\begin{align}
    \theta _{o}[\tau]= \arccos \left( -\sqrt{\hat{u}_{+}} \nu_{\theta} sn\left[ \tau \sqrt{-\hat{u}_{-} a^{2}} - \nu_{\theta} F(\hat{\psi}_{s} |\hat{k}) |\hat{k}\right] \right). \label{127}
\end{align}
\Cref{127} represents $\theta$ component of the trajectory in terms of Mino parameter. The equation is in terms of the Jacobi elliptic sine function which oscillates between $-1$ and $+1$ with a period of $4K(\hat{k})$. Thus, regardless of the assumption that a turning point has not yet been reached, the equation goes through turning points providing complete trajectories.
\subsubsection{Radial Motion}
In this part, we will begin our investigation by analysing the condition for the existence of bound photon orbits. We will obtain the constants of motion that characterize these orbits, as well as the precise radii at which they are confined. Furthermore, we will derive the equation for the radial component of photons on unbound orbits.

Bound photon orbits in \ac{rkds} spacetime satisfy the condition $\hat{R}(r)=\hat{R}'(r)=0$. The simultaneous solution of this condition leads to the two distinct parameter cases,
\begin{enumerate}
    \item \begin{align}
      \hat{\eta}=-\frac{r^4}{a^2},  \hat{\lambda}= \frac{a^2+r^2}{a} .\label{110}
    \end{align}
    \item \begin{align}
        \hat{\eta}=-\frac{3 r^3 \left(4 a^2 \left(\Lambda  r^3-3 M\right)+3 r (r-3 M)^2\right)}{a^2 \left(3 M+2 \Lambda  r^3-3 r\right)^2},\quad  \hat{\lambda}=\frac{3 a^2 M+2 a^2 \Lambda  r^3+3 a^2 r-9 M r^2+3 r^3}{a \left(3 M+2 \Lambda  r^3-3 r\right)}.\label{111}
    \end{align}
\end{enumerate}
In the first case, it is evident that $\hat{\eta}$ takes on a negative value. Upon substituting the parameters of this case into \cref{109b}, the resulting expression, $-2r^{2}$, violates the requirement that \cref{109b} must be strictly positive. Consequently, for bound photon orbits, vortical motion is not physically admissible. However, it should be noted that vortical motion is possible for unbound orbits in the \ac{rkds} spacetime.\\
The second case is relevant for $\hat{\eta}>0$ and in the limit $\hat{\eta}=0$, the photon is confined to the equatorial plane. Furthermore, it attains a maximum at $r=3M$. The zeros of $\hat{\eta}$ will then give the radii of equatorial circular photon orbits in the exterior of a \ac{rkds} black hole \cite{omwoyo2022remarks},
\begin{align}
\hat{r}_{ph+} = \frac{6 M}{4 a^2 \Lambda +3} +6 \sqrt{\frac{M^2 \left(1-4 a^2 \Lambda \right)}{\left(4 a^2 \Lambda +3\right)^2}} \cos \left(\frac{\tilde{\kappa} }{3}+\frac{4 \pi }{3}\right), \label{112} \\
\hat{r}_{ph-} = \frac{6 M}{4 a^2 \Lambda +3} +6 \sqrt{\frac{M^2 \left(1-4 a^2 \Lambda \right)}{\left(4 a^2 \Lambda +3\right)^2}} \cos \left(\frac{\tilde{\kappa} }{3}\right). \label{113}
\end{align}
$\tilde{\kappa}$ is defined via the expression,
\begin{align}
\tilde{\kappa} = \arccos\left( \frac{9 M^2-2 a^2 \left(2 \Lambda  \left(8 a^4 \Lambda +12 a^2-27 M^2\right)+9\right)}{9 M \left(4 a^2 \Lambda -1\right) \left(4 a^2 \Lambda +3\right) \sqrt{\frac{M^2 \left(1-4 a^2 \Lambda \right)}{\left(4 a^2 \Lambda +3\right)^2}}}\right). \label{114}
\end{align}
Hence, the bound photon orbits in the exterior of a \ac{rkds} black hole exist in the range $\hat{r}_{ph+}<r<\hat{r}_{ph-}$. The angular momentum ($\hat{\lambda}$) of the photons on these orbits, monotonically decreases from positive values at  $\hat{r}_{ph+}$ to negative values at $\hat{r}_{ph-}$ and vanishes at a radial coordinate,
 \begin{multline}
        r_{zamo,RKdS}= 2 \sqrt{\frac{-2 a^4 \Lambda -3 a^2+9 M^2}{\left(2 a^2 \Lambda +3\right)^2}} \cos \left( \frac{1}{3} \arccos\left[ \frac{3 M \sqrt{\frac{\left(2 a^2 \Lambda +3\right)^2}{-2 a^4 \Lambda -3 a^2+9 M^2}} \left(2 a^6 \Lambda ^2+9 a^4 \Lambda +9 a^2-9 M^2\right)}{\left(2 a^2 \Lambda +3\right) \left(2 a^4 \Lambda +3 a^2-9 M^2\right)}\right] \right)\\+\frac{3 M}{2 a^2 \Lambda +3}. \label{116}
    \end{multline}
In the case of unbound photon orbits, we will derive the radial components of the trajectories by considering the nature of the roots given in \cref{35ra} and (\ref{35r}). Similar to the situation in the \ac{kds} case, we will focus on roots of the form $(\hat{r}_{1}<\hat{r}_{2}<\hat{r}_{3}<\hat{r}_{4}<\hat{r}_s)$ and $\hat{r}_{1}<\hat{r}_{2}<\hat{r}_{s}$, where $\hat{r}_{3}=\Bar{\hat{r}}_{4}$. These two cases correspond to escape and plunge orbits, respectively. By utilizing \cref{C71} and (\ref{C74}), we obtain the $r$ component of the trajectories as follows,
\begin{itemize}
    \item \underline{Escape orbits:}\\
Making use of \cref{C71} gives,
\begin{align}
    \hat{r}_{o}^{E}[\tau]=\frac{\hat{r}_{3} \hat{r}_{1,4} \text{sn}^2\left(\left.\frac{\tau }{\hat{g}_{E}}+\nu_{r} F(\hat{\varphi}_{E,s}|\hat{k}_{E})\right|\hat{k}_{E}\right)+\hat{r}_{4} \hat{r}_{3,1}}{\hat{r}_{1,4} \text{sn}^2\left(\left.\frac{\tau }{\hat{g}_{E}}+\nu_{r} F(\hat{\varphi}_{E,s}|\hat{k}_{E})\right|\hat{k}_{E}\right)-\hat{r}_{1}+\hat{r}_{3}},\label{173}
\end{align}
where the various parameters in this equation are to be obtained from \cref{C73} by setting $i$ to $s$ or $o$ and $\hat{r}_{a,b}=\hat{r}_a-\hat{r}_b$.
\item \underline{Plunge orbits:}\\
For plunge orbits, we make use of \cref{C74} for which we obtain,
\begin{align}
	    \hat{r}_{o}^{P}[\tau]=\frac{(\hat{r}_{2} \hat{B}+\hat{A} \hat{r}_{1}) \text{cn}\left(\left.\nu_{r} F(\hat{\varphi}_{P,s}|\hat{k}_{P}) +\tau\sqrt{\hat{A}\hat{B}}\right|\hat{k}_{P}\right)+\hat{r}_{2} \hat{B}-\hat{A} \hat{r}_{1}}{(\hat{A}+\hat{B}) \text{cn}\left(\left.\nu_{r} F(\hat{\varphi}_{P,s}|\hat{k}_{P}) +\tau\sqrt{\hat{A}\hat{B}}\right|\hat{k}_{P}\right)-\hat{A}+\hat{B}} . \label{177}
	\end{align}
The various parameters can be obtained from the definitions in \cref{C76A}-(\ref{C78}).
\end{itemize}
As for the bound motion, the radial potential has double roots and hence, the $r$ component of the trajectory needs to be evaluated at a constant radial coordinate.
\subsubsection{Azimuthal motion}
As evidenced by \cref{100}, the equation governing motion in $\phi$ direction encompasses both the radial and angular integrals. In this subsection we will obtain the complete solutions for these integrals which will then be substituted into \cref{100} to obtain the photons full parameterized trajectories in $\phi$ direction.

From \cref{120} we obtain the solution of the angular integral in the azimuthal direction as,
\begin{align}
    \hat{I}_{\phi}= \dfrac{1}{\sqrt{-u_{-}a^{2}}} \left(  \nu_{\theta} \Pi(\hat{u}_{+}; \hat{\psi}_{s} |\hat{k} ) - \nu_{\theta} \Pi(\hat{u}_{+}; \hat{\psi}_{\tau} |\hat{k} ) \right), \hat{\psi}_{\tau}= -\nu_{\theta} am\left[ \tau \sqrt{-\hat{u}_{-} a^{2}} - \nu_{\theta} F(\hat{\psi}_{s} |\hat{k}) |\hat{k}\right].\label{128}
\end{align}
Using  \cref{C72}, we obtain the radial integral $\hat{I}_x$ for escape orbits as,
\begin{align}
    \hat{I}_{x,E}= -\frac{2}{\hat{r}_{x,3}. \sqrt{\hat{r}_{4,2} \hat{r}_{3,1}}} \left( F\left(\hat{\varphi}_{E,o}|\hat{k}_{E}\right)-F\left(\hat{\varphi}_{E,s}|\hat{k}_{E}\right)+\frac{\hat{r}_{4,3}(\hat{\Pi}_{o}-\hat{\Pi}_{s})}{\hat{r}_{x,4}} \right).\label{175}
\end{align}
The parameters $\Pi_o$, $\Pi_s$ and $\varphi_{E,s}$ can be directly obtained from \cref{C73} by replacing $i$ with $s$ or $o$. We define $\varphi_{E,o}$ as,
\begin{align}
    \hat{\varphi}_{E,o}=\arcsin \sqrt{\frac{\hat{r}_{3,1}\hat{r}_{o,4}}{\hat{r}_{4,1}\hat{r}_{o,3}}}=am(\frac{ \tau+\nu_{r}\hat{g}_{E}F(\hat{\varphi}_{E,s} |\hat{k}_{E})}{\hat{g}_{E}} |\hat{k}_{E}).
\end{align}

For plunge orbits we utilize \cref{C75} which gives,
\begin{align}
    \hat{I}_{x,P}=\frac{((\hat{B}-\hat{A})) \left((\hat{\alpha} -\hat{\alpha}_{2}) (\hat{R}_{1,o}-\nu_{r}\hat{R}_{1,s})+\hat{\alpha}_{2}( F(\hat{\varphi}_{P,o} \mid \hat{k}_{P})-\nu_{r}F(\hat{\varphi}_{P,s} \mid \hat{k}_{P}))\right)}{\sqrt{\hat{A}\hat{B}}(\hat{r}_{2} \hat{B}+\hat{r}_{1}\hat{A} -\hat{A} \hat{r}_{x}-\hat{B} \hat{r}_{x})}. \label{179}
\end{align}
Similarly, the parameters $\hat{R}_{1,s},\hat{R}_{1,o},\hat{\varphi}_{P,s}$ are directly obtained from \cref{C76}-(\ref{C78}) by replacing $i$ with $s$ or $o$. We define $\hat{\varphi}_{P,o}$ as,
\begin{align}
  \hat{\varphi}_{P,o}=\arccos\left[ \frac{(\hat{A}-\hat{B})r_{o}+\hat{r}_{2} \hat{B}-\hat{r}_{1} A}{(\hat{A}+\hat{B})r_{o}-\hat{r}_{2} \hat{B}-\hat{r}_{1}\hat{A}} \right]=am\left[\tau \sqrt{\hat{A}\hat{B}}+\nu_{r}F(\hat{\varphi}_{P,s} \mid \hat{k}_{P}) \right].
\end{align}
\subsubsection{Motion in  $t$ direction}
The motion in $t$ direction also entails both the radial and angular integrals. The radial integrals is still $\hat{I}_x$ for which we have obtained the solutions for both escaping and plunge orbits in \cref{175} and (\ref{179}) respectively.
The angular integral in $t$ direction is obtained through \cref{121} which results in,
\begin{align}
    \hat{I}_{t} = \dfrac{\hat{u}_{+}}{\hat{k}\sqrt{-u_{-}a^{2}}} \left( \nu_{\theta} [F(\hat{\psi}_{s} |\hat{k}) -E(\hat{\psi}_{s} |\hat{k})] - \nu_{\theta} [F(\hat{\psi}_{\tau} |\hat{k}) -E(\hat{\psi}_{\tau} |\hat{k})] \right), \label{129}
\end{align}
where $\varphi_\tau$ has been defined in \cref{128}.

\Cref{129} and \cref{175} or (\ref{179}) can then be substituted in \cref{171} in order to get the $t$ component of a photon on an escape or plunge orbit.

\begin{figure}[ht]
  \centering
  \begin{tabular}{ccccc}
    \begin{subfigure}[b]{0.2\textwidth}
      \includegraphics[width=\textwidth]{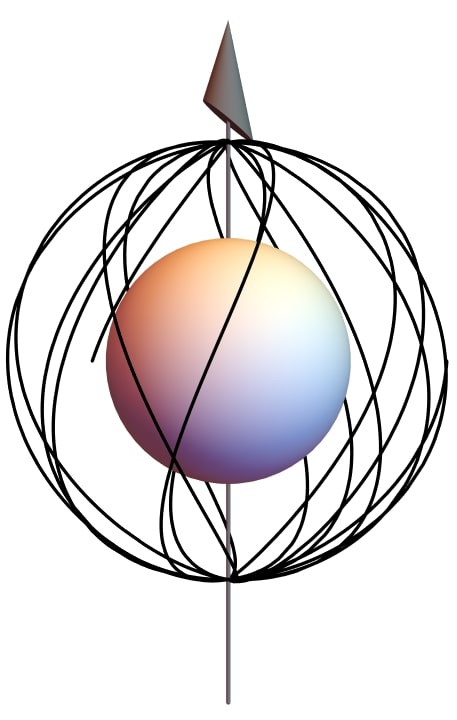}
      \caption{A bound photon orbit at $r=2.56M$ with $\hat{\lambda}=0$ and $\hat{\eta} = 23.3576792$.}
      \label{fig:zamorkds}
    \end{subfigure} & \quad \quad \quad \quad \quad
    \begin{subfigure}[b]{0.2\textwidth}
      \includegraphics[width=\textwidth]{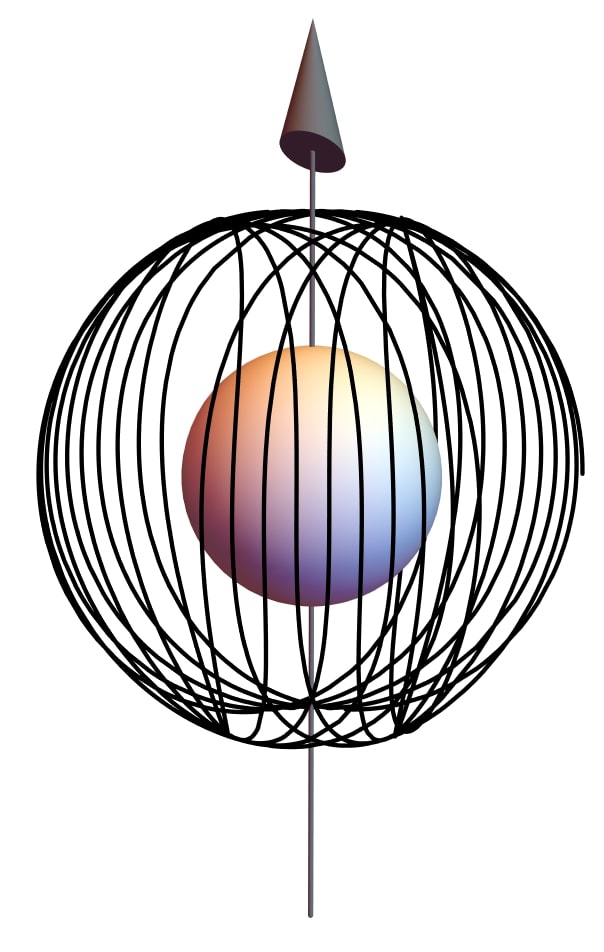}
      \caption{A bound photon orbit at $r=3.00M$ with $\hat{\lambda}=-1.8$ and $\hat{\eta}=27$. }
      \label{fig:etamaxrkds}
    \end{subfigure}
  \end{tabular}
   \begin{tabular}{ccc}
    \begin{subfigure}[b]{0.4\textwidth}
      \includegraphics[width=\textwidth]{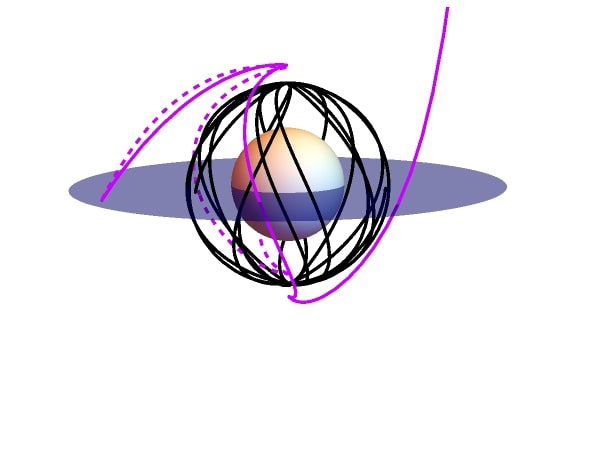}
      \caption{The escape orbit has $r_{4} = 2.749 M$, $\hat{\lambda}=0$ and $\hat{\eta} = 23.591256$ whereas the plunge orbit has $\hat{\lambda}=0$ and $\hat{\eta} = 23.124102$.}
      \label{fig:Rnbc1p1}
    \end{subfigure} & \quad \quad 
    \begin{subfigure}[b]{0.4\textwidth}
      \includegraphics[width=\textwidth]{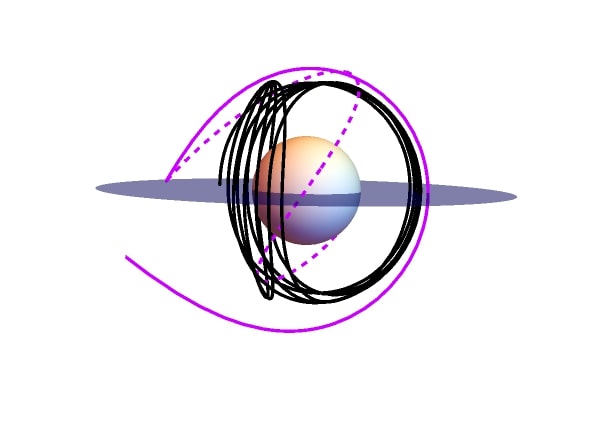}
      \caption{The escape orbit is such that $r_{4}=3.2206 M$,$\hat{\lambda}=-1.818$ and $\hat{\eta}=27.27$ while the plunge orbit has $\hat{\lambda}=-1.782$ and $\hat{\eta}=26.73$.}
      \label{fig:Rnbc2p2}
    \end{subfigure} 
  \end{tabular}
   \begin{tabular}{ccc}
    \begin{subfigure}[b]{0.4\textwidth}
      \includegraphics[width=\textwidth]{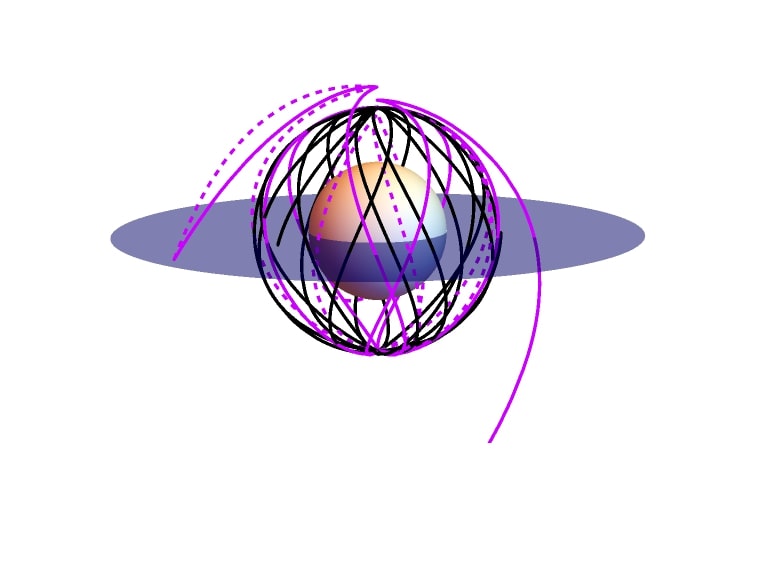}
      \caption{The escape orbit is such that $r_{4}=2.560052765 M$, $\hat{\lambda}=0$, $\hat{\eta} = 23.3576794$ while the plunge orbit has $\hat{\lambda}=0$ and $\hat{\eta} =23.3576789$.}
      \label{fig:Rnbc1p8}
    \end{subfigure} & \quad \quad 
    \begin{subfigure}[b]{0.4\textwidth}
      \includegraphics[width=\textwidth]{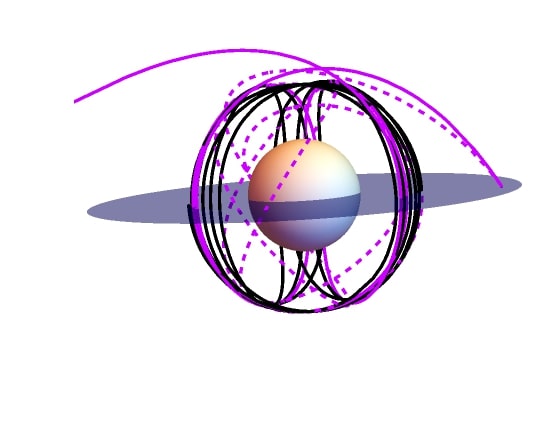}
      \caption{The escape orbit is such that $r_{4}=3.0000647 M$,$\hat{\lambda}=-1.800000018$ and $\hat{\eta}=27.00000027$. The plunge orbit is such that $\hat{\lambda}=-1.799999982$ and $\hat{\eta}=26.99999973$.}
      \label{fig:Rnbc2p8}
    \end{subfigure} 
  \end{tabular}
   \caption{ Illustration of bound, escape (solid magenta) and plunge (dotted magenta) orbits in \ac{rkds} spacetime with $a=0.9$ and $\Lambda=1.11 \times 10^{-52} m^{-2}$. The blue disk is an equatorial source. }
  \label{fig:rkdsillustration}
\end{figure}

Examples of bound, escape and plunge orbits in \ac{rkds} spacetime are as shown in \cref{fig:rkdsillustration}. The photons on escape and plunge geodesics in the first column are orbiting close to a bound orbit at $r=r_{zamo,RKdS}=2.56M$. Photons on the escape and plunge geodesics in the second column on the other hand are orbiting close to a bound orbit at $r=3.00M$. The photons have been emitted from a latitude $\theta_s=\pi/2$ and at a radial coordinate $r_s=5M$ with the observer located in the domain of outer communication. \\
In (\cref{fig:zamokds}, \cref{fig:etamaxkds}) and (\cref{fig:zamorkds}, \cref{fig:etamaxrkds}) we give examples of selected bound orbits around a \ac{kds} and \ac{rkds} black hole. Bound orbits are characterized by specific constants of motion that determine their behavior
and nearly bound orbits can be constructed by making deviations from these constants. For the rest of the diagrams in \cref{fig:kdsbillustration} and \cref{fig:rkdsillustration}, we examine escape and plunge orbits for different deviations on constants of motion of the bound orbits.
The photons on these orbits are all emitted from an equatorial source (blue disk on the diagrams) at a radial coordinate $r_s=5M$. A relatively large deviation from the constants of motion is made for the illustrations in (\cref{fig:nbc1p1}, \cref{fig:nbc2p2}) and (\cref{fig:Rnbc1p1}, \cref{fig:Rnbc2p2}). As can be seen, the emitted photons make fewer orbits close to the bound orbit before escaping or plunging into the black hole. In (\cref{fig:nbc1p8}, \cref{fig:nbc2p8}) and (\cref{fig:Rnbc1p8}, \cref{fig:Rnbc2p8}), we make a very small deviation on the constants of motion. Here, the photons make many orbits close to the bound orbit before escaping or plunging into the black hole. Decreasing the deviation even further would result in the photons making even more orbits close to the bound orbit before escaping or plunging into the black hole. The rate of exponential deviation experienced by the photons as they orbit close to the bound orbits is measured by the Lyapunov exponent. Additionally, it is evident that photons on escape orbits have higher constants of motion than the photons on the corresponding bound photon orbits, while plunge orbits have less. This can be attributed to the fact that the photon region  constitutes the boundary of escape and plunge orbits. These nearly bound orbits result in subrings of varying thickness, contingent upon the number of half orbits they complete and the Lyapunov exponent. 

\clearpage
\newpage
\clearpage
\newpage
 \section{Direct images, lensing rings and photon rings. } \label{lensingandphotonrings}

 In \cref{sectionII} we have gained insights into the general structure of bound and nearly bound photon orbits. We have seen that the number of orbits executed by photons before escaping or plunging into the black hole varies depending on the proximity of their trajectory to a bound orbit. Consequently, these diverse photon paths give rise to distinct images. The central objective of this section is to analyze these images in detail, namely direct images, lensing rings, and photon rings. 
 Direct images emerge from photons following weakly lensed orbits, while lensing rings result from photons completing one half orbit around the black hole. As for the photon ring, it arises when photons undergo two or more half orbits. It is important to note that in \cref{sectionII}, our primary focus was to gain insights into the general structure of bound and nearly bound orbits. Consequently, the assumption that a turning point had not yet been reached was efficient and simple to achieve that objective. However, in this section, our primary objective is to investigate the images of a disk surrounding the black hole through tracing the photon trajectories analytically from an observer to various source points on the disk. Therefore, it becomes essential to precisely account for the multiple times that photons intersects the disk in order to accurately obtain the corresponding images. Consequently, in this section we will utilize our solutions in \cref{sectionII} and incorporate the turning points into the solutions. Specifically, the solutions will be in terms of the number of angular turning points $m$ that the photons encounter as they orbit around the black hole. Moreover, we will see that this quantity is directly linked to the number of distinct images that can be generated for each specific value of $m$.
 
 We will consider locally static observers located in a frame of the form \cite{Li:2020drn}:
\begin{align}
\hat{e}_{t}=\sqrt{\frac{g_{\phi \phi}}{g_{t\phi}^2-g_{tt}g_{\phi\phi}}}\left(\partial_t-\frac{g_{t\phi}}{g_{\phi \phi}}\partial_\phi\right), \hat{e}_{r}=\frac{1}{\sqrt{g_{r r}}}\partial_r,\hat{e}_{\theta}=\frac{1}{\sqrt{g_{\theta \theta}}}\partial_\theta,\hat{e}_{\phi}=\frac{1}{\sqrt{g_{\phi \phi}}}\partial_\phi, \label{pr1}
\end{align}
where $\hat{e}_{\mu}$ are basis vectors, with $\hat{e}_{t}$ being the timelike vector within the domain of outer communication. This vector can be assigned as the four-velocity of the observer. The remaining three space-like vectors correspond to the observer's motion in the three spatial dimensions. Additionally, the observer has zero angular momentum, since $\hat{e}_{t} \partial_\phi=0$, hence the frame defined by \cref{pr1} is also referred to as the zero angular momentum frame.

The four-momentum $p^\mu$ can be projected onto the basis defined by \cref{pr1}. This projection gives the quantities that can be directly measured by the observer in their local frame \cite{Li:2020drn},
\begin{align}
    p^t=-p_{\mu }\hat{e}_{t}^\mu,  \quad  p^\mu=-p_{\mu }\hat{e}_{i}^\mu,\quad i=r, \theta,\phi. \label{pr2}
\end{align}
 \begin{figure}
     \centering
		\includegraphics[width=0.7\linewidth]{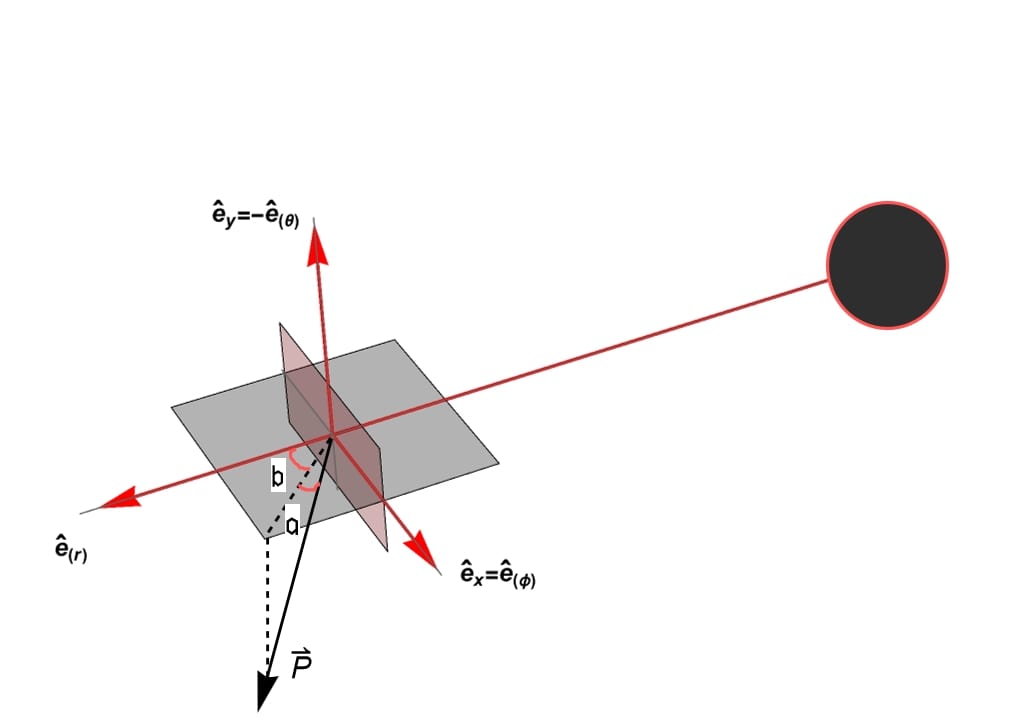}
		\caption{ The projection of the photon's four-momentum onto the observer's frame and the solid angles $(\mathfrak{a},\mathfrak{b})$ }
		\label{fig:observer}
 \end{figure}
The observation angles $(\mathfrak{a},\mathfrak{b})$, see \cref{fig:observer}, are defined as (\cite{Li:2020drn},\cite{cunha2018shadows}),
\begin{align}
    p^r=|\vec{P}|\cos (\mathfrak{a})\cos (\mathfrak{b}),\quad p^{\theta }=|\vec{P}|\sin (\mathfrak{a}), \quad p^{\phi }=|\vec{P}|\cos (\mathfrak{a})\sin (\mathfrak{b}). 
    \label{pr3} 
\end{align}
Where, $|\vec{P}|^2=\left(p^{\phi }\right)^2+\left(p^{\theta }\right)^2+\left(p^r\right)^2$ for which $|\vec{P}|=p^t$ owing to the photon being massless.

The Cartesian coordinates $(x,y)$ that represent the apparent position of an observer located at a distance $r_o$ from a black hole on the plane of the sky are mathematically defined as \cite{Bardeen:1973tla},
\begin{align}
 x=-r_o \cos (\mathfrak{a})\sin (\mathfrak{b}), \quad   y=r_o \sin (\mathfrak{a}). \label{pr4}
\end{align}
Making use of \cref{pr3}, we have that, $\cos(\mathfrak{a})\sin(\mathfrak{b}) = p^\phi/p^t$ and $\sin(\mathfrak{a}) = p^\theta/p^t$ which when substituted into \cref{pr4} gives $x = -r_o p^\phi/p^t$ and $y = r_o p^\theta/p^t$. For the case of \ac{kds} we make use of \cref{22} and (\ref{23}) to obtain,
\begin{align}
    x=\frac{r_o \left(a \Delta _{\theta } \left(a^2-a \lambda +r^2\right)+\Delta _r \left(\lambda  \csc ^2(\theta )-a\right)\right)}{a \Delta _r \left(a \sin ^2(\theta )-\lambda \right)-\left(a^2+r^2\right) \Delta _{\theta } \left(a^2-a \lambda +r^2\right)}, \label{pr5}\\
    y=\frac{r_o \sqrt{\Delta _{\theta } \left(L^2 (a-\lambda )^2+\eta \right)-L^2 (\lambda  \csc (\theta )-a \sin (\theta ))^2}}{\mathbb{T}}. \label{pr6}
\end{align}
Here, we have defined,
\begin{align}
    \mathbb{T}=L^2 \left(\frac{\left(a^2+r^2\right) \left(a^2-a \lambda +r^2\right)}{\Delta _r}+\frac{a \left(\lambda -a \sin ^2(\theta )\right)}{\Delta _{\theta }}\right).
\end{align}
Inserting equations (\ref{44}) and (\ref{45}) into (\ref{pr5}) and (\ref{pr6}), gives the critical curve of \ac{kds} spacetime, which forms a boundary on the image plane delineating the region of escape and plunge orbits. Meaning that photons that escape the black hole region have points outside the critical curve on an observer's image plane, whereas photons that fall into the black hole will have points inside the critical curve. It is convenient to express $\lambda$ and $\eta$ in terms of $x$ and $y$ to obtain such points.

By inverting (\ref{pr5}) and (\ref{pr6}), $\eta$ and $\lambda$, are then expressed in terms of the Cartesian coordinates ($x,y$) as,
\begin{align}
    \lambda=\frac{\sin ^2(\theta ) \left(\left(a^2+r^2\right) \Delta _{\theta } \left(x \left(a^2+r^2\right)+a r_o\right)-a \Delta _r \left(a x \sin ^2(\theta )+r_o\right)\right)}{a \Delta _{\theta } \sin ^2(\theta ) \left(x \left(a^2+r^2\right)+a r_o\right)-\Delta _r \left(a x \sin ^2(\theta )+r_o\right)},\label{4mom4} \\
    \eta= \frac{L^2 r_o^2 \left((\lambda  \csc (\theta )-a \sin (\theta ))^2-(a-\lambda )^2 \Delta _{\theta }\right)+\mathbb{T}^2 y^2}{\Delta _{\theta } r_o^2}. \label{4mom5}
\end{align}
Performing the same calculation for \ac{rkds} spacetime, we obtain $\hat{\lambda}$ and $\hat{\eta}$ in terms of the Cartesian coordinates as,
\begin{align}
  \hat{\lambda} = \frac{\sin ^2(\theta ) \left(a r_o \left(a^2-\Delta +r^2\right)+x \left(\left(a^2+r^2\right)^2-a^2 \Delta  \sin ^2(\theta )\right)\right)}{r_o \left(a^2 \sin ^2(\theta )-\Delta \right)+a x \sin ^2(\theta ) \left(a^2-\Delta +r^2\right)}, \label{pr82} \\
  \hat{\eta}=-a^2 \cos ^2(\theta )+\frac{y^2 \left(2 a^4+a^2 \Delta  \cos (2 \theta )-a^2 \Delta -2 a \lambda  \left(a^2-\Delta +r^2\right)+4 a^2 r^2+2 r^4\right)^2}{4 \Delta ^2 r_o^2}+\lambda ^2 \cot ^2(\theta ). \label{pr82b}
\end{align}
 
Given a locally static observer located at ($r_o, \theta_o,\phi_o,t_o$), our focus lies in finding photon orbits connecting $(x,y)$ on the observers screen to distinct source points ($r_s, \theta_s,\phi_s,t_s$) on a disk around the black hole. This task involves solving the null geodesic equations, starting from an observer $o$ to the source $s$. Since our goal is to do this analytically, we will directly use the solutions that we obtained in \cref{sectionII} and modify the integrals by simply swapping the roles of the source and observer.

Furthermore, we will express the solutions in terms of the number of turning points that the photons encounter in the $\theta$ direction. We deem it useful to include them so as to precisely track the emission paths of photons and enable the generation of accurate source images. Besides, as we proceed with our analysis, we will see the close connection between these turning points and the multiplicity of images that can be produced for a given source. 

In the case that the radial potential has four real roots, the radial component of a trajectory traced from an observer into the \ac{kds} black hole geometry is given by,
\begin{align}
     r_{s}^{E}=\frac{r_3 \left(r_1-r_4\right) \text{sn}^2\left(\left.\frac{\mathbb{I}_\theta }{g_{E}}+\nu_{r} F(\varphi_{E,o}|k_{E})\right|k_{E}\right)+\left(r_3-r_1\right) r_4}{(r_1-r_4) \text{sn}^2\left(\left.\frac{\mathbb{I}_\theta }{g_{E}}+\nu_{r} F(\varphi_{E,o}|k_{E})\right|k_{E}\right)-r_{1}+r_{3}}, \label{pr14}
\end{align}
while for the case of two real and two complex roots, the radial component of the trajectory will be given by ,
\begin{align}
	    r_{s}^{P}=\frac{(r_{2} B+A r_{1}) \text{cn}\left(\left.\nu_{r} F(\varphi_{P,o}|k_{P}) +\frac{\mathbb{I}_\theta }{g_{P}}\right|k_{P}\right)+r_{2} B-A r_{1}}{(A+B) \text{cn}\left(\left.\nu_{r} F(\varphi_{P,o}|k_{P}) +\frac{\mathbb{I}_\theta }{g_{P}}\right|k_{P}\right)-A+B}. \label{pr15}
	\end{align}
The parameter $\mathbb{I}_\theta$ has been defined in \cref{pr10} and is in terms of the number of turning points $m$ in the $\theta$ direction. Further, all other parameters that depend on $\eta$ and $\lambda$ will be calculated through \cref{4mom4} and (\ref{4mom5}). The $\phi$ and $t$ components of the trajectory will be obtained by swapping $s$ and $o$ in \cref{156} and \cref{157}. Furthermore, the angular integrals as a function of the number of angular turning points are expressed in \cref{pr10}-(\ref{pr13}). 

 Likewise, in \ac{rkds}, the radial component for the two cases will be given by,
 \begin{align}
      \hat{r}_{s}^{E}=\frac{\hat{r}_{3} \hat{r}_{1,4} \text{sn}^2\left(\left.\frac{\hat{\mathbb{I}}_\theta }{\hat{g}_{E}}+\nu_{r} F(\hat{\varphi}_{E,o}|\hat{k}_{E})\right|\hat{k}_{E}\right)+\hat{r}_{4} \hat{r}_{3,1}}{\hat{r}_{1,4} \text{sn}^2\left(\left.\frac{\hat{\mathbb{I}}_\theta }{\hat{g}_{E}}+\nu_{r} F(\hat{\varphi}_{E,o}|\hat{k}_{E})\right|\hat{k}_{E}\right)-\hat{r}_{1}+\hat{r}_{3}},\label{pr16} \\
  \hat{r}_{s}^{P}=\frac{(\hat{r}_{2} \hat{B}+\hat{A} \hat{r}_{1}) \text{cn}\left(\left.\nu_{r} F(\hat{\varphi}_{P,o}|\hat{k}_{P}) +\hat{\mathbb{I}}_\theta\sqrt{\hat{A}\hat{B}}\right|\hat{k}_{P}\right)+\hat{r}_{2} \hat{B}-\hat{A} \hat{r}_{1}}{(\hat{A}+\hat{B}) \text{cn}\left(\left.\nu_{r} F(\hat{\varphi}_{P,o}|\hat{k}_{P}) +\hat{\mathbb{I}}_\theta\sqrt{\hat{A}\hat{B}}\right|\hat{k}_{P}\right)-\hat{A}+\hat{B}}. \label{pr17}
 \end{align}
$\hat{\mathbb{I}}_\theta$ has been defined in \cref{d5} and all the parameters depending on $\hat{\lambda}$ and $\hat{\eta}$ are to be obtained through \cref{pr82} and (\ref{pr82b}). In the same way, the $t$ and $\phi$ component of the trajectory is obtained by swapping $s$ and $o$ in \cref{170} and (\ref{171}). The respective angular integrals in terms of the photon's number of angular turning points are expressed in \cref{d5}-(\ref{d7}).

 The $r$, $\phi$, and $t$ components will play a crucial role in mapping observer directions to spacetime points where light rays intersect a disk around a black hole, hence, providing a means of generating and analyzing images. These functions have been referred to as transfer functions \cite{gralla2020lensing}, \cite{Cardenas-Avendano:2022csp}. We will focus on equatorial sources and that the photons have a radial coordinate turning point, which implies that $\theta_s=\pi/2$ and $\nu_r=-1$. Furthermore, $\mathbb{I}_\theta/\hat{\mathbb{I}}_\theta$, has a discontinuity along the $x$ axis thus rendering the image curves discontinuous. To remedy this, we reparametrize $m=n+H(y)$, where $H$ represents the Heaviside function, see also [\cite{gralla2020lensing}, \cite{Cardenas-Avendano:2022csp}]. This transformation eliminates the aforementioned discontinuity. 
 
\begin{figure}[ht]
  \centering
  \begin{tabular}{ccc}
    \begin{subfigure}[b]{0.3\textwidth}
      \includegraphics[width=\textwidth]{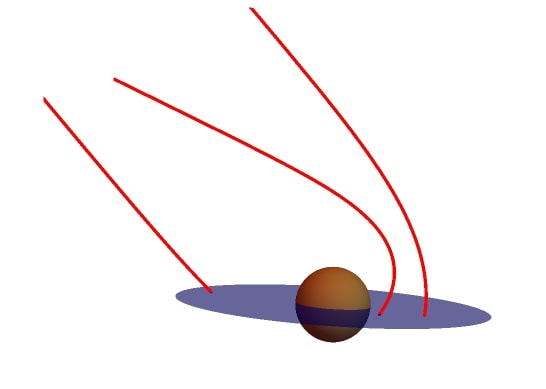}
      \caption{$n=0$: These photons have made zero half orbits and are weakly lensed. They terminate on the front side of the disk resulting in one image of the front side of the disk.}
      \label{fig:n0photons}
    \end{subfigure} & \quad \quad \quad \quad
    \begin{subfigure}[b]{0.3\textwidth}
      \includegraphics[width=\textwidth]{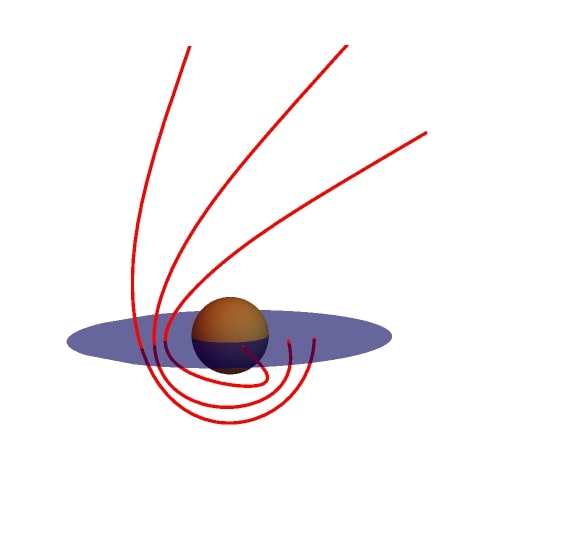}
      \caption{$n=1$: The photons on this diagram have made one half orbit around the black hole. Their first equatorial crossing is from the front of the disk, while the second is from the back, resulting in two images of the equatorial disk being detected by an observer (the front and back image).}
      \label{fig:n1photons}
    \end{subfigure} 
  \end{tabular}
   \begin{tabular}{ccc}
    \begin{subfigure}[b]{0.4\textwidth}
      \includegraphics[width=\textwidth]{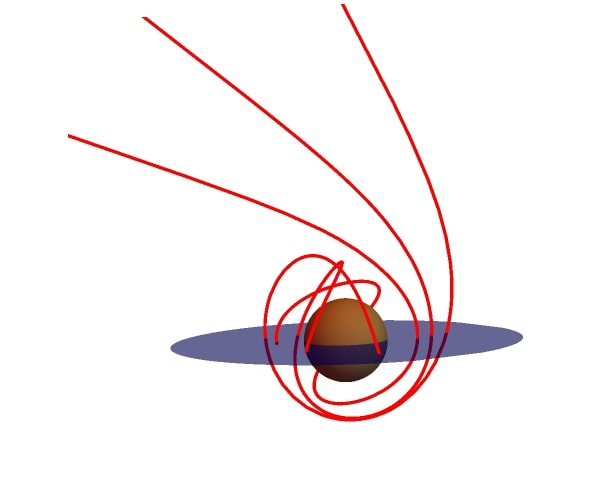}
      \caption{$n=2$: The photons on these orbits have made two half orbits around the black hole. Their first equatorial crossing is from the front of the disk, the second is from the back of the disk and finally the photons terminate on the front side of the disk. Thus an observer will detect two images of the front of the disk and one image of the back side.}
      \label{fig:n2photons}
    \end{subfigure}
  \end{tabular}
   \caption{
Illustration of the equatorial crossings of the photons traced from an observer back into the black hole geometry for even and odd values of $n$. It is evident that for every $n$, the photon makes contact with the equatorial disk precisely $n+1$ times. As such, each value of $n$ corresponds to $n+1$ distinct images of the equatorial disk.}
  \label{fig:nphotons}
\end{figure}

 For even values of $n$, the photons cross the disk from the front and for odd values of $n$, on the other hand, the photons cross the disk from the back. The front of the disk is the side of the disk facing the observer while the back is the side opposite to the observer.  An illustration of this is as we show in \cref{fig:nphotons}. Therefore, as $n$ varies, the photons alternate between showing the front and back side of the disk. 

\begin{figure}[ht]
  \centering
  \begin{tabular}{ccc}
     \begin{subfigure}[b]{0.4\textwidth}
      \includegraphics[width=\textwidth]{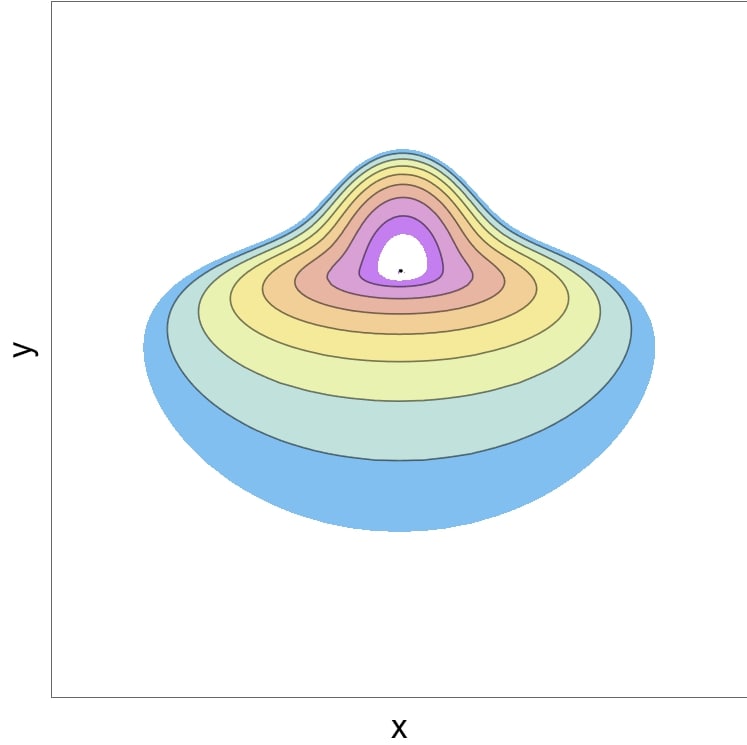}
      \caption{Direct ($n=0$) image of the equatorial disk around a \ac{kds} black hole.}
      \label{fig:m080kds}
    \end{subfigure}  & \quad \quad \quad \quad
     \begin{subfigure}[b]{0.4\textwidth}
      \includegraphics[width=\textwidth]{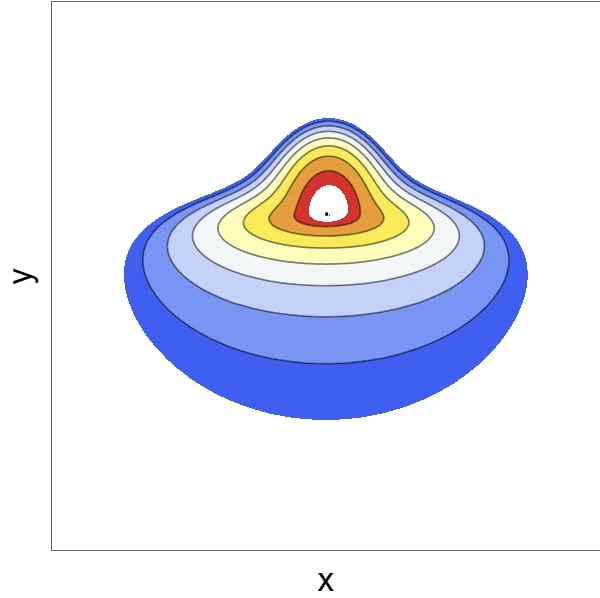}
      \caption{Direct ($n=0$) image of the equatorial disk around a \ac{rkds} black hole.}
      \label{fig:m080rkds}
    \end{subfigure} 
  \end{tabular}
  \caption{ The observer is at $r_o=50$, $\theta_o=80^\circ$ with black hole spin $a=0.5$ and $\Lambda=1.11\times 10^{-52}m^{-2}$. Each contour represents a constant radial coordinate $r_s$ increasing from the event horizon outwards. The photons generating these images have made zero half orbits around the black hole and are weakly lensed.}
   \caption*{ }
  \label{fig:directimages}
\end{figure}

The $n=0$ image will be a direct image of the equatorial disk's front side. In \cref{fig:directimages}, we show direct images ($n=0$) of an equatorial disk around a \ac{kds} and \ac{rkds} black hole. Each of the contours on these plots is a ring with a constant Boyer-Lindquist radial coordinate $r_s$, and this produces a single image for each emission from the equatorial disk. In addition, $r_s$ increases monotonically outwards, with the central region in the plots representing the region beneath the event horizon. Direct images are strongly sensitive to the astrophysical source profile surrounding the black hole since they are generated by photons on weakly lensed orbits. 

\begin{figure}[ht]
  \centering
  \begin{tabular}{ccc}
     \begin{subfigure}[b]{0.4\textwidth}
      \includegraphics[width=\textwidth]{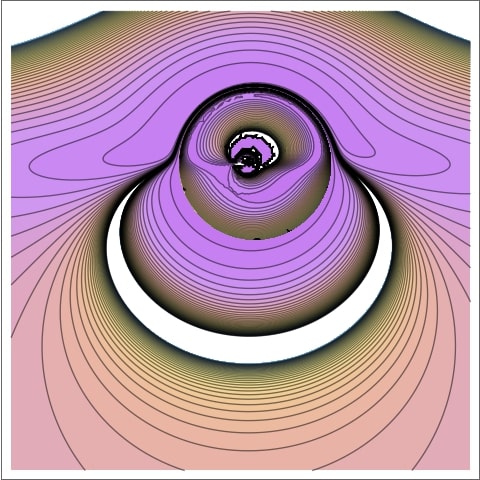}
      \caption{Images of the equatorial disk around a \ac{kds} black hole for $n=1$.}
      \label{fig:n1kds0}
    \end{subfigure}  & \quad \quad \quad \quad
     \begin{subfigure}[b]{0.4\textwidth}
      \includegraphics[width=\textwidth]{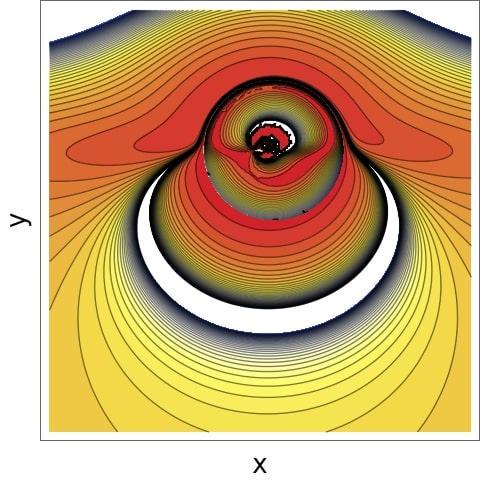}
      \caption{Images of the equatorial disk around a \ac{rkds} black hole for $n=1$.}
      \label{fig:n1rkds0}
    \end{subfigure} 
  \end{tabular}
   \begin{tabular}{ccc}
     \begin{subfigure}[b]{0.4\textwidth}
      \includegraphics[width=\textwidth]{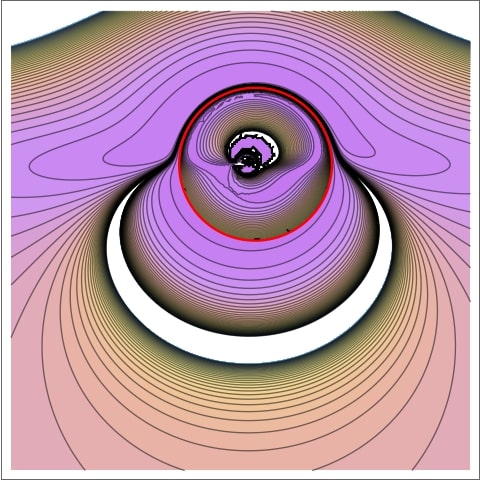}
      \caption{Images of the equatorial disk around a \ac{kds} black hole merged with the critical curve (the red solid curve).}
      \label{fig:n1kds}
    \end{subfigure}  & \quad \quad \quad \quad
     \begin{subfigure}[b]{0.4\textwidth}
      \includegraphics[width=\textwidth]{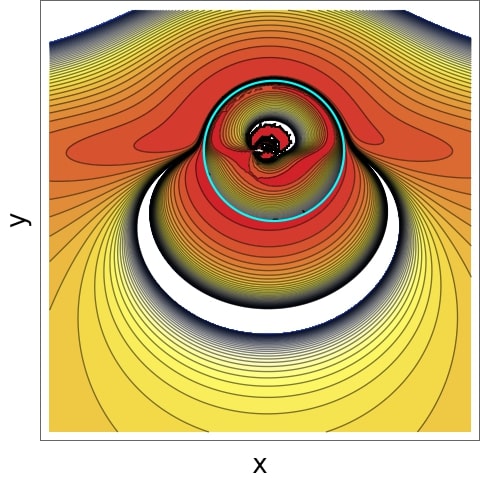}
      \caption{Images of the equatorial disk around a \ac{rkds} black hole merged with the critical curve (the cyan solid curve).}
      \label{fig:n1rkds}
    \end{subfigure} 
  \end{tabular}
  \caption{ The observer is at $r_o=50$, $\theta_o=80^\circ$ with black hole spin $a=0.5$ and $\Lambda=1.11\times 10^{-52}m^{-2}$.}
   \caption*{ }
  \label{fig:n1images}
\end{figure}

When $n=1$, photons cross the equatorial plane twice, resulting in two images for each source on the equatorial disk. These images will show the disk's front side and backside. The second image will be more lensed in comparison to the first because it appears after the photons have completed an additional half orbit around the black hole. \Cref{fig:n1images} is an illustration of $n=1$ images. It is apparent that in this instance, two regions emerge, implying that there are two images for each source point on the equatorial disk. The first image has been wrapped on itself, and the second lensed image has been superposed on this first image. The second image is oftentimes referred to as the lensing ring \cite{ma2022black}, \cite{Hou:2022gge}. When the images are merged with the critical curve, the lensed ring gets closer to the critical curve slightly. The region beneath the critical curve corresponds to the captured photons, which are then traced back to the event horizon. Hence, the contours inside the critical curve are an infinite unfolding of the event horizon. Contours outside the critical curve correspond to sources outside the photon region, whereas the critical curve corresponds to points within the photon region. 

\begin{figure}[ht]
  \centering
  \begin{tabular}{ccc}
     \begin{subfigure}[b]{0.4\textwidth}
      \includegraphics[width=\textwidth]{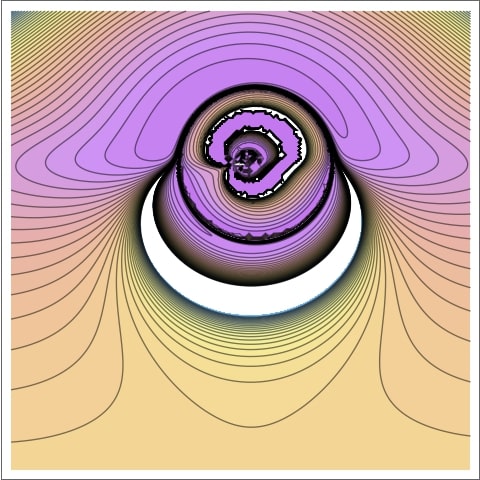}
      \caption{$n=2$ images of the equatorial disk around a \ac{kds} black hole.}
      \label{fig:n2kds0}
    \end{subfigure}  & \quad \quad \quad \quad
     \begin{subfigure}[b]{0.4\textwidth}
      \includegraphics[width=\textwidth]{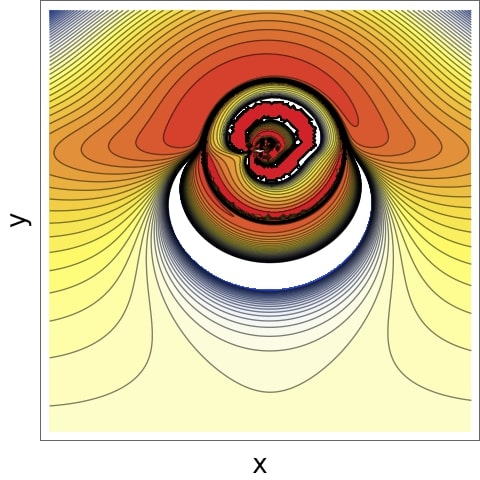}
      \caption{$n=2$ images of the equatorial disk around a \ac{rkds} black hole.}
      \label{fig:n2rkds0}
    \end{subfigure} 
  \end{tabular}
   \begin{tabular}{ccc}
     \begin{subfigure}[b]{0.4\textwidth}
      \includegraphics[width=\textwidth]{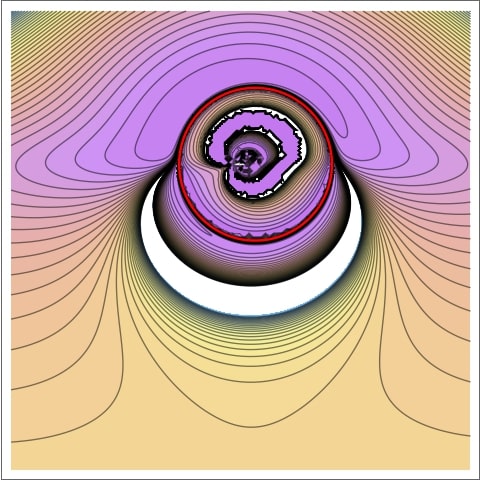}
      \caption{$n=2$ images of the equatorial disk around a \ac{kds} black hole merged with the critical curve.}
      \label{fig:n2kds}
    \end{subfigure}  & \quad \quad \quad \quad
     \begin{subfigure}[b]{0.4\textwidth}
      \includegraphics[width=\textwidth]{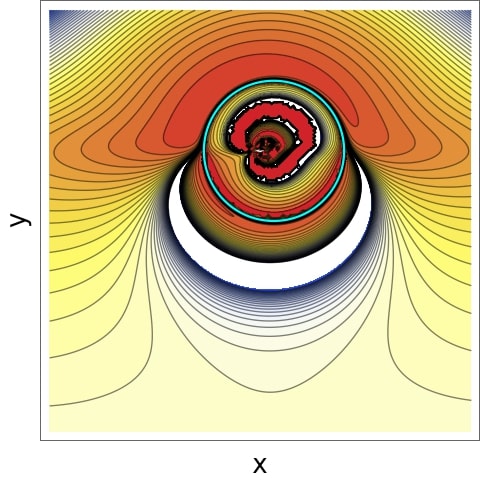}
      \caption{$n=2$ images of the equatorial disk around a \ac{rkds} black hole merged with the critical curve.}
      \label{fig:n2rkds}
    \end{subfigure} 
  \end{tabular}
  \caption{ The observer is at $r_o=50$, $\theta_o=80^\circ$ with black hole spin $a=0.5$ and $\Lambda=1.11\times 10^{-52}m^{-2}$.}
   \caption*{ }
  \label{fig:n2images}
\end{figure}

For $n=2$, the photons cross the equatorial plane thrice hence there will be a nested sequence of three images for each source. The third image will be extremely lensed as compared to the first and second. Furthermore, the third image will be superposed on the first and second image and it will move even more closer to the critical curve. This image and other higher order images constitute the photon ring. In \cref{fig:n2images}, we illustrate images for $n=2$ in a \ac{kds} and \ac{rkds} black hole. It is now clear from \cref{fig:n2kds0} and (\ref{fig:n2rkds0}) that there is an extra black curve superposed within these images as compared to the $n=1$ images. The extremely lensed region (black curves) is the third image in both plots. Merging these images with the critical curve shows that the third image is very close to the critical curve and tends to take the shape of the critical curve as compared to the first and second images. 

\begin{figure}[ht]
  \centering
  \begin{tabular}{ccc}
     \begin{subfigure}[b]{0.4\textwidth}
      \includegraphics[width=\textwidth]{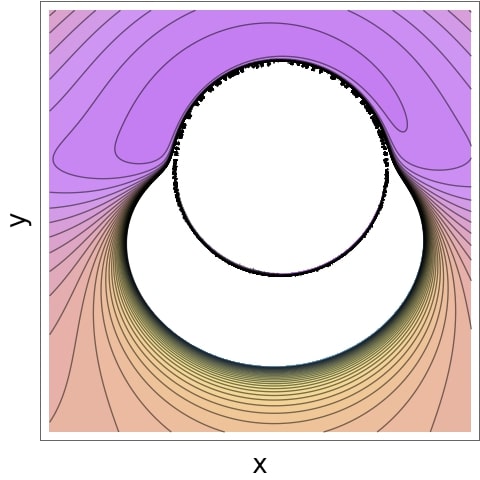}
      \caption{Two images (the colored region and the thin black curve) of the front side of the equatorial disk}
      \label{fig:m280kds0}
    \end{subfigure}  & \quad \quad \quad \quad
     \begin{subfigure}[b]{0.4\textwidth}
      \includegraphics[width=\textwidth]{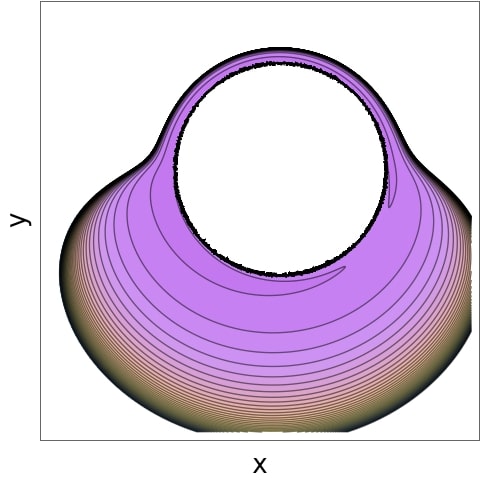}
      \caption{Two images (the colored region and the thin black curve in the inner boundary) for the backside of the disk}
      \label{fig:m380kds0}
    \end{subfigure} 
  \end{tabular}
   \begin{tabular}{ccc}
     \begin{subfigure}[b]{0.4\textwidth}
      \includegraphics[width=\textwidth]{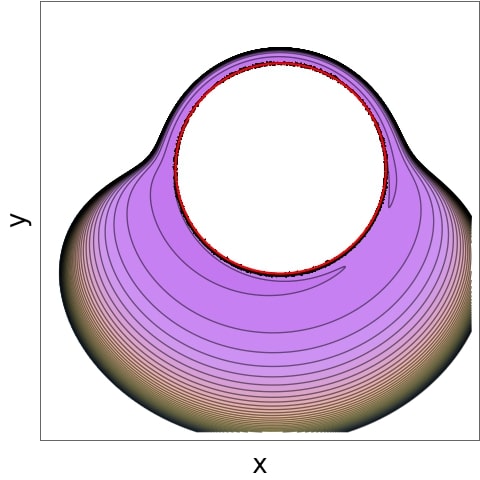}
      \caption{Merging the forth image with the critical curve.}
      \label{fig:m380kds}
    \end{subfigure} 
  \end{tabular}
  \caption{Illustration of $n=3$ images for a \ac{kds} equatorial disk. The observer is at $r_o=50$, $\theta_o=80^\circ$ with black hole spin $a=0.5$ and $\Lambda=1.11\times 10^{-52}m^{-2}$.}
   \caption*{ }
  \label{fig:m3kds}
\end{figure}

\begin{figure}[ht]
  \centering
  \begin{tabular}{ccc}
     \begin{subfigure}[b]{0.4\textwidth}
      \includegraphics[width=\textwidth]{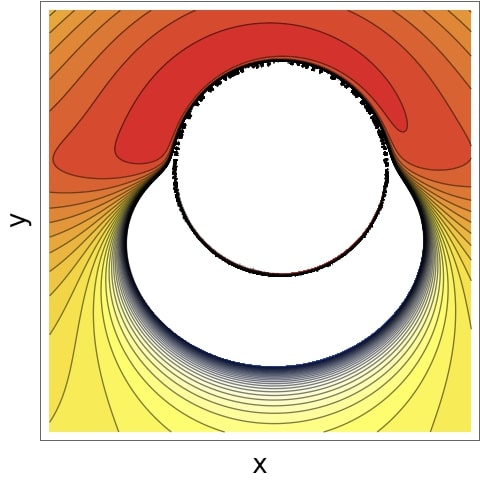}
      \caption{Two images (the colored region and the thin black curve) of the front side of the equatorial disk}
      \label{fig:m280rkds0}
    \end{subfigure}  & \quad \quad \quad \quad
     \begin{subfigure}[b]{0.4\textwidth}
      \includegraphics[width=\textwidth]{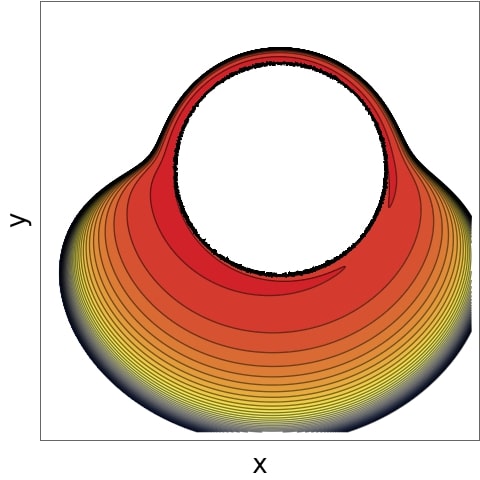}
      \caption{Two images (the colored region and the thin black curve in the inner boundary) for the backside of the disk}
      \label{fig:m380rkds0}
    \end{subfigure} 
  \end{tabular}
   \begin{tabular}{ccc}
     \begin{subfigure}[b]{0.4\textwidth}
      \includegraphics[width=\textwidth]{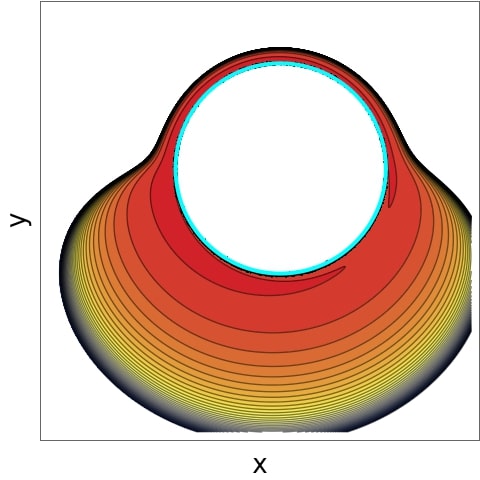}
      \caption{Merging the forth image with the critical curve.}
      \label{fig:m380rkds}
    \end{subfigure} 
  \end{tabular}
  \caption{Illustration of $n=3$ images for a \ac{rkds} equatorial disk. The observer is at $r_o=50$, $\theta_o=80^\circ$ with black hole spin $a=0.5$ and $\Lambda=1.11\times 10^{-52}m^{-2}$.}
   \caption*{ }
  \label{fig:m3rkds}
\end{figure}

Finally, we analyze the images for $n=3$, which will be a sequence of four images, two from the front and two from the backside of the disk. 
Because the extremely lensed images are very close to each other and hard to distinguish in a single plot, we have chosen to separate the front and backside images as illustrated in \cref{fig:m3kds} and \cref{fig:m3rkds}. Furthermore, we have extracted out the the region of photon capture to make our illustration clear. The two images on the front are related to the photons' first and third equatorial crossings, whilst the two images on the back correspond to the photons' second and fourth equatorial crossings. Thus, there will be four images for a particular source. Notably, these higher-order images exhibit extreme lensing, manifesting as thin black curves in the plots. We already saw in the $n=2$ images that the third image was extremely close to the critical curve. For $n=3$ we also merge the forth image with the critical curves. It is clear that the fourth image merges with the critical curve, resulting in an indistinguishable appearance in both the \ac{kds} and \ac{rkds} spacetimes (see Figures \ref{fig:m380kds} and \ref{fig:m380rkds}). It is evident that the images composing the photon ring exhibit a striking resemblance to the critical curve.

Concluding our analysis with $n=3$, we state that the pattern remains consistent as $n$ increases: each increment of $n$ generates $n+1$ images, all of which exhibit extreme lensing effects. Furthermore, the higher-order images continue to appear in close proximity to the critical curve of their respective spacetimes. 

\begin{figure}[ht]
  \centering
  \begin{tabular}{ccc}
     \begin{subfigure}[b]{0.4\textwidth}
      \includegraphics[width=\textwidth]{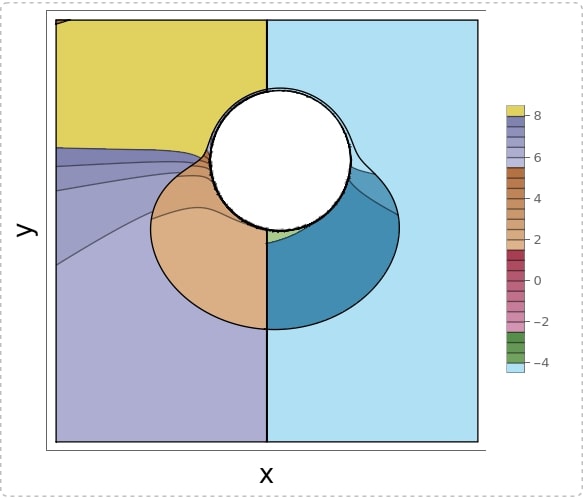}
      \caption{Change in azimuthal angle in images of a \ac{kds} black hole equatorial disk.}
      \label{fig:azm}
    \end{subfigure}  & \quad \quad \quad \quad
     \begin{subfigure}[b]{0.4\textwidth}
      \includegraphics[width=\textwidth]{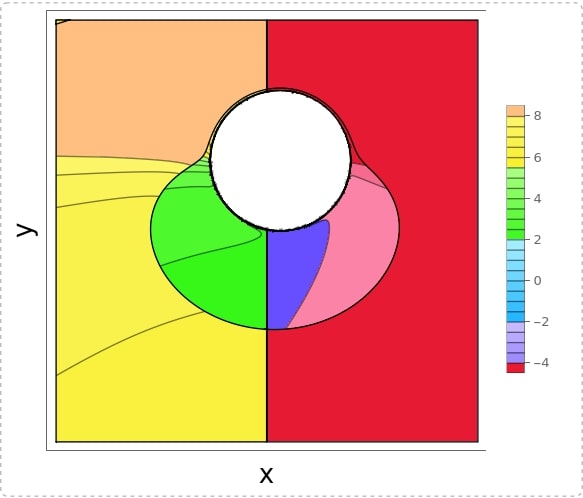}
      \caption{Change in azimuthal angle in images of a \ac{rkds} black hole equatorial disk.}
      \label{fig:Razm}
    \end{subfigure} 
  \end{tabular}
  \caption{ The observer is at $r_o=50$, $\theta_o=80^\circ$ with black hole spin $a=0.5$ and $\Lambda=1.11\times 10^{-52}m^{-2}$.}
   \caption*{ }
  \label{fig:azmchange}
\end{figure}

Moreover, as $n$ increases, the photons undergo a change in azimuthal angle. Consequently, subsequent images exhibit a rotation. In \cref{fig:azmchange}, we illustrate the azimuthal angle shift for images corresponding to $n=1$. Additionally, we provide a color palette indicating the measured radians for each segment. Remarkably, in both spacetimes, it is evident that the variation in azimuthal angle across the outer region (representing the first image) differs from that observed in the second image. Hence, the second image manifests a distinct rotation relative to the first image. To enhance clarity, we have extracted out the region of photon capture in our illustration. 

In summary, this section has provided a comprehensive analysis of images for with equatorial disks around \ac{kds} and \ac{rkds} black holes. Subsequent images are extremely lensed and approach the critical curve exponentially $(\propto \exp(n \gamma))$. $\gamma$ is the Lyapunov exponent and will be analyzed in details in the next section. Furthermore, through the change in azimuthal angle, subsequent images undergo a rotation. The amount of this rotation due to change in azimuthal angle is calculated by $(\Delta \phi_{m+1}-\Delta \phi_{m})$. Additionally, the amount of delay in detection of subsequent images is also obtained through $(\Delta t_{m+1}-\Delta t_{m})$. We have obtained explicit forms of ($\Delta \phi_{m+1}-\Delta \phi_{m}$) and ($\Delta t_{m+1}-\Delta t_{m}$) in \cref{subsimages}. In the next section, we shall see that they are related to the critical parameters controlling the structure of the photon ring.


\section{Analysis of Critical Parameters}\label{sectionIII}
In \cref{lensingandphotonrings}, we have analyzed direct images, lensed rings and photon rings of the equatorial disks around a \ac{kds} and \ac{rkds} blackhole. By varying the number of half orbits undertaken by the photons around the black hole, we have observed a nested sequence of images, each undergoing significant lensing effects. Notably, the images comprising the photon ring exhibit a remarkable proximity and resemblance to the critical curve. Additionally, we have observed the change in the azimuthal angle between the first and second images. Furthermore, as the photons make additional half orbits around the black hole, subsequent images exhibit a delay in detection. Consequently, the photon ring encompasses a series of subsequently demagnified images that appear rotated and are detected at different time intervals. Our focus in this section is to examine the critical parameters governing this demagnification, rotation, and delay in detection of successive images. These critical parameters encompass the Lyapunov exponent, azimuthal angle parameter, and time delay and are evaluated on the radial coordinate of bound photon orbits.
\subsection{Time Delay}
The critical parameter time delay is defined as the amount of time photons take to complete half orbits around the black hole. This parameter controls the arrival time of subsequent images. Form \cref{zeta1} and \cref{zeta2} we obtain the relations for time delay in \ac{kds} and \ac{rkds} as,
\begin{align}
\zeta=\left(\dfrac{L^{2}}{\Delta_{r}}((r^{2}+a^{2})^{2}-a\lambda(a^{2}+r^{2}))\Gamma_{\theta}+a^2 L^2 \Gamma_{t}-aL^2(a-\lambda)\Gamma_{\phi} \right), \label{94}\\
    \hat{\zeta}= \left( \frac{(r^{2}+a^{2})(r^{2}+a^{2}-a \hat{\lambda})}{\hat{\Delta}} +a \hat{\lambda}-a^{2} \right) \hat{\Gamma}_{\theta } +a^{2} \hat{\Gamma}_{t}, \label{133}
\end{align}
Where $\Gamma_{\phi}, \Gamma_t,  \hat{\Gamma}_{t}$ have been derived in \cref{202}-(\ref{205}) while $\hat{\Gamma}_\theta$ and $\hat{\Gamma}_t$ have been derived in \cref{124a} and (\ref{124b}), respectively.

Upon comparing these equations with \cref{tkds} and \cref{trkds}, a clear correspondence emerges: $ \Delta t_{m+1}-\Delta t_{m} \approx \zeta$ and $ \hat{\Delta}\phi_{m+1}-\hat{\Delta}\phi_{m} \approx \hat{\zeta}$. Thus the delay in the detection of successive images in the photon ring is approximately equal to the time delay evaluated on radial points within the photon region. We then proceed to analyze \cref{94} and (\ref{133}).

\begin{figure}[ht]
  \centering
  \begin{tabular}{ccc}
     \begin{subfigure}[b]{0.4\textwidth}
      \includegraphics[width=\textwidth]{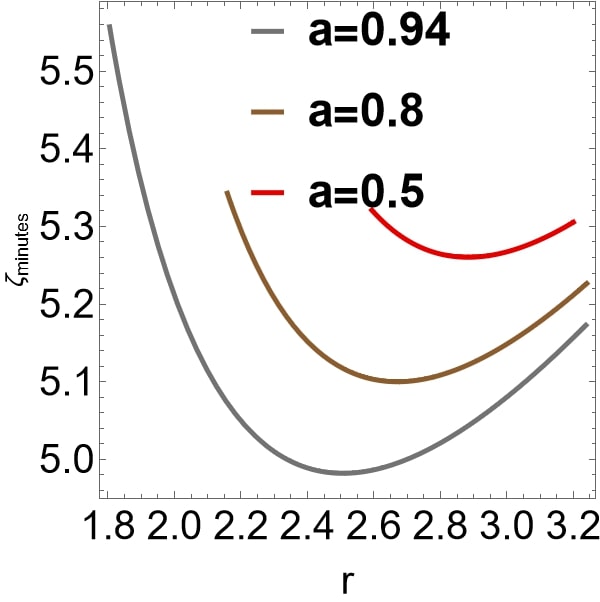}
      \caption{Time delay in \ac{sgrA}.}
      \label{fig:TDSA}
    \end{subfigure}  & \quad \quad \quad \quad
     \begin{subfigure}[b]{0.4\textwidth}
      \includegraphics[width=\textwidth]{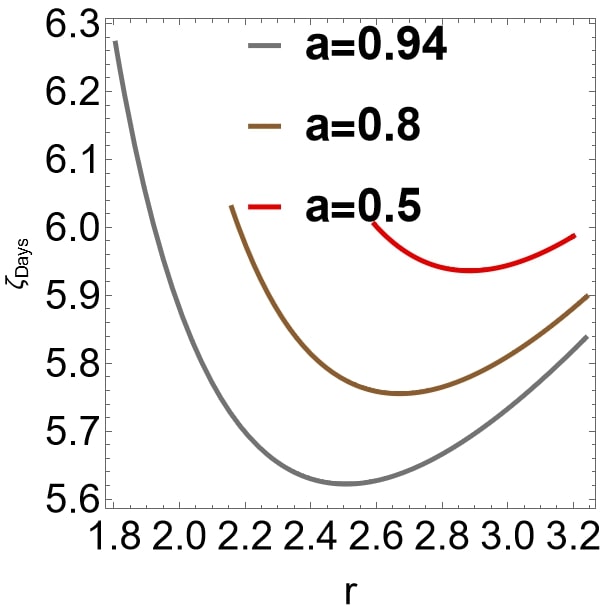}
      \caption{Time delay in M87.}
      \label{fig:TDM87}
    \end{subfigure} 
  \end{tabular}
     \caption*{Time delay for a \ac{kds} black hole applied to \ac{sgrA} and M87}
   \begin{tabular}{ccc}
     \begin{subfigure}[b]{0.4\textwidth}
      \includegraphics[width=\textwidth]{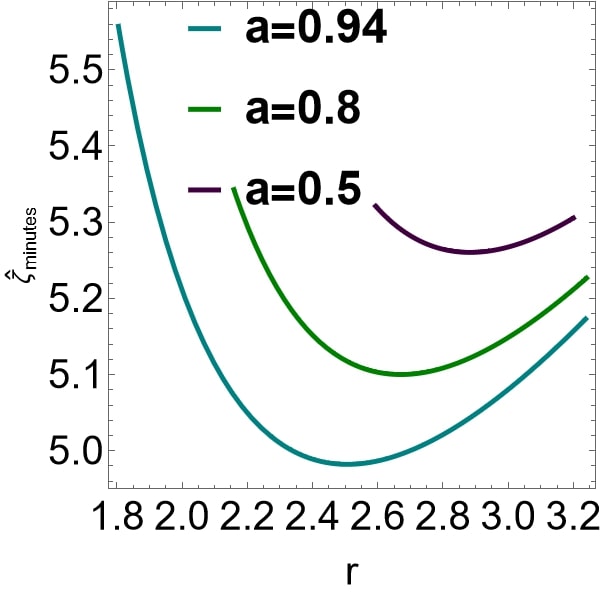}
      \caption{Time delay in \ac{sgrA}.}
      \label{fig:TDSAR}
    \end{subfigure}  & \quad \quad \quad \quad
     \begin{subfigure}[b]{0.4\textwidth}
      \includegraphics[width=\textwidth]{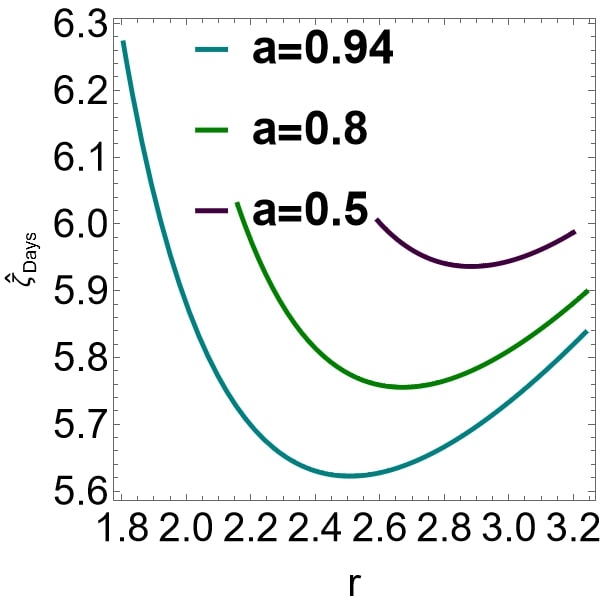}
      \caption{Time delay in M87}
      \label{fig:TDM87R}
    \end{subfigure} 
  \end{tabular}
   \caption*{Time delay for a \ac{rkds} black hole applied to \ac{sgrA} and M87}
   \caption{ Time delay in \ac{kds} and \ac{rkds} applied to \ac{sgrA} and M87 black holes for different values of spin. The observer is inclined at $30^\circ$ and $\Lambda=1.11\times 10^{-52}m^{-2}$.}
  \label{fig:timedelays}
\end{figure}

 Each point in the photon region is delayed differently such that time delay decreases monotonically from \\$r_{ph+}/\hat{r}_{ph+}$ to $r_{zamo, KdS}/r_{zamo, RKdS}$ where it begins to increase monotonically towards $r_{ph-}/\hat{r}_{ph-}$.  This behaviour can be seen in the plots of \cref{fig:timedelays}. As a result, there are two local maxima for time delay, at $r_{ph+}/\hat{r}_{ph+}$ and $r_{ph-}/\hat{r}_{ph-}$. The maxima at $r_{ph+}/\hat{r}_{ph+}$, however, is greater than that at $r_{ph-}$/$\hat{r}_{ph-}$, hence, the equatorial circular prograde photon orbit experiences the greatest amount of time delay. Time delay obtains a minima at $r_{zamo, KdS}/r_{zamo, RKdS}$, hence, the zero angular momentum orbit experiences the least amount of delay. We attribute this small delay to the fact that the photon lacks angular momentum at this radial coordinate, and thus its orbit is solely due to the dragging of inertial frames around the black hole. Furthermore, as $a \longrightarrow 1$ and $\theta \longrightarrow \pi/2$, the equatorial circular prograde photon orbit starts to experience very large time delays. We further express the time delay in standard units  $(\zeta_{dim}=t_{g} \zeta)$ so as to have dimensional estimates, with $t_{g}=\frac{G M}{c^{3}}$.\\
 \Cref{fig:timedelays} illustrates the time delays of our results applied to  \ac{sgrA} and M87. In \ac{sgrA}, when $a=0.94, 0.8,0.5$, the time delay range within $\zeta_{dim} \in [4.98352,5.5589], [5.10068,5.34345]$ and $ [5.26041,5.32105]$ minutes respectively. In M87, for the same values of black hole spin, the time delays range within $\zeta_{dim}= [5.62376,6.27306], [5.75597,6.02994]$ and $[5.93623,6.00465]$ days respectively. As a result, \ac{sgrA} exhibits very rapid changes as compared to M87, which can remain unchanged for many days. The \ac{eht} collaboration's recent observation of M87 also indicates that the image features of this black hole are equivalent across all four days \cite{akiyama2019first}. Furthermore, unlike M87, the \ac{eht} results of \ac{sgrA} show that the image features of \ac{sgrA} revolve in a $4-30$ minutes range \cite{akiyama2022first}.
 
\begin{figure}
     \centering
     	\begin{subfigure}{0.35\textwidth}
		\includegraphics[width=0.7\linewidth]{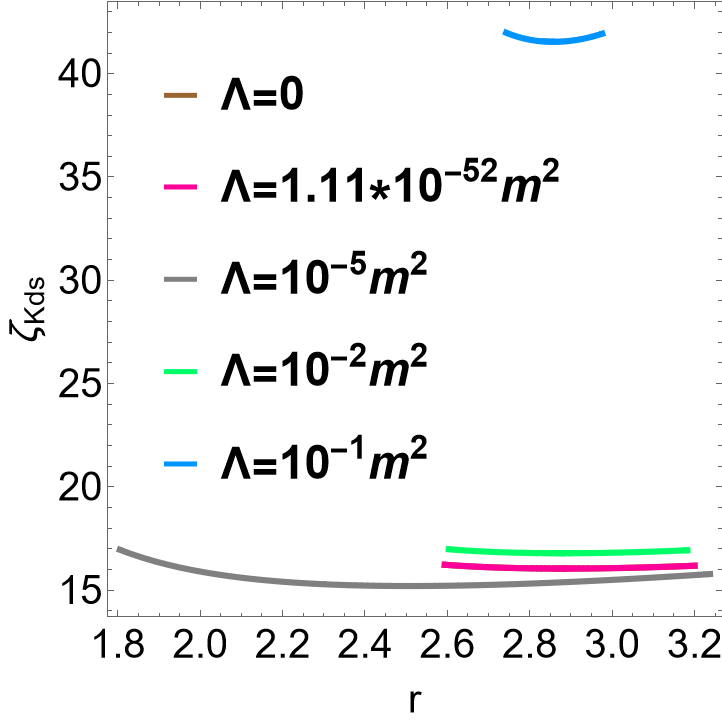}
		\caption{ Time delay for a \ac{kds} black hole }
		\label{fig:TDsgrAL}
	\end{subfigure}\hfil
	\begin{subfigure}{0.35\textwidth}
		\includegraphics[width=0.7\linewidth]{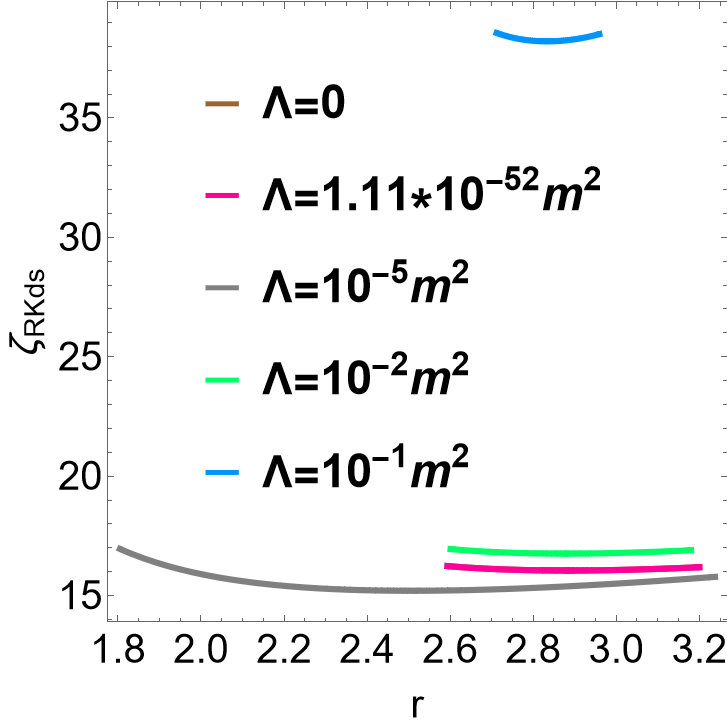}
		\caption{Time delay for a \ac{rkds} black hole.}
		\label{fig:TDrkdsL}
	\end{subfigure}\hfil
     \caption{Time delay in both \ac{kds} and \ac{rkds} for different values of $\Lambda$ with $a=0.5$ and the observers are inclined at $30 ^{\circ}$. The curves for $\Lambda=0, 1.11 \times 10^{-52}m^{-2},10^{-5}m^{-2}$ are all indistinguishable. 
      }
     \label{fig:timedelaysL}
 \end{figure}

\begin{figure}
     \centering
     	\begin{subfigure}{0.35\textwidth}
		\includegraphics[width=0.7\linewidth]{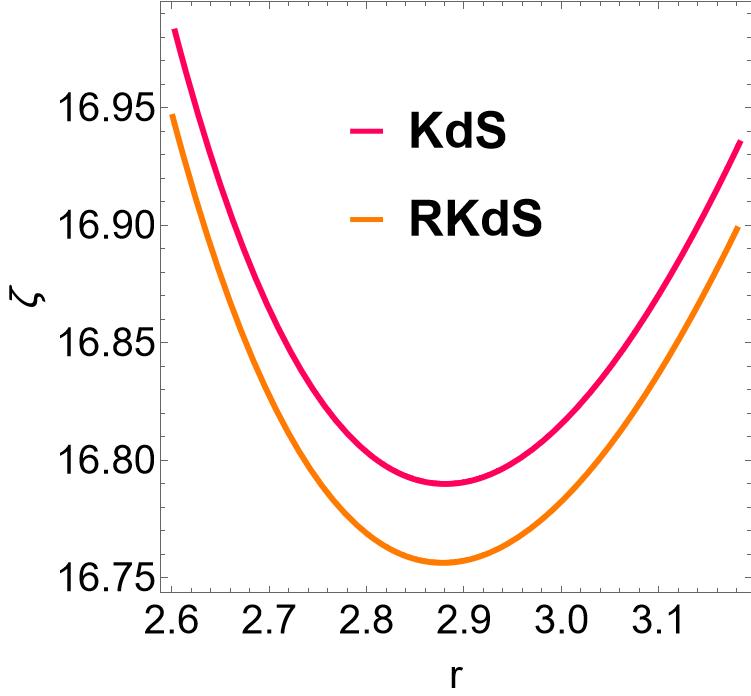}
		\caption{ Time delay for $\Lambda=0.02 m^{-2}$. }
		\label{fig:timedelayboth02}
	\end{subfigure}\hfil
	\begin{subfigure}{0.35\textwidth}
		\includegraphics[width=0.7\linewidth]{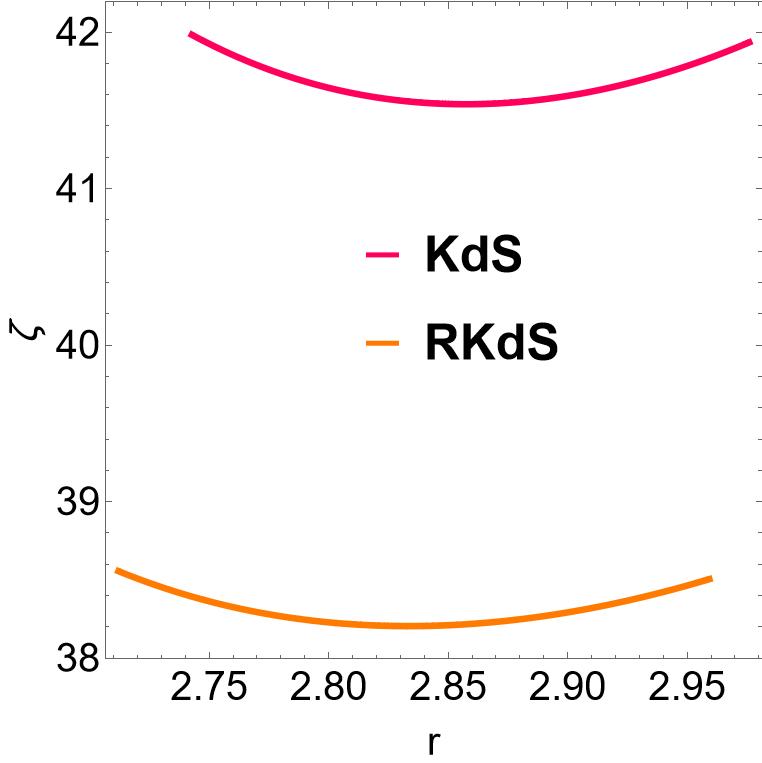}
		\caption{Time delay for $\Lambda=0.01 m^{-2}$.}
		\label{fig:timedelayboth01}
	\end{subfigure}\hfil
     \caption{Comparison of time delay in \ac{kds} and \ac{rkds} with $a=0.5$ and the observers are inclined at $30 ^{\circ}$.
      }
     \label{fig:timedelayboth}
 \end{figure}

 In the case of a rapidly rotating black hole, such as $a=0.99999$ with an observer located in the equatorial plane, the time delay for \ac{sgrA} ranges within $\zeta_{dim}=[4.92024,463.102] $ minutes, whereas M87 will be $\zeta_{dim}=[5.55235,522.598] $ days. As a result, in such a scenario, \ac{sgrA}'s equatorial circular prograde photon orbit experiences a time delay of approximately $463.102$ minutes, while M87's experiences a time delay of approximately $522.598$ days. Thus, for very high values of black hole spin and large angles of inclination, M87 can remain unchanged even up to a period of a year. However, such large values of spin and inclinations have been ruled out in recent observations\cite{omwoyo2022remarks}.\\
 For the case of vanishing $\Lambda$, which represents a Kerr black hole, we find that when $a=0.94, 0.8,0.5$, the time delay applied to \ac{sgrA} gives $\zeta_{dim,Kerr} \in [4.98213,5.55707], [5.10001,5.34276]$ and $ [5.26032,5.32098]$ minutes respectively. Moreover, considering M87, for the same values of black hole spin we obtain a time delay of $\zeta_{dim,Kerr} \in [5.62219,6.271], [5.75522,6.02916]$ and $[5.93612,6.00457]$ days respectively. Thus, we observe that the time delays when $\Lambda=1.11 \times 10^{-52}m^{-2}$ and $\Lambda=0$ in both \ac{sgrA} and M87 are approximately the same. \\
 Increasing the value of $\Lambda$ begins to have a noticeable effect on $\zeta$ when $\Lambda \approx 10^{-2} m^{-2}$ where the values of time delay begins to increase. This is evident from \cref{fig:timedelaysL} where the curves representing for $\Lambda=0, 1.11 \times 10^{-52}m^{-2}, 10^{-5}m^{-2}$ are indistinguishable. However, as $\Lambda$ begins to have a noticeable effect we observe that an increase in $\Lambda$ results in an increase on the amount of time photons take to complete their respective half orbits in both spacetimes.\\
 Comparing time delays in \ac{kds} and \ac{rkds} for large values of the cosmological constant, we observe that they are no longer indistinguishable but rather the amount of time photons take to complete half orbits is larger in \ac{kds} black hole. This can is also illustrated in \cref{fig:timedelayboth}.\\
In the limit $a=0$, which represents a \ac{sds} black hole, we obtain $\zeta_{SdS}$ $=3 \sqrt{3} \pi  \sqrt{\frac{M^2}{1-9 \Lambda  M^2}}$. For $\Lambda=0$, this relation reduces to  $3 \sqrt{3} \pi  M=16.3242$ which is the time delay for a Schwarzschild black hole. For $\Lambda=1.11 \times 10^{-52}m^{-2}, 10^{-5}m^{-2}, 10^{-2}m^{-2}, 10^{-1}m^{-2}$ we obtain the values of time delay as $16.3242,16.3249,17.1124$ and $51.6216$ respectively. We then observe that just as the rotating case, the cosmological constant begins to have a significant effect when $\Lambda \approx 10^{-2}m^{-2}$. Furthermore, when $\Lambda$ has a significant effect, we observe that time delay increases with the increasing value of $\Lambda$.

 \subsection{Lyapunov Exponent}
\begin{figure}[ht]
  \centering
  \begin{tabular}{ccc}
     \begin{subfigure}[b]{0.4\textwidth}
      \includegraphics[width=\textwidth]{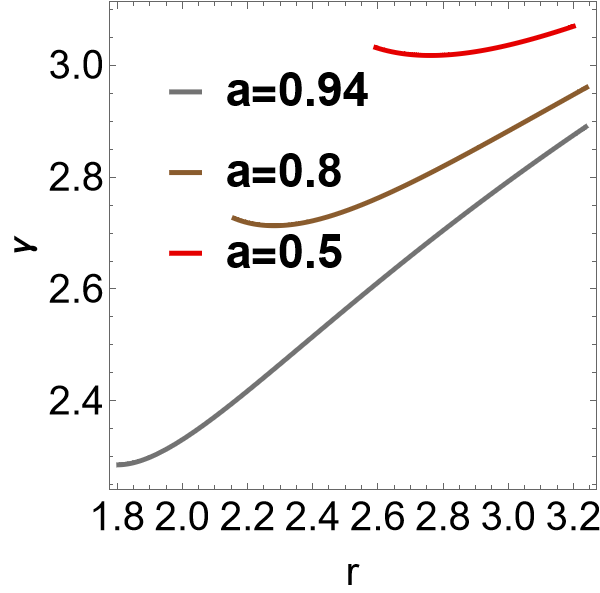}
      \caption{\ac{kds} Lyapunov exponent for observers inclined at $30^{\circ}$}
      \label{fig:LyaKdSdifa30}
    \end{subfigure}  & \quad \quad \quad \quad
     \begin{subfigure}[b]{0.4\textwidth}
      \includegraphics[width=\textwidth]{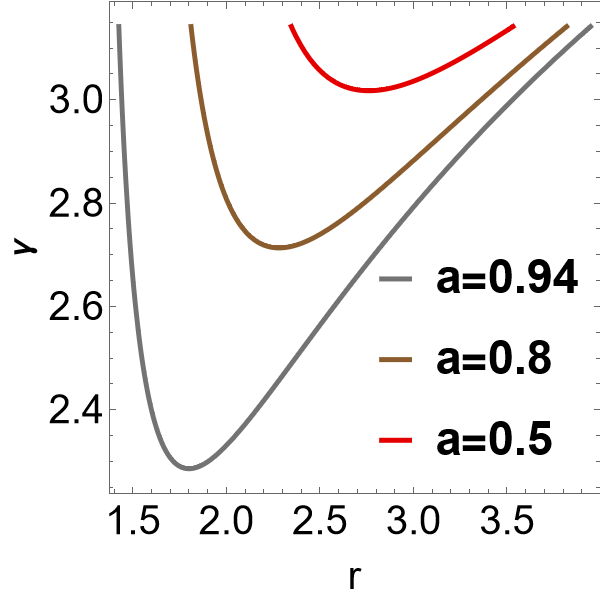}
      \caption{\ac{kds} Lyapunov exponent for observers inclined at $90^{\circ}$}
      \label{fig:LyaKdSdifa90}
    \end{subfigure} 
  \end{tabular}
   \begin{tabular}{ccc}
     \begin{subfigure}[b]{0.4\textwidth}
      \includegraphics[width=\textwidth]{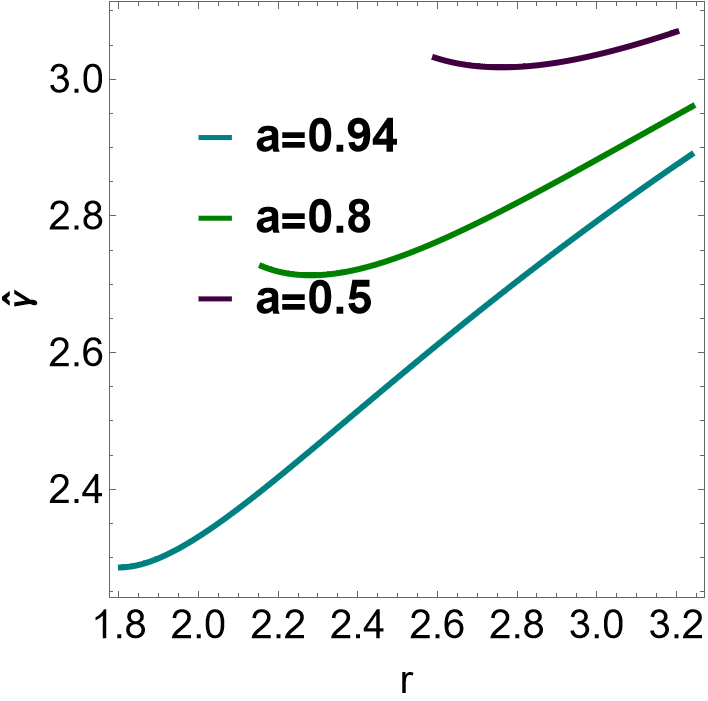}
      \caption{\ac{rkds} Lyapunov exponent for observers inclined at $30^{\circ}$}
      \label{fig:LyaRKdSdifa30}
    \end{subfigure}  & \quad \quad \quad \quad
     \begin{subfigure}[b]{0.4\textwidth}
      \includegraphics[width=\textwidth]{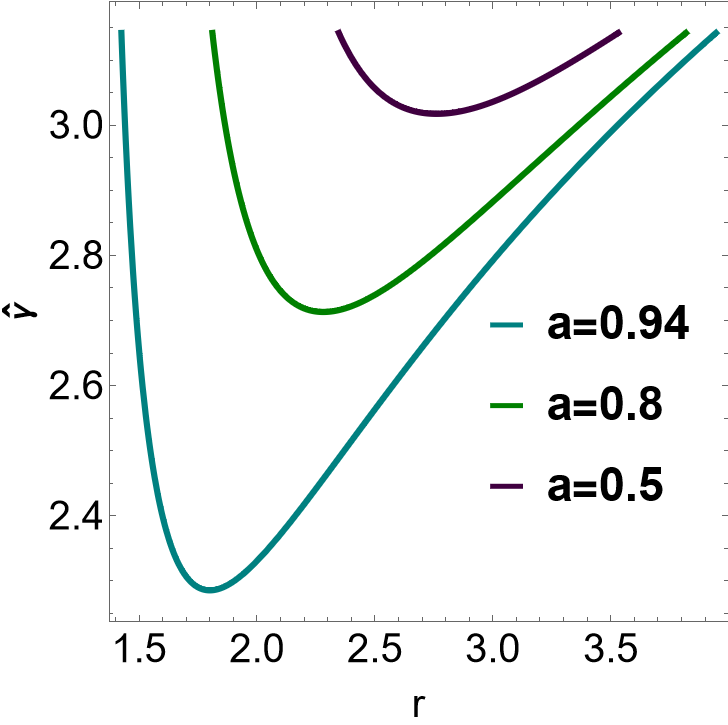}
      \caption{\ac{rkds} Lyapunov exponent for observers inclined at $90^{\circ}$}
      \label{fig:LyaRKdSdifa90}
    \end{subfigure} 
  \end{tabular}
  \caption{ Lyapunov exponent for different values of black hole spin with $\Lambda=1.11 \times 10^{-52} m^{-2}$.}
   \caption*{ }
  \label{fig:LyaKdSdifa}
\end{figure}

Small changes in the initial conditions can cause the distances between two orbits to diverge or converge exponentially. Our reference orbits in this work are bound photon orbits, and our goal is to investigate the rate of exponential deviation experienced by photons that come very close to the photon region. The Lyapunov exponents measure the rate of this exponential divergence/convergence. The Lyapunov exponent thus embodies the rate of exponential deviation of nearly bound photons from bound photon orbits after a number of half orbits around the black hole. Besides, the Lyapunov exponent controls the exponential demagnification of successive images. From \cref{radtaylor5} and (\ref{124c}) we obtain the Lyapunov exponent in \ac{kds} and \ac{rkds} spacetime as, 
\begin{align}
    \gamma =\sqrt{\frac{1}{3} \Lambda  \left(a^2+6 r^2\right) \left(L^2 (a-\lambda )^2+\eta \right)+L^2 \left(a^2-\lambda ^2+6 r^2\right)-\eta} \Gamma_{\theta}, \label{95b}\\
     \hat{\gamma}= \sqrt{a^2+2 \Lambda  \text{r}^2 \left((a-\hat{\lambda} )^2+\hat{\eta} \right)-\hat{\eta} -\hat{\lambda} ^2+6 \text{r}^2}\hat{\Gamma}_{\theta }.\label{134b}
\end{align}
Analysis of \cref{95b} and (\ref{134b}) show that the Lyapunov exponents are always positive in both spacetimes. This occurs because bound photon orbits around \ac{kds} and \ac{rkds} black holes are unstable to radial perturbations. As a result, the distance between these orbits and any nearly bound photon orbit will diverge exponentially with time before escaping or plunging into the black hole. Furthermore, from \cref{fig:LyaKdSdifa}, we observe that the general behaviour of Lyapunov exponent greatly depends on the observer's angle of inclination. That is to say, for observers inclined at $90^{\circ}$, the Lyapunov exponent decreases from $\pi$ at $r_{ph+}$ towards a minimum and begins to increase to $\pi$ at $r_{ph-}$. Thus at this angle of inclination \cref{fig:LyaKdSdifa90}, orbits close to the retrograde and prograde equatorial circular orbits experience the same rate of exponential deviation. The rate of exponential deviation from equatorial circular photon orbits is no longer equal as the angle of inclination decreases, i.e. at $30^{\circ}$ \cref{fig:LyaKdSdifa30}. Rather, the equatorial retrograde circular orbit exhibits more instability than the prograde orbit. 
Furthermore, as black hole spin increases, the rate of exponential deviation decreases. Nearly bound photon orbits that escape the black hole will then generate a series of subrings whose thickness is controlled by these Lyapunov exponents.

\begin{figure}
     \centering
     	\begin{subfigure}{0.35\textwidth}
		\includegraphics[width=0.7\linewidth]{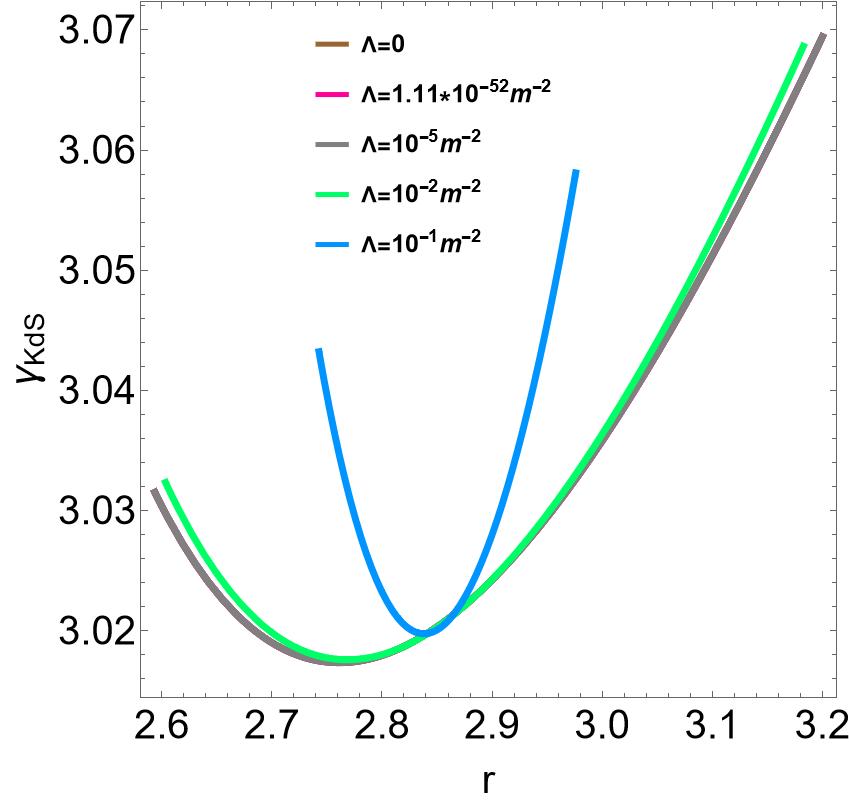}
		\caption{ Lyapunov exponent for \ac{kds} black hole for an observer inclined at $30^{\circ}$.}
		\label{fig:LyakdsL}
	\end{subfigure}\hfil
	\begin{subfigure}{0.35\textwidth}
		\includegraphics[width=0.7\linewidth]{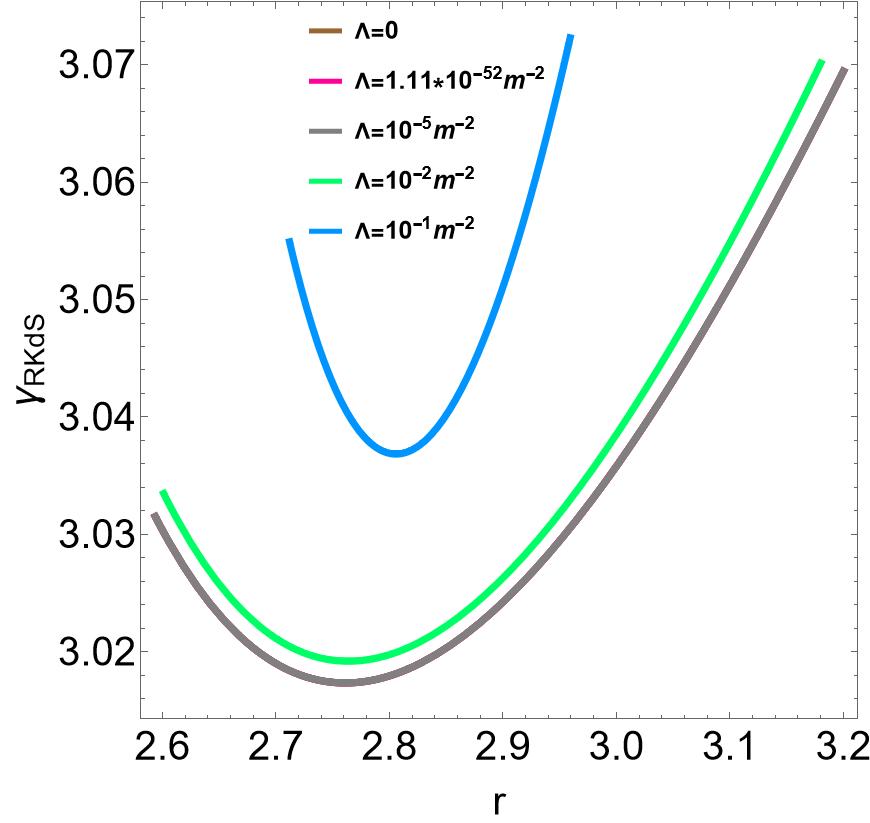}
		\caption{ Lyapunov exponent for \ac{rkds} black hole for an observer inclined at $30^{\circ}$. }
		\label{fig:LyarkdsL}
	\end{subfigure}\hfil
 \begin{subfigure}{0.35\textwidth}
		\includegraphics[width=0.7\linewidth]{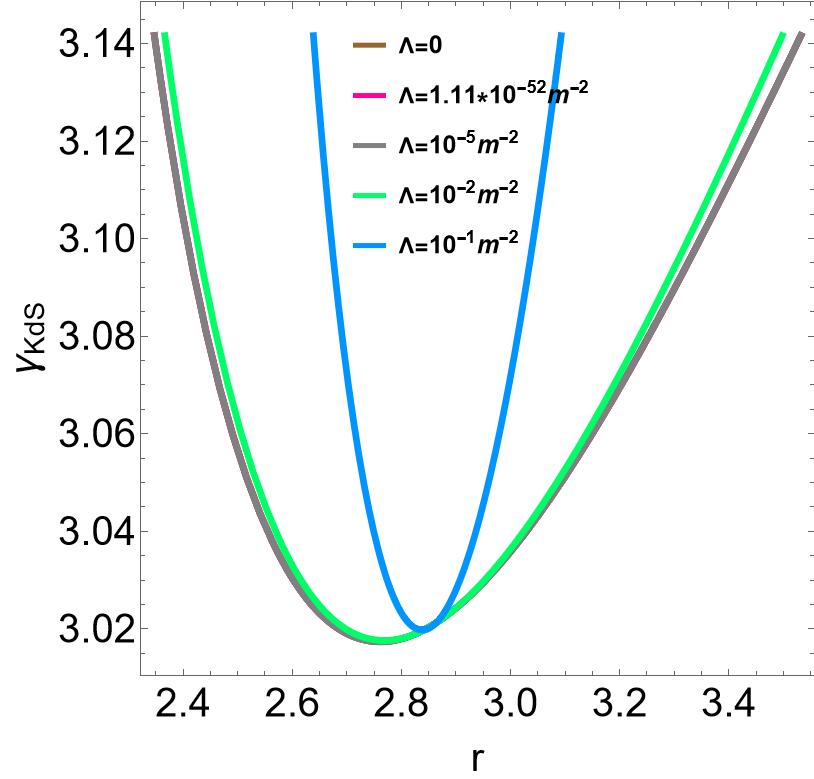}
		\caption{ Lyapunov exponent for \ac{kds} black hole for an observer inclined at $90^{\circ}$. }
		\label{fig:LyakdsL90}
	\end{subfigure}\hfil
	\begin{subfigure}{0.35\textwidth}
		\includegraphics[width=0.7\linewidth]{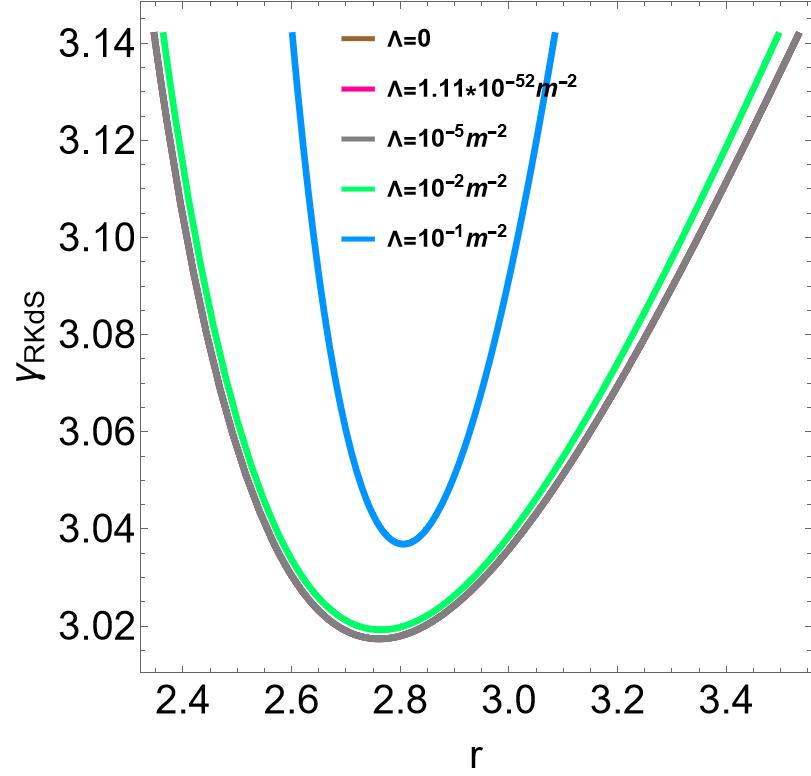}
		\caption{ Lyapunov exponent for \ac{rkds} black hole for an observer inclined at $90^{\circ}$. }
		\label{fig:LyarkdsL90}
	\end{subfigure}\hfil
     \caption{Lyapunov exponent for different values of $\Lambda$ in \ac{kds} and \ac{rkds} black hole spin with $a=0.5$. The curves for $\Lambda=0, 1.11 \times 10^{-52}m^{-2},10^{-5}m^{-2}$ cannot be distinguished from each other. }
     \label{fig:LyakdrkdL}
 \end{figure}

\begin{figure}
     \centering
     	\begin{subfigure}{0.35\textwidth}
		\includegraphics[width=0.7\linewidth]{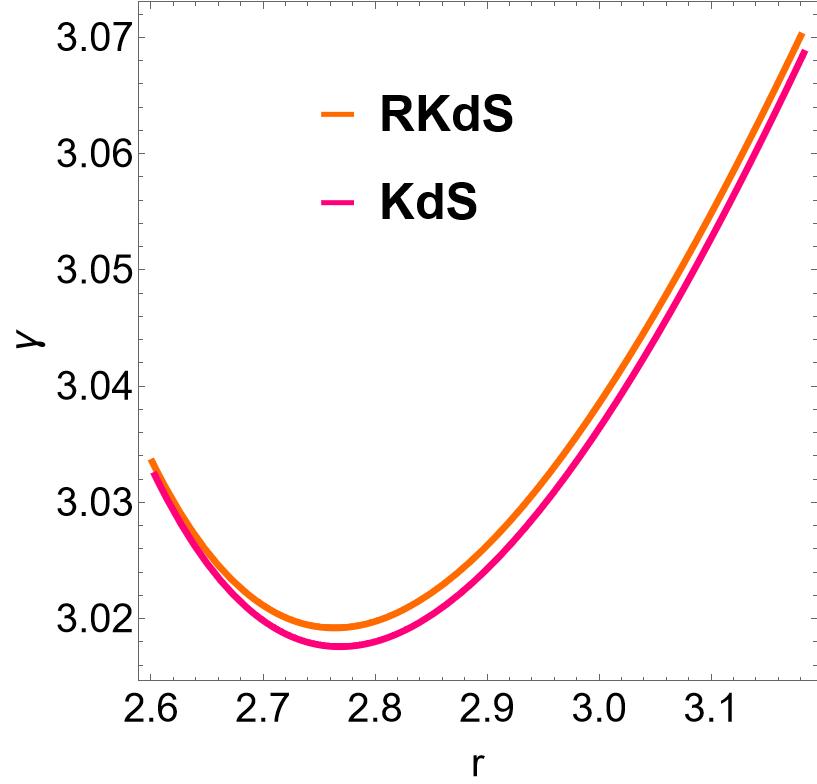}
		\caption{ Lyapunov exponent for both \ac{kds} and \ac{rkds} black hole with $\Lambda=0.02 m^{-2}$}
		\label{fig:Lyaboth02}
	\end{subfigure}\hfil
	\begin{subfigure}{0.35\textwidth}
		\includegraphics[width=0.7\linewidth]{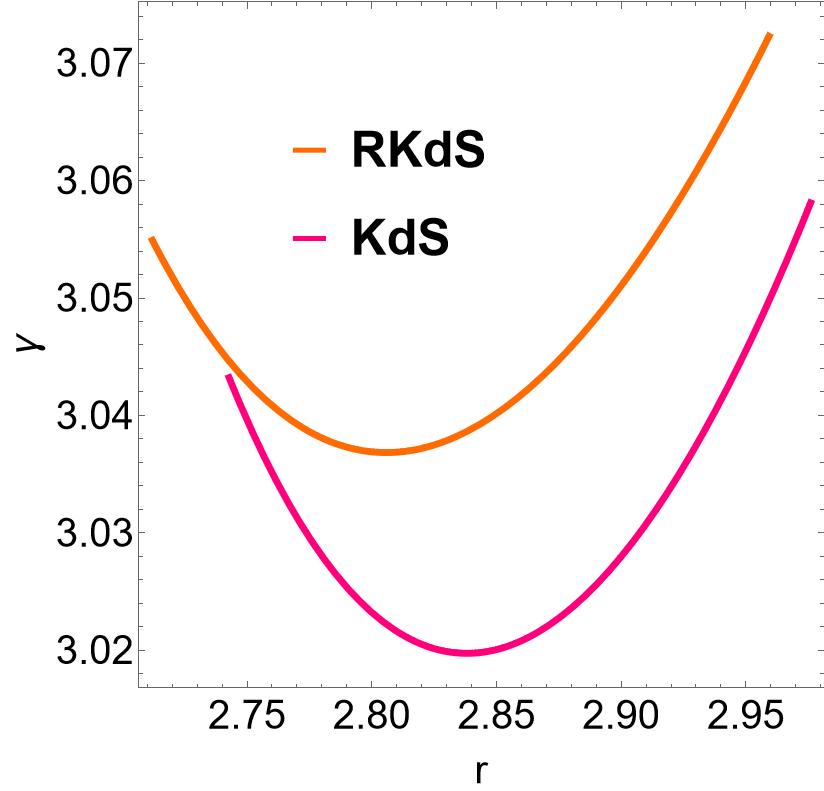}
		\caption{ Lyapunov exponent for both \ac{kds} and \ac{rkds} black hole with $\Lambda=0.01 m^{-2}$ }
		\label{fig:Lyaboth01}
	\end{subfigure}\hfil
     \caption{ Comparing the Lyapunov exponent in \ac{kds} and \ac{rkds} black hole with $a=0.5$ and observers are inclined at $30^{\circ}$.}
     \label{fig:LyabothL}
 \end{figure}

In the case of a Kerr black hole with $\Lambda=0$, the Lyapunov exponent values are approximately equal to those obtained with an extremely small cosmological constant of $\Lambda=1.11 \times 10^{-52} m^{-2}$. However, as the cosmological constant increases, a significant impact on the Lyapunov exponent becomes apparent when $\Lambda \approx 10^{-2}m^{-2}$. This can be seen in \cref{fig:LyakdrkdL}, where the curves for $\Lambda=0,1.11\times 10^{-52} m^{-2}, 10^{-5} m^{-2}$ cannot be distinguished in both spacetimes. However, when the effect of the cosmological constant becomes noticeable, larger values result in the rate of exponential deviation from equatorial circular prograde orbits to increase, and leads to a decrease on the rate of deviation from equatorial circular retrograde orbits in a \ac{kds} black hole, as demonstrated in \cref{fig:LyakdsL} for observers inclined at $30^{\circ}$.
In contrast, for \ac{rkds} \cref{fig:LyarkdsL}, an increase in the cosmological constant leads to an increase in the rate of exponential deviation from both the equatorial circular prograde and retrograde orbits. We attribute this difference in the general behavior of the Lyapunov exponent in \ac{rkds} to the presence of warped curvature in the vicinity of this black hole, which can cause changes in the behavior of the rate of exponential deviation of orbits.
As demonstrated in \cref{fig:LyakdsL90} and \cref{fig:LyarkdsL90}, when observers are located in the equatorial plane, the Lyapunov exponent displays a similar overall behavior in both spacetimes. The Lyapunov exponent in both spacetimes exhibits a monotonic decrease from $\pi$ at the equatorial circular prograde orbit to a local minimum, followed by a monotonic increase to a value of $\pi$ at the equatorial circular retrograde orbit. It is now evident that the general behavior of the rate of exponential deviation is similar in both spacetimes. We attribute this similarity to the fact that the \ac{rkds} solution has constant curvature on the equatorial plane, resulting in a similar structure of the general behavior of the rate of exponential deviation as that of \ac{kds}, which also has constant curvature.\\
Furthermore, a comparison between \ac{rkds} and \ac{kds} black holes, \cref{fig:LyabothL}, shows that for large values of the cosmological constant, the rate of exponential deviation for photons about \ac{rkds} is higher than that experienced by photons about \ac{kds} black hole. \\
In the static case, we obtain that the Lyapunov exponent is equal to $\pi$, indicating that the cosmological constant has no effect on the rate of exponential deviation from \ac{sds} photon spheres unlike in \ac{kds}/\ac{rkds} where $\Lambda$ has an effect on the rate of exponential deviation.
Negative Lyapunov exponents may indicate stability, and thus exponential convergence of orbits. This scenario can be possible in black hole spacetimes which have stable bound photon orbits.\\
\subsection{Change in azimuthal angle}
The change in azimuthal angle parameter measures the amount with which the azimuthal angle changes per half orbit. This effect is reflected on the photon ring structure as it causes a rotation on the subsequent subrings. Thus, the change in azimuthal angle parameter governs the rotation of subrings. 
Using \cref{del1} and (\ref{del2}) we obtain the change in azimuthal angle parameter in both \ac{kds} and \ac{rkds} as,
\begin{align}
     \delta=\left(\frac{\left(a L^2\right) \left(a (a-\lambda )+\text{r}^2\right)}{\text{$\Delta_{r} $}} \Gamma_{\theta} -L^{2}a \Gamma_{\phi}+ L^{2}\lambda \Bar{\Gamma}_{\phi} \right),\hat{\delta}= \left( \frac{a r^{2}+a^{3}-a \hat{\Delta}-a^{2}\hat{\lambda}}{\hat{\Delta}} \right) \hat{\Gamma}_{\theta }+ \hat{\lambda} \hat{\Gamma}_{\phi}. \label{93}
\end{align}

Upon comparing these equations with \cref{D8c} and (\ref{D10}), it becomes evident that $\Delta \phi_{m+1}-\Delta \phi_{m} \approx \delta$ and $ \hat{\Delta}\phi_{m+1}-\hat{\Delta}\phi_{m} \approx \hat{\delta}$. Thus, the change in azimuthal angle parameter between successive images of the photon ring is approximately equivalent to evaluating the change in azimuthal angle at the critical radial coordinate. We then proceed to analyze \cref{99}.

\begin{figure}[ht]
  \centering
  \begin{tabular}{ccc}
     \begin{subfigure}[b]{0.4\textwidth}
      \includegraphics[width=\textwidth]{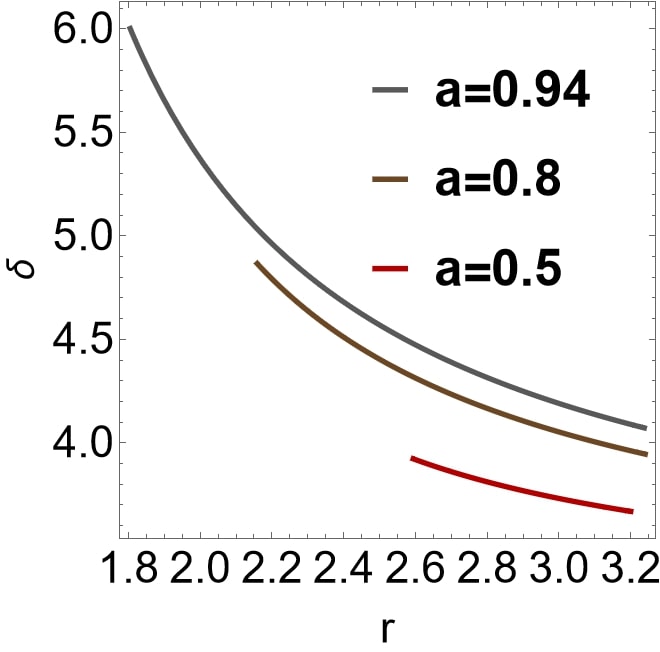}
      \caption{\ac{kds} Change in azimuthal angle for observers inclined at $30^{\circ}$}
      \label{fig:delphikds30}
    \end{subfigure}  & \quad \quad \quad \quad
     \begin{subfigure}[b]{0.4\textwidth}
      \includegraphics[width=\textwidth]{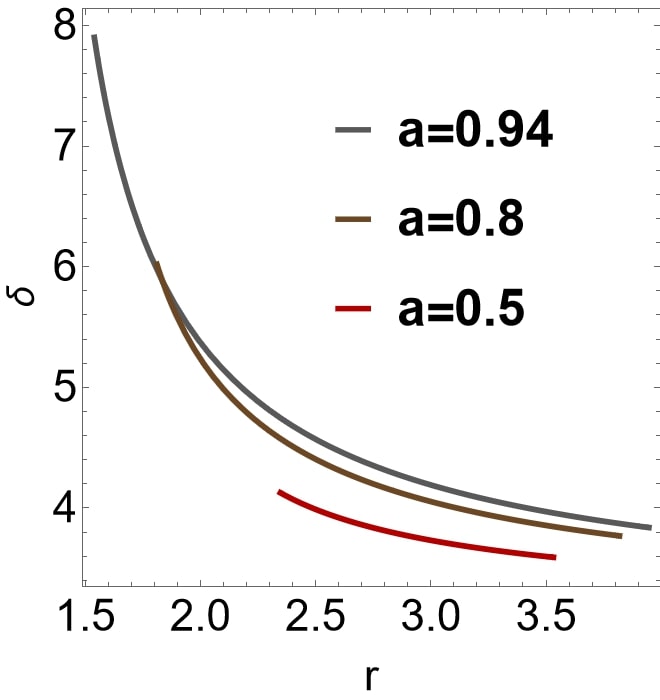}
      \caption{\ac{kds} Change in azimuthal angle for observers inclined at $90^{\circ}$}
      \label{fig:delphikds90}
    \end{subfigure} 
  \end{tabular}
   \begin{tabular}{ccc}
     \begin{subfigure}[b]{0.4\textwidth}
      \includegraphics[width=\textwidth]{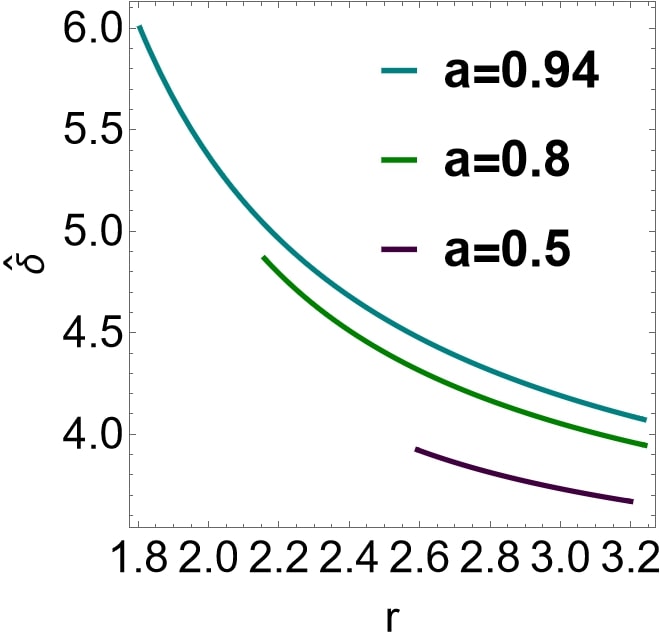}
      \caption{\ac{rkds} Change in azimuthal angle for observers inclined at $30^{\circ}$}
      \label{fig:delphirkds30}
    \end{subfigure}  & \quad \quad \quad \quad
     \begin{subfigure}[b]{0.4\textwidth}
      \includegraphics[width=\textwidth]{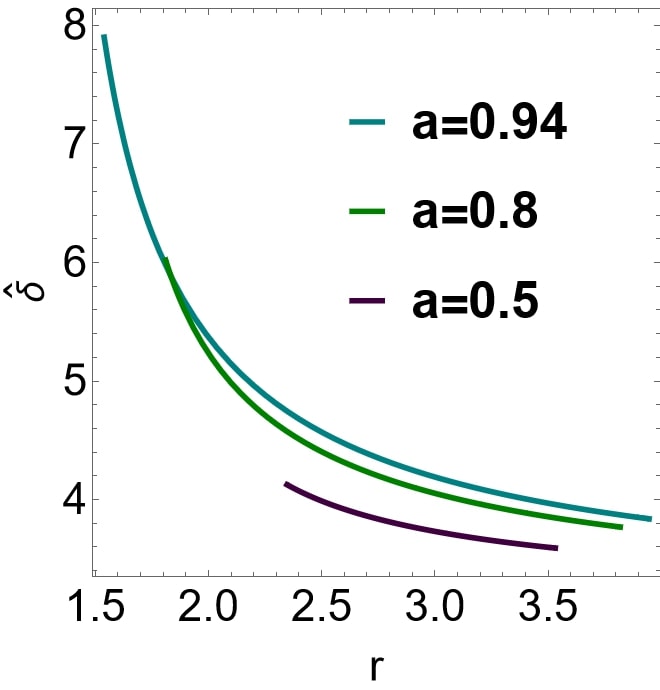}
      \caption{\ac{rkds} Change in azimuthal angle for observers inclined at $90^{\circ}$}
      \label{fig:delphirkds90}
    \end{subfigure} 
  \end{tabular}
  \caption{ Change in azimuthal angle with $\Lambda=1.11 \times 10^{-52} m^{-2}$.}
   \caption*{ }
  \label{fig:delphikds}
\end{figure}

This change in azimuthal angle parameter decreases from $r_{ph+}$/$\hat{r}_{ph+}$  to $r_{ph-}$/$\hat{r}_{ph-}$, with a discontinuity at $r_{zamo,KdS}$/$\hat{r}_{zamo}$. The discontinuity occurs when the orbits are transitioning from prograde to retrograde and can be discarded by adding $2 \pi H(r-r_{zamo})/2 \pi H(r-\hat{r}_{zamo})$  \cite{gralla2020lensing} to \cref{93} ,where $H$ is a Heaviside function. This parameter specifies the amount of azimuthal lapse for the orbits and, as a result, encodes the factor through which subsequent images will appear rotated on the observer's screen.\\
From \cref{fig:delphikds}, the change in azimuthal angle parameter is monotonic. Besides that, we have done away with the discontinuity that appears at $r_{zamo,KdS}$. The equatorial prograde circular orbits experience the greatest azimuthal shift, while the retrograde orbits have the least. Furthermore, as the black hole spin increases, so does the change in azimuthal angle parameter. Thus, the subrings for black holes with larger spin will experience greater rotation as compared to those with small spin.

\begin{figure}
     \centering
     	\begin{subfigure}{0.35\textwidth}
		\includegraphics[width=0.9\linewidth]{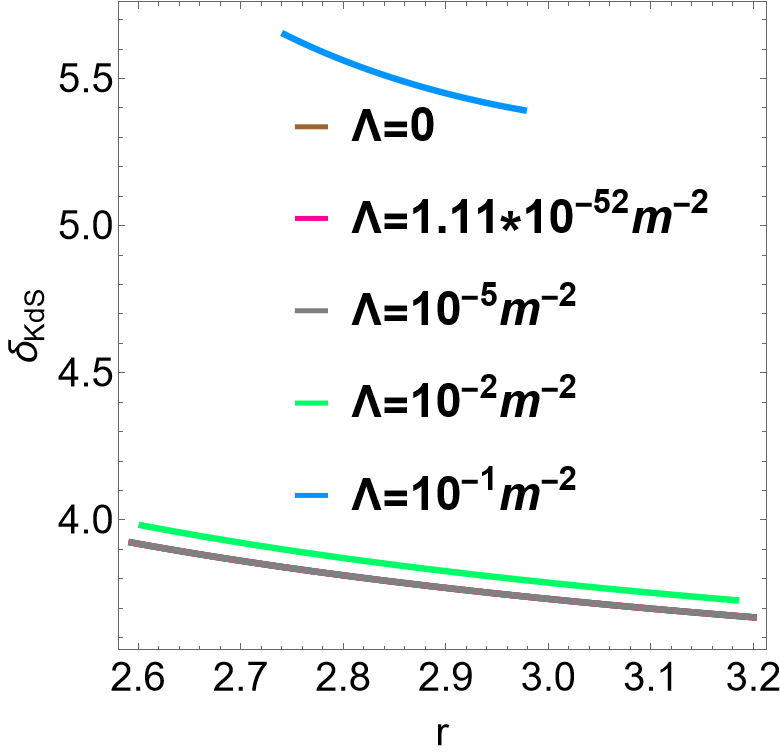}
		\caption{ Change in azimuthal angle for \ac{kds} black hole. }
		\label{fig:delkdsL}
	\end{subfigure}\hfil
	\begin{subfigure}{0.35\textwidth}
		\includegraphics[width=0.8\linewidth]{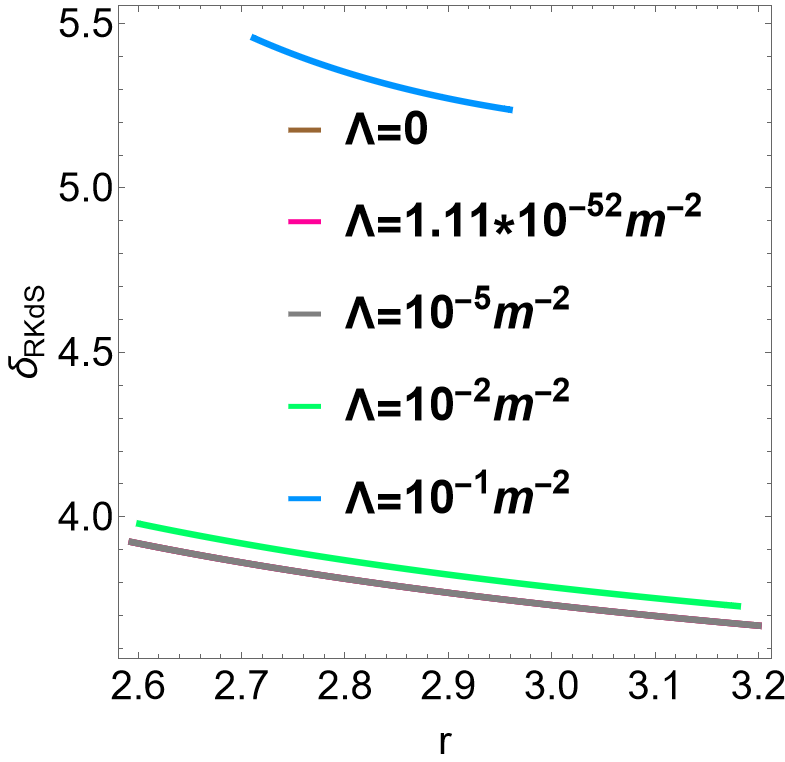}
		\caption{ Change in azimuthal angle for \ac{rkds} black hole. }
		\label{fig:delrkdsL}
	\end{subfigure}\hfil
     \caption{Change in azimuthal angle for different values of the cosmological constant with $a=0.5$ and observers inclined at $30^{\circ}$. The curves for $\Lambda=0, 1.11 \times 10^{-52}m^{-2},10^{-5}m^{-2}$ are all indistinguishable.}
     \label{fig:delbothL}
 \end{figure}

\begin{figure}
     \centering
     	\begin{subfigure}{0.35\textwidth}
		\includegraphics[width=0.9\linewidth]{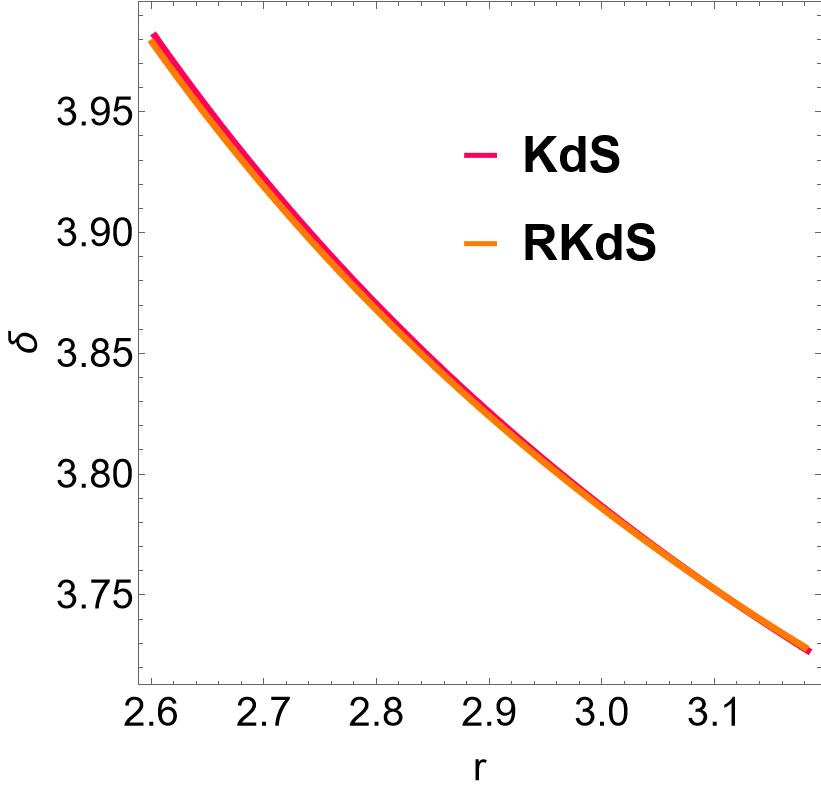}
		\caption{ Change in azimuthal angle for $\Lambda=0.02 m^2$. }
		\label{fig:delbothL02L}
	\end{subfigure}\hfil
	\begin{subfigure}{0.35\textwidth}
		\includegraphics[width=0.8\linewidth]{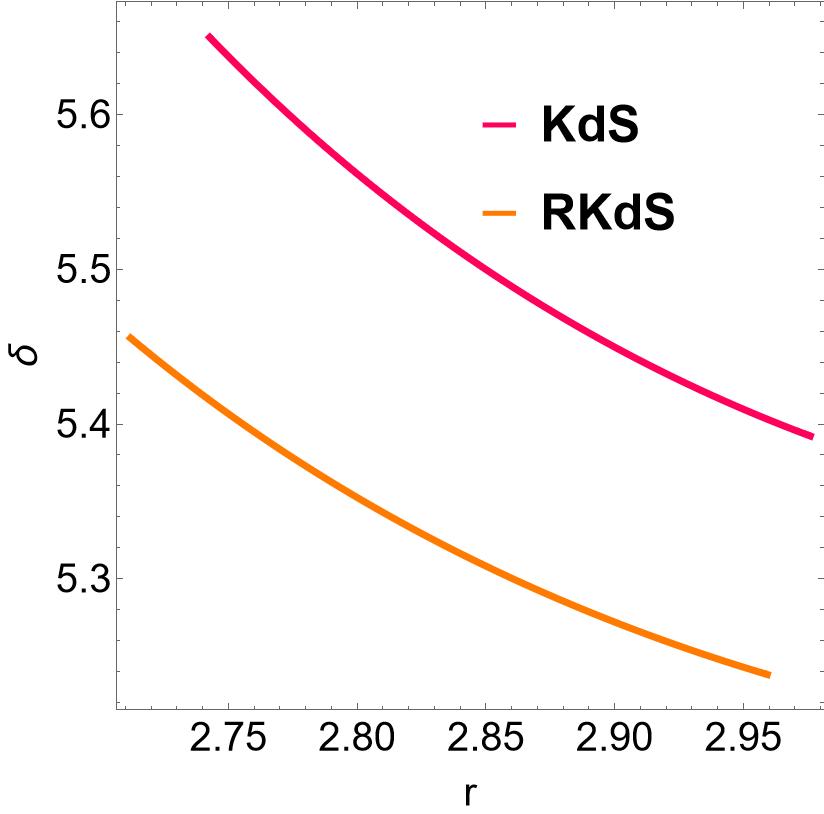}
		\caption{ Change in azimuthal angle for $\Lambda=0.1 m^2$. }
		\label{fig:delbothL01}
	\end{subfigure}\hfil
     \caption{Comparing the change in azimuthal angle in \ac{kds} and \ac{rkds} black hole with $a=0.5$ and observers inclined at $30^{\circ}$.}
     \label{fig:delbothL1}
 \end{figure}

When $\Lambda=0$, we observe that the change in azimuthal angle parameter is still indistinguishable with the result of $\Lambda=1.11 \times 10^{-52}m^{-2}$. Moreover, increasing $\Lambda$ starts to have a considerable effect on the change in azimuthal angle when $\Lambda \approx 10^{-2} m^{-2}$. This is apparent from \cref{fig:delbothL} where we observe that the curves for $\Lambda=0,1.11 \times  10^{-52} m^{-2},  10^{-5} m^{-2}$ overlap. However, when $\Lambda$ is approximately equal or greater than $10^{-2} m^{-2}$, the change in azimuthal angle takes on larger values. Comparing this parameter in a \ac{kds} and \ac{rkds} for larger values of $\Lambda$, \cref{fig:delbothL1}, we observe that photons around a \ac{kds} black hole undergo a greater change is azimuthal angle than those around a \ac{rkds} black hole.

In the static case, we find that the change in azimuthal angle equals $\pi$ with the exception of the bound photon orbits going through the poles which experience zero azimuthal shift.


\section{Conclusion} \label{sectionIV}
This work is structured into three sections, each building upon the preceding one. We commence by deriving analytical solutions for bound and nearly bound photon orbits, strategically chosen as the starting point to comprehend the overall structure of these orbits in both \ac{kds} and \ac{rkds} spacetimes. Given that these orbits serve as the fundamental basis for the analysis in this work, gaining a comprehensive understanding of their general structure facilitates more intuitive insights into the subsequent sections. Leveraging our obtained solutions, we have  conducted an analytic ray-tracing in \ac{kds} and \ac{rkds} spacetimes. We have embarked on a detailed analysis of equatorial disk images, considering the varying number of half orbits taken by photons around the black hole. An increase in the number of half orbits results in the emergence of additional lensed images, with each successive image exhibiting greater lensing effects and approaching the critical curve. Notably, the $n\geq2$ images demonstrates a closer approximation to the critical curve compared to the $n=1$ image, affirming their potential as robust testing ground for \ac{gr}. Moreover, an analysis of the azimuthal angle variations in the $n=1$ images has showcased the rotation between the first and second images. Our observations among other works in the literature, provide compelling evidence for the profound capabilities of using the photon ring to test \ac{gr} in the strong gravity regimes as the critical curve itself is not experimentally observable. Subsequently, we have proceeded to derive and analyze the critical parameters that govern the intricate structure of the photon ring.

We observe that time delay initially decreases from the equatorial circular prograde photon orbit to the \ac{zamo}, and then increases towards the equatorial circular retrograde orbit. Photons in equatorial circular prograde orbits experience the longest delays, while those in \ac{zamo} orbits experience the shortest delays. The shorter delays in \ac{zamo} orbits can be attributed to the Lense-Thirring effect, which arises from the dragging of inertial frames. We have applied our findings to the black holes \ac{sgrA} and M87, considering various spin values and a cosmological constant of $\Lambda=1.11 \times 10^{-52}m^{-2}$. For example, in \ac{sgrA}, for spin values of $a=0.94, 0.8, 0.5$, the time delay is approximately 5 minutes. In contrast, for the same spin values in M87, the time delay is on the scale of 5 days. This implies that M87 can appear unchanged for days, while \ac{sgrA} may exhibit noticeable changes within minutes. For large values of black hole spin and angle of inclination (e.g., $a=0.99999$ and $\theta=\pi/2$), the time delay for \ac{sgrA} ranges from approximately $5-460$ minutes, while for M87, it ranges from about $5-522$ days. However, large values of black hole spin and angle of inclination have been shown not to pass the constraints of recent \ac{eht} observations. We observe that these results are approximately equal to those of a \ac{rkds} and Kerr black hole, $\Lambda=0$. Increasing the cosmological constant initiates a discernible influence on time delay, primarily when $\Lambda$ values approach or exceed $10^{-2}m^{-2}$. For such large values, the time delay increases as $\Lambda$ increases. In addition, in circumstances where $\Lambda$ is considerable, photons orbiting a \ac{kds} black hole will be delayed more than those orbiting a \ac{rkds} black hole.

Photons on the equatorial circular prograde orbit undergo the maximum azimuthal shift, while those following the equatorial circular retrograde orbit exhibit the least azimuthal shift per half-orbit. For $\Lambda=1.11 \times 10^{-52}m^{-2}$, the change in azimuthal angle is indistinguishable in \ac{kds}, \ac{rkds}  and Kerr ($\Lambda=0$). Additionally, the change in the azimuthal angle increases with an increase in black hole spin. Notably, when the value of the cosmological constant approaches or exceeds $10^{-2}m^{-2}$, it starts to manifest an observable effect on the change in the azimuthal angle. For such significant values, increasing the cosmological constant results in an increase in the azimuthal shift experienced by photons per half-orbit. Furthermore, for these large values of $\Lambda$, photons traversing a \ac{kds} black hole experience a more substantial azimuthal shift than those orbiting a \ac{rkds} black hole.

The Lyapunov exponent, on the other hand, does not reveal a global behavior but rather depends on the observer's inclination. This parameter exhibits positive values in both spacetimes, indicating that the trajectories approaching the bound photon orbits tend to exponentially deviate with time. For $\Lambda=1.11 \times 10^{-52}m^{-2}$, both the \ac{kds} and \ac{rkds} exhibit the same instability rate for the circular prograde and retrograde orbits, when observed from the equatorial plane. However, this behavior drastically changes when observed away from the equatorial plane, where the rate of exponential deviation from the equatorial circular prograde orbit is less than the rate of deviation from the equatorial circular retrograde orbit. As we increase the value of $\Lambda$, we begin to observe a notable impact on the Lyapunov exponent, particularly when $\Lambda$ approaches and exceeds $10^{-2}m^{-2}$. For such values, the behavior of the Lyapunov exponent in \ac{kds} and \ac{rkds} manifests distinct general behaviors for observers inclined away from the equatorial plane. For \ac{kds}, large values of $\Lambda$ increase the rate of exponential deviation from the equatorial circular prograde photon orbit, while decreasing the rate of exponential deviation from the equatorial circular retrograde photon orbit. On the other hand, for \ac{rkds}, increasing $\Lambda$ increases the rate of exponential deviation from both the equatorial circular prograde and retrograde photon orbits. However, when the observer is inclined at $\theta=\pi/2$, the Lyapunov exponent in both spacetime exhibits a similar general behavior, where it decreases from a value of $\pi$ at the equatorial circular photon orbit to a local minimum, where it again increases to $\pi$ on the equatorial circular retrograde photon orbit. We attribute this difference in the general behavior of the Lyapunov exponent to the warped curvature of the \ac{rkds} spacetime  in the vicinity of the black hole. However, on the equatorial plane, the warped curvature vanishes, and the solution becomes a constant curvature solution, resulting in a similar general behavior of the rate of exponential deviation to that of a \ac{kds} black hole. Overall, this work highlights the Lyapunov exponent as the sole critical parameter that exhibits different general behaviors in \ac{kds} and \ac{rkds} black holes. 

Thus, the accepted cosmological value $\Lambda \sim 10^{-52} m^{-2}$ plays no relevant role in the physical situations studied here. Even though large values of the cosmological constant may not be astrophysically relevant, we note that in \ac{rkds} spacetime, such large values are feasible, as $\Lambda$ has the interpretation of vacuum energy and can be deformed in the vicinity of this black hole \cite{ovalle2021kerr},\cite{ovalle2022warped}. Furthermore, it is worth mentioning that in strong gravitational field environments there is the possibility of an amplification of the $\Lambda$ value due to the quantum effects \cite{Lima:2010na}. This kind of mechanism may justify a high $\Lambda$ value in the vicinity of a black hole compared with its asymptotic value. From our work, we observe that detection of the critical parameters analysed in this work might still not distinguish a Kerr black hole from \ac{kds} or \ac{rkds} since such parameters are approximately the same for the typical cosmological relevant value of the cosmological constant.

\appendix \label{app}
\section{ Calculating roots using Ferrari's method} \label{ferari}
The radial potential, \cref{24}, can be re-expressed as, 
\begin{align}
  R(r)=  r^{4}+\mathcal{E}r^{2}+\mathcal{F}r+\mathcal{G}, \label{33r}
\end{align}
for which we have defined the coefficients as,
\begin{align}
    \mathcal{E}=a^2-\frac{3 \left(\eta +\lambda ^2 L^2\right)}{L^2 \left(\Lambda  (a-\lambda )^2+3\right)+\eta  \Lambda }, \mathcal{F}=\frac{6 M \left(L^2 (a-\lambda )^2+\eta \right)}{L^2 \left(\Lambda  (a-\lambda )^2+3\right)+\eta  \Lambda },\mathcal{G}=-\frac{3 a^2 \eta }{L^2 \left(\Lambda  (a-\lambda )^2+3\right)+\eta  \Lambda }.
\end{align}
\Cref{33r} is already a depressed quartic and the general form of its roots are given by  Ferrari's method \cite{tignol2015galois} as,
\begin{align}
    r=  \pm_{1}\sqrt{ \frac{p_{1}}{2} }\pm_{2}\sqrt{-\left(\frac{\mathcal{E}}{2}+ \frac{p_{1}}{2}\pm_{1}\sqrt{\frac{2}{p_{1}}}\frac{\mathcal{F}}{4}\right)}, \label{F1}
\end{align}
$p_{1}$ is a root of the resolvent cubic of this quartic polynomial. The resolvent cubic is obtained as follows. Firstly, \cref{33r} can be re-written as,
\begin{align}
    \left( r^2+\frac{\mathcal{E}}{2} \right)^2=-\mathcal{F}r-\mathcal{G}+\frac{\mathcal{E}^2}{4}, \label{F2}
\end{align}
A parameter $p$ is then introduced into \cref{F2} by adding $2 r^2 p+\mathcal{E}p+p^2$ on the left and right hand side such that the right side is perfect square. This results in,
\begin{align}
    \left( r^2+p+\frac{\mathcal{E}}{2}  \right)^2=-\mathcal{F}r-\mathcal{G}+\frac{\mathcal{E}^2}{4}+2 r^2 p+\mathcal{E}p+p^2. \label{F3}
\end{align}
Recall that a quadratic equation is a perfect square if and only if it's discriminant equal zero. Thus, obtaining the discriminant in $r$ for the right hand side of \cref{F3} and equating to zero results in,
\begin{align}
    8p^3+8\mathcal{E}p^2+(2\mathcal{E}^2-8\mathcal{G})p-\mathcal{F}^2=0. \label{F4}
\end{align}
\Cref{F4} is the resolvent cubic of \cref{33r} whose roots are obtained by first expressing it as a depressed cubic through the relation $p=v-\mathcal{E}/3$. This results in,
\begin{align}
    v^3+\mathcal{P}t+\mathcal{U}=0,  \mathcal{U}=\mathcal{G}-\frac{\mathcal{E}^{3}}{108}-\frac{\mathcal{F}^{2}}{8}, \mathcal{T}=-\mathcal{G}-\frac{\mathcal{E}^{2}}{12}  \label{F5}
\end{align}
Cardano's method \cite{tignol2015galois} gives the roots of \cref{F5}, whereby we judiciously select the real root,
\begin{align}
    v=\mathcal{Z}_++\mathcal{Z}_-, \mathcal{Z}_{\pm}=\sqrt[3]{\pm \sqrt{\frac{\mathcal{U}^{2}}{4}+\frac{\mathcal{T}^{3}}{27}}-\frac{\mathcal{U}}{2}}.
\end{align}
Recall that we used the substitution $p=v-\mathcal{E}/3$. Plugging $v$ into this relation gives,
\begin{align}
    p_1=\mathcal{Z}_++\mathcal{Z}_--\mathcal{E}/3, \label{F6}
\end{align}
Substituting \cref{F6} into (\ref{F1}) and distributing the signs $\pm_1$/$\pm_2$ results in the four roots of the radial potential,
\begin{align}
    r_{1}= -\sqrt{\frac{p_{1}}{2}}-\sqrt{-\frac{\mathcal{E}}{2}+\frac{\mathcal{F}}{2\sqrt{2 p_1}}-\frac{p_{1}}{2}},
    r_{2}=-\sqrt{\frac{p_{1}}{2}}+\sqrt{-\frac{\mathcal{E}}{2}+\frac{\mathcal{F}}{2\sqrt{2 p_1}}-\frac{p_{1}}{2}}, \label{34ra}\\
    r_{3}= -\sqrt{\frac{p_{1}}{2}}-\sqrt{-\frac{\mathcal{E}}{2}-\frac{\mathcal{F}}{2\sqrt{2 p_1}}-\frac{p_{1}}{2}},
    r_{4}= \sqrt{\frac{p_{1}}{2}}+\sqrt{-\frac{\mathcal{E}}{2}+\frac{\mathcal{F}}{2\sqrt{2 p_1}}-\frac{p_{1}}{2}}. \label{34r}
\end{align}
The roots in \cref{34r} can be analyzed using the discriminant of \cref{33r},
\begin{align}
    \Delta_{0}=16 \mathcal{E}^4 \mathcal{G}-4 \mathcal{E}^3 \mathcal{F}^2-128 \mathcal{E}^2 \mathcal{G}^2+144 \mathcal{E} \mathcal{F}^2 \mathcal{G}-27 \mathcal{F}^4+256 \mathcal{G}^3. \label{36r}
\end{align}
$\Delta_{0}<0$ results in two real $(r_{1}<r_{2}<r_{-}<r_{+})$ and two complex conjugate roots $(r_{3}=\Bar{r}_{4})$. Further, $\Delta_{0}>0$ gives four real roots $(r_{1}<r_{2}<r_{3}<r_{4})$ if $\mathcal{E}<0$ and only complex roots $(r_{1}=\Bar{r}_{2}, r_{3}=\Bar{r}_{4})$ if $\mathcal{E}>0$. $\Delta_{0}=0$ produces double roots $(r_{1}<r_{2}<r_{3}=r_{4})$. Generally, irrespective of the nature of the roots, the following relations are satisfied:  $r_{1}+r_{2}+r_{3}+r_{4}=0$ and $r_{1}r_{2}r_{3}r_{4}=\mathcal{G}$. \\
The same approach applies to the radial potential of the \ac{rkds} metric which results in,
\begin{align}
    \hat{r}_{1}= -\sqrt{\frac{\hat{p}_{1}}{2}}-\sqrt{-\frac{\hat{\mathcal{E}}}{2}+\frac{\hat{\mathcal{F}}}{2\sqrt{2 \hat{p}_1}}-\frac{\hat{p}_{1}}{2}},
    \hat{r}_{2}=-\sqrt{\frac{\hat{p}_{1}}{2}}+\sqrt{-\frac{\hat{\mathcal{E}}}{2}+\frac{\hat{\mathcal{F}}}{2\sqrt{2 \hat{p}_1}}-\frac{\hat{p}_{1}}{2}}, \label{35ra}\\
    \hat{r}_{3}= -\sqrt{\frac{\hat{p}_{1}}{2}}-\sqrt{-\frac{\hat{\mathcal{E}}}{2}-\frac{\hat{\mathcal{F}}}{2\sqrt{2 \hat{p}_1}}-\frac{\hat{p}_{1}}{2}},
    \hat{r}_{4}= \sqrt{\frac{\hat{p}_{1}}{2}}+\sqrt{-\frac{\hat{\mathcal{E}}}{2}+\frac{\hat{\mathcal{F}}}{2\sqrt{2 \hat{p}_1}}-\frac{\hat{p}_{1}}{2}}, \label{35r}
\end{align}
where we have defined,
\begin{align}
    \hat{\mathcal{E}}=\frac{3 \left(a^2-\hat{\eta} -\hat{\lambda }^2\right)}{\Lambda  \left((a-\hat{\lambda } )^2+\hat{\eta} \right)+3}, \hat{\mathcal{F}}=\frac{6 M \left((a-\hat{\lambda } )^2+\hat{\eta} \right)}{\Lambda  \left((a-\hat{\lambda } )^2+\hat{\eta} \right)+3},\hat{\mathcal{G}}=-\frac{3 a^2 \hat{\eta} }{\Lambda  \left((a-\hat{\lambda } )^2+\hat{\eta}\right)+3}.
\end{align}
To avoid crowding equations, we state that the parameter $\hat{p}_1$ takes the same form as \cref{F6} and only need to add the hat symbol to $p_1$ and its components.
\section{ Definition of the elliptic integrals} \label{apB}
The functions' $F(x|k)$, $E(x|k)$, and $\Pi(n,x|k)$, respectively, represent incomplete elliptic integrals of the first, second, and third kinds, where $0\leq k<1$ is the elliptic modulus and $x$ is the argument of the integral. These functions' equations are as follows:

The incomplete elliptic integral of the first kind, $F(x|k)$, is defined as follows,

\begin{align}
    F(x|k) = \int_0^x \frac{1}{\sqrt{1-k\sin^2 t}} \text{dt}.
\end{align}

The incomplete elliptic integral of the second kind, $E(x|k)$, is defined as:

\begin{align}
     E(x|k) = \int_0^x \sqrt{1-k\sin^2 t} \text{dt}.
\end{align}
The incomplete elliptic integral of the third kind, $\Pi(n,x|k)$, is defined as:
\begin{align}
    \Pi(n,x|k) = \int_0^x \frac{1}{(1-n\sin^2 t)\sqrt{1-k\sin^2 t}} \text{dt}.
\end{align}
where $n\in\mathbb{R}$ is the characteristic of the elliptic integral.

The incomplete elliptic integrals become complete when $x=\frac{\pi}{2}$. As a result, we have:
\begin{align}
     F\left(\frac{\pi}{2}|k\right) = K(k), E\left(\frac{\pi}{2}|k\right) = E(k),\Pi(n,\frac{\pi}{2}|k) = \Pi(n,k).
\end{align}
Moreover, a linear combination of the the elliptic integral of first and third kind results in the so called associate incomplete elliptic integral of the third kind \cite{fukushima2012precise}, $J(n,x|k)$,
\begin{align}
    \Pi(n;x|k)-F(x|k)= n J(n,x | k),
\end{align}
where $J(x,n|k)$ is related to Carlson's $R_{j}$ through the relation,
\begin{align}
    J(n,x|k)= \dfrac{1}{3} \sin^{3}(x) R_{J}\left(\cos^{2} (x),1-k\sin^{2}(x),1,1-n\sin^{2}(x) \right).
\end{align}
Carlson's $R_{j}$ is a Carlson's elliptic integral of the first kind and is expressed as;
\begin{align}
    R_J(x,y,z,\rho) = \frac{3}{2}\int_0^\infty \frac{\text{dt}}{(t+x)^{-1/2}(t+\mathcal{Y})^{-1/2}(t+z)^{-1/2}(t+\rho)^{-1}}.
\end{align}

\section{Using Jacobi's Elliptic Functions to Solve Differential Equations} \label{apI}
In this section we give the general approach that we use to obtain solutions of the integrals in terms of Jacobi elliptic functions.
\subsection{Kerr de Sitter} \label{kdsAppendix}
We express the integral form of \cref{22} and (\ref{23}) as,
\begin{align}
  I_{\theta} = \int_{\theta_{s}}^{\theta_{o}} \frac{\text{d$\theta $}}{\pm_{\theta} \sqrt{\Theta(\theta)}} , \label{a30}\\
   I_{r}= \int_{r_{s}}^{r_{o}} \frac{\text{dr}}{\pm_{r} \sqrt{R(r)}}, \label{30a}\\
    \phi_{o}-\phi_{s}= \int_{r_{s}}^{r_{o}} \frac{\left(a L^2\right) \left(a (a-\lambda )+\text{r}^2\right) \text{dr}}{\pm_{r} \text{$\Delta_{r} $} \sqrt{R(r)}}  -L^{2}a I_{\phi}+ L^{2}\lambda \Bar{I}_{\phi}, \label{30}\\
    t_{o}-t_{s}=\int_{r_{s}}^{r_{o}} \dfrac{L^{2}((r^{2}+a^{2})^{2}-a\lambda(a^{2}+r^{2})) \text{dr}}{\pm_{r} \Delta_{r} \sqrt{R(r)}} +a^2 L^2 I_{t}-a L^2(a-\lambda)I_{\phi}, \label{31}
\end{align}
with,
\begin{align}
    I_{\phi}= \int_{\theta_{s}}^{\theta_{o}} \dfrac{\text{d$\theta $}}{\pm_{\theta} \sqrt{\Theta(\theta)} \Delta_{\theta}} , \quad
\Bar{I_{\phi}}= \int_{\theta_{s}}^{\theta_{o}} \dfrac{\text{d$\theta $}}{\pm_{\theta} \sqrt{\Theta(\theta)} \Delta_{\theta}\sin^{2}\theta} , \quad
I_{t}= \int_{\theta_{s}}^{\theta_{o}} \dfrac{\cos^2\theta \text{d$\theta $}}{\pm_{\theta} \sqrt{\Theta(\theta)} \Delta_{\theta}}, \label{32} 
\end{align}
where the subscripts $o,s$ denote the observer and the source respectively. The integral form  and Mino parameter approach are related by \cite{gralla2020null},
\begin{align}
    \tau=I_{\theta}=I_{r}.\label{32b}
\end{align}
We proceed to obtain general solutions of the angular integrals, \cref{32}, as follows.

 To begin with, we express the angular potential in terms of its roots,$u_{+}$ and $u_{-}$ as,
\begin{align}
    \Theta(u)=\frac{1}{1-u}[\Xi (u_{+}-u)(u-u_{-})]. \label{51}
\end{align}
The parameter $\Xi$ has been defined as,
\begin{align}
\Xi = \left( \frac{1}{3} a^2 \eta  \Lambda +\frac{1}{3} a^2 \lambda ^2 \Lambda  L^2-\frac{2}{3} a^3 \lambda  \Lambda  L^2+\frac{1}{3} a^4 \Lambda  L^2+a^2 L^2 \right). \label{52}
\end{align}


Recall that $u=\cos^2 \theta$, hence, $\text{\text{du}}=-2\sqrt{u}\sqrt{1-u}\text{d$\theta $}$. We use this substitution to transform the integrals,
\begin{enumerate}
    \item 
    \begin{subequations}
    \begin{align}
    \left| \int_{\pi/2}^{\theta_i} \frac{\text{d$\theta $}}{\sqrt{\Theta(\theta)}}\right| =  \int_{0}^{u_i} \frac{\sqrt{1-u}\text{du}}{2\sqrt{u}\sqrt{1-u}\sqrt{\Xi (u_+-u)(u-u_-)}},\label{56} \\= \frac{1}{2\sqrt{\Xi}}\int_{0}^{u_i} \frac{\text{du}}{\sqrt{ u (u_+-u)(u-u_-)}} 
    = \frac{1}{2\sqrt{\Xi}}\int_{0}^{x_{i}} \frac{2 u_{+}\cos x\sin x \text{dx}}{\sqrt{-u_{-}u_{+}^{2}\sin^{2}x(1-\sin^{2}x)(1-\frac{u_{+}}{u_{-}}\sin^{2}x)}},\label{57}\\= \dfrac{1}{\sqrt{-u_{-}\Xi}}\int_{0}^{x_{i}} \dfrac{ \text{dx}}{\sqrt{1-k\sin^{2}x }} =  \dfrac{1}{\sqrt{-u_{-}\Xi}} F\left( x_{i}| k\right). \label{58}
\end{align}
\end{subequations}
In \cref{57} we have utilized the relation $u=u_{+}\sin^2{x}$.
We define $x_{i}$ and $k$ in \cref{58} as,
\begin{align}
    x_i=\arcsin \sqrt{\frac{u_i}{u_+}}=\arcsin \left[\frac{\cos{\theta_{i}}}{\sqrt{u_{+}}}\right], \quad k=\frac{u_{+}}{u_{-}}. \label{59}
\end{align}
\item \begin{align}
     \left| \int_{\pi/2}^{\theta_i} \frac{\text{d$\theta $}}{\Delta_ {\theta} \sqrt{\Theta_{\theta}}}\right|= \dfrac{1}{\sqrt{-u_{-}\Xi}}\int_{0}^{x_{i}} \dfrac{\text{dx}}{(1+u_+ \mathcal{Y} \sin^{2}x)\sqrt{1-k\sin^{2}x }} = \dfrac{1}{\sqrt{-u_{-}\Xi}}\Pi \left( -u_{+} \mathcal{Y};x_i|k\right).
    \label{61}
\end{align}
\item 
\begin{subequations}
\begin{align}
    \left| \int_{\pi/2}^{\theta_{i}}\dfrac{\text{d$\theta$}}{ \sqrt{\Theta(\theta)} \Delta_{\theta}\sin^{2}\theta} \right|= \dfrac{1}{\sqrt{-u_{-}\Xi}}\int_{0}^{x_{i}} \dfrac{\text{dx}}{(1-u_{+}\sin^{2}x)(1+u_+ \mathcal{Y} \sin^{2}x)\sqrt{1-k\sin^{2}x }}, \label{63}\\
    =\dfrac{1}{\sqrt{-u_{-}\Xi}}\int_{0}^{t_{i}} \dfrac{\text{dt}}{(1-u_{+}t^{2})(1+u_+ \mathcal{Y} t^{2})\sqrt{(1-t^{2})(1-kt^{2}) }},\label{64}\\
    =\dfrac{1}{\sqrt{-u_{-}\Xi}} \dfrac{1}{1+\mathcal{Y}}[\Pi(u_{+};\arcsin t_i|k)+\mathcal{Y} \Pi( -u_+ \mathcal{Y};\arcsin t_i|k) ]
    . \label{66}
\end{align}
\end{subequations}
Where $t_{i}$ is given by the relation,
\begin{align}
    t_{i}=\sqrt{\frac{u_{i}}{u_{+}}}.\label{67}
\end{align}
\item 
\begin{subequations}
\begin{align}
    \left| \int_{\pi/2}^{\theta_i} \frac{\cos^2 \theta \text{d$\theta $}}{\Delta_ {\theta} \sqrt{\Theta_{\theta}}}\right|=\dfrac{1}{\sqrt{-u_{-}\Xi}}\int_{0}^{x_{i}} \dfrac{u_{+}\sin^{2}x\text{dx}}{(1+u_+ \mathcal{Y} \sin^{2}x)\sqrt{1-k\sin^{2}x }}\label{68}\\= \dfrac{u_+}{\sqrt{-u_{-}\Xi}}\int_{0}^{t_{i}} \dfrac{ t^2 \text{dt}}{(1+u_+ \mathcal{Y} t^2)\sqrt{(1-t^{2})\left( 1-kt^{2}\right) }} = \dfrac{3}{a^2 \Lambda\sqrt{-u_{-}\Xi}} (F(\arcsin t_i|k)-\Pi( -u_+ \mathcal{Y};\arcsin t_i|k)).
    \label{69pa}
\end{align}
\end{subequations}
We have observed the equivalence of $F(x|k)$ and $\Pi(n;x|k)$ when $n=0$. Thus, when the value of $n$ is extremely small, this combination may not be suitable due to potential round-off errors. For example, in \cref{69pa}, considering the astrophysically relevant value of the cosmological constant, $\mathcal{Y}=\frac{a^2 \Lambda}{3}$ becomes very small. Consequently, in our numerical computations, the term $(F(\arcsin t_i|k)-\Pi(\frac{-u_+a^2 \Lambda}{3};\arcsin t_i|k))$ suffers round off errors arising from this smallness. To circumvent the round off errors, we employ an alternative representation of this linear combination using the so called associate incomplete elliptic integral of third kind, $J(n,x|k)$ \cite{fukushima2012precise},
\begin{align}
  \dfrac{3}{a^2 \Lambda\sqrt{-u_{-}\Xi}} (F(\arcsin t_i|k)-\Pi( -u_+ \mathcal{Y};\arcsin t_i|k))=  \frac{u_{+} J(-u_+ \mathcal{Y},\arcsin t_i|k)}{\sqrt{-u_{-}\Xi}}. \label{69}
\end{align}
\end{enumerate}
The angular component of the photon trajectories can then be determined through utilizing the above solutions by substituting the subscript $i$ with either $s$ or $o$, representing the source or observer respectively, along with the corresponding transformed coordinate.

To derive the solutions for the integrals used in determining critical parameters, we adopt the following approach. Critical parameters are evaluated on the critical radius which is the radial coordinate of bound photon orbits. As a consequence, the integral over the radial coordinate is treated as a constant. Moreover, these parameters are defined over half orbits. A photon is deemed to have completed half an orbit around the black hole when it transitions between its maximum and minimum inclinations with respect to the equatorial plane, or vice versa. This corresponds to integrating the angular integrals from one turning point to the next. Our analysis focuses on ordinary motion, leading us to evaluate the integrals for half orbits over $\theta_{1}$ to $\theta_{4}$ or vice versa.

As a result, the change in azimuthal angle parameter and time delay will be obtained by evaluating the integrals in \cref{30} and \cref{31} over half orbits and at constant $r$,
\begin{align}
    \delta= \frac{\left(a L^2\right) \left(a (a-\lambda )+\text{r}^2\right) }{ \text{$\Delta_{r} $} }  \left| \int_{\theta_1}^{\theta_4} \frac{\text{d$\theta $}}{\sqrt{\Theta(\theta)}}\right| -L^{2}a  \left| \int_{\theta_1}^{\theta_4} \frac{\text{d$\theta $}}{\Delta_ {\theta} \sqrt{\Theta_{\theta}}}\right|+ L^{2}\lambda  \left| \int_{\theta_1}^{\theta_{4}}\dfrac{\text{d$\theta $}}{ \sqrt{\Theta(\theta)} \Delta_{\theta}\sin^{2}\theta} \right|,\label{del1}\\
    \zeta=\dfrac{L^{2}((r^{2}+a^{2})^{2}-a\lambda(a^{2}+r^{2}))}{ \Delta_{r} } \left| \int_{\theta_1}^{\theta_4} \frac{\text{d$\theta $}}{\sqrt{\Theta(\theta)}}\right| +a^2 L^2  \left| \int_{\theta_1}^{\theta_4} \frac{\cos^2 \theta \text{d$\theta $}}{\Delta_ {\theta} \sqrt{\Theta_{\theta}}}\right|-a L^2(a-\lambda) \left| \int_{\theta_1}^{\theta_4} \frac{\text{d$\theta $}}{\Delta_ {\theta} \sqrt{\Theta_{\theta}}}\right|. \label{zeta1}
\end{align}
Due to the constancy of the integral over the radial coordinate, we have employed \cref{32b} to replace the Mino parameter with the angular integral in the first part of \cref{del1} and (\ref{zeta1}).

Utilizing \cref{58} and the definition of $x_i$ in \cref{59}, we have that,
\begin{subequations}
\begin{align}
    \left| \int_{\theta_1}^{\theta_4} \frac{\text{d$\theta $}}{\sqrt{\Theta(\theta)}}\right|= \left|\dfrac{1}{\sqrt{-u_{-}\Xi}} F\left(\arcsin \left[\frac{\cos{\theta_{4}}}{\sqrt{u_{+}}}\right]| k\right)- \dfrac{1}{\sqrt{-u_{-}\Xi}} F\left(\arcsin \left[\frac{\cos{\theta_{1}}}{\sqrt{u_{+}}}\right]| k\right)\right|,\label{200}\\
    = \left|-\dfrac{1}{\sqrt{-u_{-}\Xi}} F\left(\arcsin \left[\frac{\sqrt{u_{+}}}{\sqrt{u_{+}}}\right]| k\right)- \dfrac{1}{\sqrt{-u_{-}\Xi}} F\left(\arcsin \left[\frac{\sqrt{u_{+}}}{\sqrt{u_{+}}}\right]| k\right)\right|,\label{201}\\
    =\dfrac{2}{\sqrt{-u_{-}\Xi}} F\left(\frac{\pi}{2}| k \right)=\dfrac{2}{\sqrt{-u_{-}\Xi}} K(k)=\Gamma_\theta. \label{202}
\end{align}
\end{subequations}
In \cref{201}, we have used the definitions from \cref{36} and (\ref{39}) where $\sqrt{u_{+}}=\cos{\theta_1}$ and $-\sqrt{u_{+}}=\cos{\theta_4}$. Furthermore, we have made use of the property, $F(-x \mid k)=-F(x \mid k)$ and $\arcsin{1}=\pi/2$. 
If we still integrated from $\theta_{4}$ to $\theta_{1}$, the result will remain the same.

For the subsequent integrals, we make use of the general solutions \cref{61}-(\ref{69}), and directly substitute $\arcsin \left[\frac{\cos{\theta_{4}}}{\sqrt{u_{+}}}\right]=-\frac{\pi}{2}$ and $\arcsin \left[\frac{\cos{\theta_{1}}}{\sqrt{u_{+}}}\right]=\frac{\pi}{2}$ as we have explained how this comes up which transforms the incomplete integrals to be complete. This gives,
\begin{align}
     \left| \int_{\theta_1}^{\theta_4} \frac{\text{d$\theta $}}{\Delta_ {\theta} \sqrt{\Theta_{\theta}}}\right|= \dfrac{2}{\sqrt{-u_{-}\Xi}}\Pi \left( -u_{+} \mathcal{Y};|k\right)=\Gamma_\phi, \label{203} \\
      \left| \int_{\theta_1}^{\theta_{4}}\dfrac{\text{d$\theta$}}{ \sqrt{\Theta(\theta)} \Delta_{\theta}\sin^{2}\theta} \right|= \dfrac{2}{\sqrt{-u_{-}\Xi}} \dfrac{1}{1+\mathcal{Y}}[\Pi(u_{+};|k)+\mathcal{Y} \Pi(-u_+ \mathcal{Y};|k) ]=\Bar{\Gamma}_\phi,\label{204}\\
      \left| \int_{\theta_1}^{\theta_4} \frac{\cos^2 \theta \text{d$\theta $}}{\Delta_ {\theta} \sqrt{\Theta_{\theta}}}\right|= \frac{2u_{+} J(-u_+ \mathcal{Y},|k)}{\sqrt{-u_{-}\Xi}}=\Gamma_t. \label{205}
\end{align}
\Cref{202}-(\ref{205}) will then be inserted into (\ref{del1}) and (\ref{zeta1}) to obtain the change in azimuthal angle and time delay.

The Lyapunov exponent will be calculated in the following way. The radial coordinate of a nearly bound photon orbit is very close to the radial coordinate of a bound photon orbit, with just a small difference between them. We therefore write the radial coordinate of the nearly bound orbit as $r = r_b + \delta r$, where $r_b$ is the radial coordinate of the bound photon orbit and $0<\delta r <1$ is a small deviation from this radial coordinate. Performing a Taylor expansion
on the radial potential $R(r)$ with  respect to this deviation results in,
 \begin{align}
     R(r)=R(r_b)+(r-r_b) R'(r_b)+\frac{1}{2} (r-r_b)^2 R''(r_b)+\frac{1}{6} (r-r_b)^3 R^{(3)}(r_b)+O\left((r-r_b)^4\right).\label{radtaylor}
 \end{align}
 \Cref{radtaylor} is an approximation of the radial potential for a nearly bound photon orbit. We then proceed by keeping terms up to second order in the deviation $\delta r$ because higher order terms will become smaller as we move away from the bound orbit. Further, recall that for a bound photon orbit, $R(r)=R'(r)=0$, hence the first two terms of \cref{radtaylor} will vanish leaving us with only the third term. Let $\delta r_1$ be the initial deviation and $\delta r_n$ be the deviation after $n$ half orbits. Thus after $n$ half orbits, we have that,
\begin{align}
      \int_{r_b + \delta r_1}^{r_b + \delta r_n} \frac{\text{dr}}{ \sqrt{R(r)}} \approx \int_{r_b + \delta r_1}^{r_b + \delta r_n} \frac{\sqrt{2}\text{dr}}{ \sqrt{R''(r_b)(r-r_b)^2}}=\frac{\sqrt{2}(\ln{(\delta r_n)}-\ln{(\delta r_1)})}{\sqrt{R''(r_b)}}.\label{radtaylor2}
\end{align}
The Mino period for half an orbit in the latitudinal direction has been defined in \cref{202}. Therefore for $n$ half orbits the photon will have a Mino period of $n \Gamma_\theta$. \Cref{32b} gives us the equality of the radial and latitudinal integral and by this definition we have that after $n$ half  orbits,
\begin{align}
    \frac{\sqrt{2}(\ln{(\delta r_n)}-\ln{(\delta r_1)})}{\sqrt{R''(r_b)}}=n \Gamma_\theta. \label{radtaylor3}
\end{align}
Simplifying \cref{radtaylor3} gives,
\begin{align}
    \delta r_n=\delta r_1 \exp{\left( \sqrt{\frac{R''(r_b)}{2}} n \Gamma_\theta\right)}. \label{radtaylor4}
\end{align}
\Cref{radtaylor4} tells us that after $n$ half orbits, a nearly bound photon orbit exponentially deviates from a bound orbit at a rate given by,
\begin{align}
    \gamma=\sqrt{\frac{R''(r_b)}{2}} \Gamma_\theta. \label{radtaylor5}
\end{align}
Thus, \cref{radtaylor5} defines the Lyapunov exponent per half orbit. A similar relation has been obtained in  \cite{johnson2020universal} in the case of a Kerr black hole.\\
For nearly bound photon orbits we also need to obtain the general solutions of the radial integrals. Firstly,
we express the radial potential and the parameter $\Delta_{r}$ in terms of their roots,
\begin{align}
    R(r)=(r-r_{1})(r-r_{2})(r-r_{3})(r-r_{4}), \Delta_{r}=-\frac{3}{\Lambda}(r-r_{-c})(r-r_{-})(r-r_{+})(r-r_{c}). \label{158}
\end{align}
Where $r_{-c},r_{-},r_{+},r_{c}$ are the \ac{kds} horizons.
Applying the notation \cref{158},we re-express \cref{30a}- (\ref{31}) as,

\begin{align}
  I_{r}= \int_{r_{s}}^{r_{o}} \frac{\text{dr}}{\pm \sqrt{(r-r_{1})(r-r_{2})(r-r_{3})(r-r_{4})}}, \label{158b}\\
   \phi_{o}-\phi_{s}=-\frac{3 a L^2 }{\Lambda }\left(\frac{\Omega_{c}}{r_{c,-c} r_{c,-} r_{c,+}}-\frac{\Omega_{-c}}{r_{c,-c} r_{-c,-} r_{-c,+}}-\frac{\Omega_{-}}{r_{c,-} r_{-,-c} r_{-,+}}-\frac{\Omega_{+}}{r_{c,+} r_{+,-c} r_{+,-}}\right)-L^{2}a I_{\phi}+ L^{2}\lambda \Bar{I}_{\phi} ,\label{156}\\
    t_{o}-t_{s}=\frac{3 L^2}{\Lambda } \left(-\frac{\Phi_{c}}{r_{c,-c} r_{c,-} r_{c,+}}+\frac{\Phi_{-c}}{r_{c,-c} r_{-c,-} r_{-c,+}}+\frac{\Phi_{-}}{r_{c,-} r_{-,-c} r_{-,+}}+\frac{\Phi_{+}}{r_{c,+} r_{+,-c} r_{+,-}}-\tau \right)+a^2 L^2 I_{t}-a L^2(a-\lambda)I_{\phi}, \label{157}\\
          \Phi_{x}=\left(a^2+r_{x}^2\right)\Omega_{x},\Omega_{x}=\left(a^2-a \lambda +r_{x}^2\right)I_{x},I_{x}= \int_{r_{s}}^{r_{o}} \frac{\text{dr}}{\pm_{r}(r-r_{x}) \sqrt{R(r)}}.\label{159*}
\end{align}
The subscripts $x$ correspond to the horizons $-c,-,+,c$, where $r_{c}$, $r_{-}$, $r_{+}$, and $r_{-c}$ represent the cosmological horizon, Cauchy horizon, event horizon, and the dual of the cosmological horizon, respectively. Additionally, $r_{a,b}=r_{a}-r_{b}$. We will solve the radial integrals $I_r$ and $I_x$ for two scenarios; when the radial potential has four real roots and that which when the radial potential has only two real roots with the other two being complex conjugates. The solutions will be obtained by utilizing the substitutions in Ref.\cite{byrd1954handbook}.  
\begin{itemize}
    \item \underline{$r_{1}<r_{2}<r_{3}<r_{4}<r_s$}\\
    Utilizing part $258.00$ of Ref. \cite{byrd1954handbook}, an integral of the type $I_r$ has a solution of the form,
    \begin{align}
        \int_{r_{4}}^{r_{i}} \frac{\text{dr}}{ \sqrt{(r-r_{1})(r-r_{2})(r-r_{3})(r-r_{4})}}= g_E F(\varphi_{E,i} \mid k_E),\label{Ixg}
    \end{align}
    where $i$ denotes the source $s$ or observer $o$ and,
\begin{align}
    k_{E}=\frac{(r_{4}-r_{1}) (r_{3}-r_{2})}{(r_{4}-r_{2}) (r_{3}-r_{1})} ,\quad g_{E}=\frac{2}{\sqrt{(r_{4}-r_{2})(r_{3}-r_{1})}},\quad \varphi_{E,i}=\arcsin \sqrt{\frac{(r_{3}-r_{1})(r_{i}-r_{4})}{(r_{4}-r_{1})(r_{i}-r_{3})}}.\label{160}
\end{align}
Furthermore, making use of part $258.39$ and $340.01$ of Ref. \cite{byrd1954handbook}, the solution to the integral of the form $I_x$ is expressed as,
\begin{align}
    \int_{r_{4}}^{r_{i}} \frac{\text{dr}}{(r-r_{x}) \sqrt{(r-r_{1})(r-r_{2})(r-r_{3})(r-r_{4})}}=-\frac{2 [ (r_{4}-r_{3})\Pi_{i}-(r_x-r_4)F_{i}]}{(r_{x}-r_{4}) (r_{x}-r_{3}) \sqrt{(r_{4}-r_{2}) (r_{3}-r_{1})}}, \label{Irg}
\end{align}
where, 
\begin{align}
    \Pi_{i}=\Pi \left(\frac{(r_{4}-r_{1}) (r_{x}-r_{3})}{(r_{x}-r_{4}) (r_{3}-r_{1})};\varphi_{E,i}|k_E\right), F_{i}=F\left(\varphi_{E,i}|k_E\right). \label{Ixsym}
\end{align}
\item \underline{$r_{1}<r_{2}<r_{s}, r_{3}=\Bar{r}_{4}$}\\
For these nature of roots, the appropriate substitution to use in solving integrals of the form $I_r$ and $I_x$ in Ref.\cite{byrd1954handbook} are parts $260.00$ and $260.04$ respectively. Using these substitutions we obtain,
\begin{align}
     \int_{r_{2}}^{r_{i}} \frac{\text{dr}}{ \sqrt{(r-r_{1})(r-r_{2})(r-r_{3})(r-r_{4})}}=g_P F(\varphi_{P,i}\mid k_P), \label{Irgp}\\
     \int_{r_{2}}^{r_{i}} \frac{\text{dr}}{(r-r_{x}) \sqrt{(r-r_{1})(r-r_{2})(r-r_{3})(r-r_{4})}}=  \frac{(g_{P} (B-A)) \left((\alpha -\alpha_{2})R_{1,i}+\alpha_{2} F(\varphi_{P,i} \mid k_{P})\right)}{r_{2} B+r_{1}A -A r_{x}-B r_{x}}, \label{Izgp}
\end{align}
where the parameters $A, g_P,k_P$ and $B$ have been defined as,
\begin{align}
	    A=\sqrt{(r_{2}-b_{1}^{2})^{2}+a_{1}^{2}}, B=\sqrt{(r_{1}-b_{1})^{2}+a_{1}^{2}},\quad g_{P}=\frac{1}{\sqrt{AB}}, k_{P}=\frac{(A+B)^{2}-(r_{2}-r_{1})^{2}}{4A B}, \label{A}\\
	    a_{1}=\sqrt{-\frac{(r_{3}-r_{4})^{2}}{4}},\quad b_{1}=\frac{r_{3}+r_{4}}{2},\label{B}
	\end{align}
we have also defined,
\begin{align}
  R_{1,i}=\frac{1}{1-\alpha^2}\left( \Pi\left[ \frac{\alpha^2}{\alpha^2-1};\varphi_{P,i} \mid k_{P}  \right] -\alpha f_{1,i}\right),\quad \alpha=\frac{B r_{2}+A r_{1}-r_{x}(A+B)}{B r_{2}-r_{1}A+r_{x}(A+B)},\quad \alpha_{2}=\frac{A+B}{B-A}, \label{A1}\\ f_{1,i}=\sqrt{\frac{\alpha^2-1}{k_{P}+(1-k_{P})\alpha^2}}\ln{\left | \frac{\sqrt{k_{P}+(1-k_{P})\alpha^2}dn(u_{i}\mid k_{P})+\sqrt{\alpha^2-1}sn(u_{i}\mid k_{P})}{\sqrt{k_{P}+(1-k_{P})\alpha^2}dn(u_{i}\mid k_{P})-\sqrt{\alpha^2-1}sn(u_{i}\mid k_{P})} \right |},\label{A2}\\ u_{i}=cn^{-1}\left[ \cos(\varphi_{P,i}) \mid k_{P} \right],\varphi_{P,i}=\arccos\left[ \frac{(A-B)r_{i}+r_{2} B-r_{1} A}{(A+B)r_{i}-r_{2} B-r_{1}A} \right]. \label{A3}
\end{align}
\end{itemize}
The complete solutions for radial integrals will be derived using the above solutions by substituting $i$ with $s$ or $o$ which denotes the source or the observer respectively.
\subsection{Kerr de Sitter Revisited} \label{rkdsAppendix}
\Cref{96} and (\ref{97}) can then be expressed in integral forms as,
\begin{align}
    \hat{I}_{\hat{\theta}}= \int_{\hat{\theta}_{s}}^{\hat{\theta}_{o}} \frac{\text{d$\theta$}}{\sqrt{\hat{\Theta}(\hat{\theta})}},\label{100aa} \\
    \hat{I}_r=\int_{r_{s}}^{r_{o}} \frac{\text{dr}}{\sqrt{\hat{R}(r)}},\label{100a}\\
    \hat{\phi}_{o}-\hat{\phi}_{s}=  \int_{r_{s}}^{r_{o}}  \left( \frac{a r^{2}+a^{3}-a \hat{\Delta}-a^{2}\hat{\lambda}}{\hat{\Delta}} \right) \frac{\text{dr}}{\sqrt{\hat{R}(r)}}+ \hat{\lambda} \hat{I}_{\phi}, \label{100} \\
    t_{o}-t_{s}=  \int_{r_{s}}^{r_{o}} \left( \frac{(r^{2}+a^{2})(r^{2}+a^{2}-a \lambda)}{\hat{\Delta}} +a \hat{\lambda}-a^{2} \right) \frac{\text{dr}}{\sqrt{\hat{R}(r)}} +a^{2} \hat{I}_{t}, \label{101}
\end{align}
for which we have defined,
\begin{align}
    \hat{I}_{\phi}= \int_{\theta_{s}}^{\theta_{o}} \frac{\text{d$\theta$}}{\sin^{2} \theta \sqrt{\hat{\Theta}(\theta)}} , \hat{I}_{t}= \int_{\theta_{s}}^{\theta_{o}} \frac{\cos^{2}\theta \text{d$\theta $}}{\sqrt{\hat{\Theta}(\theta)}}. \label{102}
\end{align}

The subscripts $s$ and $o$ denotes the source and observer respectively. To obtain the solutions of \cref{102}, we proceed by first expressing the angular potential in terms of its roots $\hat{u_{+}}$ and $\hat{u_{-}}$,
\begin{align}
    \hat{\Theta}_{\hat{u}}= \frac{a^{2}}{1-\hat{u}}(\hat{u}_{+}-\hat{u})(\hat{u}-\hat{u}_{-}).\label{117}
\end{align}
The angular integrals still unpack as \cref{53}. We obtain the general solutions as,
\begin{enumerate}
    \item \begin{align}
         \left| \int_{\pi/2}^{\theta_i} \frac{\text{d$\theta $}}{\sqrt{\hat{\Theta}(\theta)}}\right| = \dfrac{1}{\sqrt{-\hat{u}_{-} a^{2}}} F\left( \hat{x}_{i}|\hat{k}\right), \label{118}
    \end{align}
    where,
    \begin{align}
       \hat{x}_{i}= \arcsin \left( \sqrt{\frac{\cos^{2}\theta_{i}}{\hat{u}_{+}}}\right), \hat{k}=\frac{\hat{u}_{+}}{\hat{u}_{-}} .\label{119}
    \end{align}
    \item \begin{align}
        \left| \int_{\pi/2}^{\theta_{i}} \frac{\text{d$\theta$}}{\sin^{2} \theta \sqrt{\hat{\Theta}(\theta)}}\right| =\dfrac{1}{\sqrt{-u_{-}a^{2}}}\int_{0}^{\hat{x}_{i}} \dfrac{\text{dx}}{(1-u_{+}\sin^{2}x)\sqrt{1-k\sin^{2}x }}=\dfrac{1}{\sqrt{-u_{-}a^{2}}} \Pi[\hat{u}_{+};\hat{x}_{i} |\hat{k}]. \label{120} 
    \end{align}
    \item \begin{align}
        \left| \int_{\pi/2}^{\theta_{i}} \frac{\cos^{2}\theta \text{d$\theta $}}{\sqrt{\hat{\Theta}(\theta)}}\right| = \dfrac{1}{\sqrt{-u_{-}a^{2}}}\int_{0}^{\hat{x}_{i}} \dfrac{\hat{u}_{+}\sin^{2}x \text{dx}}{\sqrt{1-k\sin^{2}x }} =\dfrac{1}{\sqrt{-u_{-}a^{2}}} \frac{\hat{u}_{+}(F(\hat{x}_{i} |\hat{k}) -E(\hat{x}_{i} |\hat{k}))}{\hat{k}}.\label{121}
    \end{align}
\end{enumerate}
Moreover, in order to obtain the time delay and azimuthal angle parameter, we employ a similar approach as in the case of \ac{kds}. This involves integrating over half orbits and treating the radial coordinate as a constant. Thus, we express \cref{100} and (\ref{101}) as,
\begin{align}
     \hat{\delta}=  \left( \frac{a r^{2}+a^{3}-a \hat{\Delta}-a^{2}\hat{\lambda}}{\hat{\Delta}} \right) \left| \int_{\theta_1}^{\theta_4} \frac{\text{d$\theta $}}{\sqrt{\hat{\Theta}(\theta)}}\right|+ \hat{\lambda} \left| \int_{\theta_1}^{\theta_{4}} \frac{\text{d$\theta$}}{\sin^{2} \theta \sqrt{\hat{\Theta}(\theta)}}\right|, \label{del2} \\
    \hat{\zeta}=  \left( \frac{(r^{2}+a^{2})(r^{2}+a^{2}-a \lambda)}{\hat{\Delta}} +a \hat{\lambda}-a^{2} \right)  \left| \int_{\theta_1}^{\theta_4} \frac{\text{d$\theta $}}{\sqrt{\hat{\Theta}(\theta)}}\right| +a^{2}  \left| \int_{\theta_1}^{\theta_{4}} \frac{\cos^{2}\theta \text{d$\theta $}}{\sqrt{\hat{\Theta}(\theta)}}\right|, \label{zeta2}
\end{align}
where we have employed the equality relation given by \cref{32b} to transform the Mino parameter into an angular integral. Making use of the general solutions \cref{118}-(\ref{121}), we obtain,
\begin{align}
     \left| \int_{\theta_1}^{\theta_4} \frac{\text{d$\theta $}}{\sqrt{\hat{\Theta}(\theta)}}\right|=\dfrac{2}{\sqrt{-\hat{u}_{-} a^{2}}} K\left(\hat{k}\right)=\hat{\Gamma}_\theta,  \left| \int_{\theta_1}^{\theta_{4}} \frac{\cos^{2}\theta \text{d$\theta $}}{\sqrt{\hat{\Theta}(\theta)}}\right|=\dfrac{2\hat{u}_{+}}{\hat{k}\sqrt{-u_{-}a^{2}}}\left( K(\hat{k})-E(\hat{k})\right)=\hat{\Gamma}_t, \label{124a}\\
      \left| \int_{\theta_1}^{\theta_{4}} \frac{\text{d$\theta$}}{\sin^{2} \theta \sqrt{\hat{\Theta}(\theta)}}\right|=\dfrac{2}{\sqrt{-u_{-}a^{2}}} \Pi(\hat{u}_{+} |\hat{k} )=\hat{\Gamma}_\phi. \label{124b}
\end{align}
Moreover, the Lyapunov exponent per half orbits is derived using the same approach in \ac{kds} through the relation,
\begin{align}
    \hat{\gamma}=\sqrt{\frac{\hat{R''}(r_b)}{2}} \hat{\Gamma}_\theta. \label{124c}
\end{align}
To obtain solutions for the radial integrals, firstly we express the radial potential and the parameter $\hat{\delta}_r$ in terms of their roots, 
\begin{align}
    \hat{R}(r)=(r-\hat{r}_{1})(r-\hat{r}_{2})(r-\hat{r}_{3})(r-\hat{r}_{4}), \hat{\Delta}=-\frac{3}{\Lambda}(r-\hat{r}_{-c})(r-\hat{r}_{-})(r-\hat{r}_{+})(r-\hat{r}_{c}). \label{168}
\end{align}
Using \cref{168} we rewrite \cref{100a}-(\ref{101}) as,
\begin{align}
    \hat{I}_{r}=\int_{r_{s}}^{r_{o}} \frac{\text{dr}}{\pm \sqrt{(r-\hat{r}_{1})(r-\hat{r}_{2})(r-\hat{r}_{3})(r-\hat{r}_{4})}},\label{167}\\
    \hat{\phi}_{o}-\hat{\phi}_{s}=a \left(\frac{\hat{\Omega}_{-c}}{r_{c,-c}r_{-,-c}r_{+,-c}}\hat{I}_{-c}+\frac{\hat{\Omega}_{-}}{r_{c,-}r_{-c,-}r_{+,-}}\hat{I}_{-}+\frac{\hat{\Omega}_{+}}{r_{c,+}r_{-c,+}r_{-,+}}\hat{I}_{+}+\frac{\hat{\Omega}_{c}}{r_{-c,c}r_{-,c}r_{+,c}}\hat{I}_{c}-\tau \right)+\hat{\lambda} \hat{I}_{\phi},\label{170}\\
     \hat{t}_{o}-\hat{t}_{s}= \frac{\hat{\Phi}_{c}}{\hat{r}_{-c,c} \hat{r}_{-,c} \hat{r}_{+,c}}+\frac{\hat{\Phi}_{-c}}{\hat{r}_{c,-c} \hat{r}_{c,-c} \hat{r}_{+,-c}}+\frac{\hat{\Phi}_{-}}{\hat{r}_{c,-} \hat{r}_{-c,-} \hat{r}_{+,-}}+\frac{\hat{\Phi}_{+}}{\hat{r}_{c,+} \hat{r}_{-c,+} \hat{r}_{-,+}}+\tau  \left(-a^2+a \hat{\lambda} -\frac{\Lambda }{3}\right), \label{171} \\
     \hat{\Phi}_{x}=\left(a^2+\hat{r}_{x}^2\right)\hat{\Omega}_{x} \hat{I}_{x}, \hat{\Omega}_{x}=\frac{3 }{\Lambda}\left(a^2-a \hat{\lambda} +\hat{r}_{x}^2\right),\hat{I}_{x}=\int_{r_{s}}^{r_{o}} \frac{\text{dr}}{(r-\hat{r}_{x}) \sqrt{\hat{R}(r)}}.\label{172}
\end{align}
The subscripts $x$ denotes the horizons $-c,-,+,c$, where $\hat{r}_{+}$, $\hat{r}_{-}$, $\hat{r}_{c}$, and $\hat{r}_{-c}$ represent the  event horizon, Cauchy horizon, the cosmological horizon and its dual, respectively. Further, $\hat{r}_{a,b}=\hat{r}_{a}-r_{b}$. Further, the subscript $o$ and $s$ denote the source and the observer respectively, where $r_o$ and $r_s$ need to be chosen within the domain of outer communication. The radial integrals $\hat{I}_r$ and $\hat{I}_x$ takes the same form as that of \ac{kds}. In the same way we obtain their solutions for the roots ($\hat{r}_{1}<\hat{r}_{2}<\hat{r}_{3}<\hat{r}_{4}<r_s$) and ($\hat{r}_{1}<\hat{r}_{2}<r_{s}, \hat{r}_{3}=\Bar{\hat{r}}_{4}$).
\begin{itemize}
    \item \underline{$\hat{r}_{1}<\hat{r}_{2}<\hat{r}_{3}<\hat{r}_{4}<r_s$ }\\
    \begin{align}
        \int_{r_{4}}^{r_{i}} \frac{\text{dr}}{\pm \sqrt{(r-\hat{r}_{1})(r-\hat{r}_{2})(r-\hat{r}_{3})(r-\hat{r}_{4})}}=\hat{g}_{E}F(\hat{\varphi}_{E,i}|\hat{k}_{E}), \label{C71}\\
        \int_{r_{4}}^{r_{i}} \frac{\text{dr}}{{\pm_r}(r-\hat{r}_x)\sqrt{(r-\hat{r}_{1})(r-\hat{r}_{2})(r-\hat{r}_{3})(r-\hat{r}_{4})}}=-\frac{2}{\hat{r}_{x,3}. \sqrt{\hat{r}_{4,2} \hat{r}_{3,1}}} \left( F\left(\hat{\varphi}_{E,i}|\hat{k}_{E}\right)+\frac{\hat{r}_{4,3}\hat{\Pi}_{i})}{\hat{r}_{x,4}} \right) ,\label{C72}
    \end{align}
    where we have defined,
    \begin{align}
    \hat{\varphi}_{E,i}=\arcsin \sqrt{\frac{\hat{r}_{3,1}\hat{r}_{i,4}}{\hat{r}_{4,1}\hat{r}_{i,3}}}, \hat{k}_{E}=\frac{\hat{r}_{4,1} \hat{r}_{3,2}}{\hat{r}_{4,2} \hat{r}_{3,1}} , \hat{g}_{E}=\frac{2}{\sqrt{\hat{r}_{4,2}\hat{r}_{3,1}}}, \hat{\Pi}_{i}=\Pi \left(\frac{\hat{r}_{4,1} \hat{r}_{x,3}}{\hat{r}_{x,4} \hat{r}_{3,1}};\hat{\varphi}_{E,i}|\hat{k}_{E}\right) \label{C73}
\end{align}
\item \underline{$\hat{r}_{1}<\hat{r}_{2}<r_{s}, \hat{r}_{3}=\Bar{\hat{r}}_{4}$}\\
In the same way as \cref{Irgp} and (\ref{Izgp}) we obtain the solution for this case in \ac{rkds} as,
\begin{align}
\int_{r_{2}}^{r_{i}} \frac{\text{dr}}{\pm \sqrt{(r-\hat{r}_{1})(r-\hat{r}_{2})(r-\hat{r}_{3})(r-\hat{r}_{4})}}=\hat{g}_P F(\hat{\varphi}_{P,i}\mid k_P), \label{C74}\\
 \int_{r_{2}}^{r_{i}} \frac{\text{dr}}{{\pm_r}(r-\hat{r}_x)\sqrt{(r-\hat{r}_{1})(r-\hat{r}_{2})(r-\hat{r}_{3})(r-\hat{r}_{4})}}=    \frac{(\hat{B}-\hat{A}) \left((\hat{\alpha} -\hat{\alpha}_{2}) \hat{R}_{1,i}+\hat{\alpha}_{2}F(\hat{\varphi}_{P,i} \mid \hat{k}_{P})\right)}{\sqrt{\hat{A}\hat{B}}(\hat{r}_{2} \hat{B}+\hat{r}_{1}\hat{A} -\hat{A} \hat{r}_{x}-\hat{B} \hat{r}_{x})}.\label{C75}
\end{align}
The parameters $\hat{A},\hat{B},\hat{k}_P$ have been defined as,
\begin{align}
    \hat{A}=\sqrt{\left(\hat{r}_{2}-\frac{\hat{r}_{3}+\hat{r}_{4}}{2}\right)^{2}-\frac{(\hat{r}_{3}-\hat{r}_{4})^{2}}{4}}, 
    \hat{B}=\sqrt{\left(\hat{r}_{1}-\frac{\hat{r}_{3}+\hat{r}_{4}}{2} \right)^{2}-\frac{(\hat{r}_{3}-\hat{r}_{4})^{2}}{4}},\\ \hat{k}_{P}=\frac{(\hat{A}+\hat{B})^{2}-(\hat{r}_{2}-\hat{r}_{1})^{2}}{4\hat{A} \hat{B}}.\label{C76A}
\end{align}
We have also defined,
\begin{align}
  \hat{R}_{1,i}=\frac{1}{1-\hat{\alpha}^2}\left( \Pi\left[ \frac{\hat{\alpha}^2}{\hat{\alpha}^2-1};\hat{\varphi}_{P,i} \mid \hat{k}_{P}  \right] -\hat{\alpha} \hat{f}_{1,i}\right),\quad \hat{\alpha}=\frac{\hat{B} r_{2}+\hat{A}r_{1}-r_{x}(A+\hat{B})}{\hat{B} r_{2}-r_{1}\hat{A}+r_{x}(\hat{A}+\hat{B})},\hat{\alpha}_{2}=\frac{\hat{A}+\hat{B}}{\hat{B}-\hat{A}},\label{C76}\\ \hat{f}_{1,i}=\sqrt{\frac{\hat{\alpha}^2-1}{\hat{k}_{P}+(1-\hat{k}_{P})\hat{\alpha}^2}}\ln{\left | \frac{\sqrt{\hat{k}_{P}+(1-\hat{k}_{P})\hat{\alpha}^2}dn(\hat{u}_{i}\mid \hat{k}_{P})+\sqrt{\hat{\alpha}^2-1}sn(\hat{u}_{i}\mid \hat{k}_{P})}{\sqrt{\hat{k}_{P}+(1-\hat{k}_{P})\hat{\alpha}^2}dn(\hat{u}_{i}\mid \hat{k}_{P})-\sqrt{\hat{\alpha}^2-1}sn(\hat{u}_{i}\mid \hat{k}_{P})} \right |},\label{C77}\\\hat{u}_{i}=cn^{-1}\left[ \cos(\hat{\varphi}_{P,i}) \mid \hat{k}_{P} \right],\hat{\varphi}_{P,i}=\arccos\left[ \frac{(\hat{A}-\hat{B})r_{i}+\hat{r}_{2} \hat{B}-\hat{r}_{1} A}{(\hat{A}+\hat{B})r_{i}-\hat{r}_{2} \hat{B}-\hat{r}_{1}\hat{A}} \right]. \label{C78}
\end{align}
\end{itemize}
\begin{subequations}
\end{subequations}
\section{Integrals in terms of angular turning points} \label{apII}
For purposes of analytic ray tracing, it becomes imperative to express the solutions in terms of the turning points experienced by photons. In this section we will express the angular integrals in terms of the number of turning points that the photons encounter in the $\theta$ direction.
Considering all possible configurations, Ref. \cite{kapec2019particle} demonstrates that angular path integrals for ordinary motion unpack as,
 \begin{align}
    \int_{\theta_s}^{\theta_o}=2m\left|\int_{\pi/2}^{\theta_\pm} \right| +\eta_s \left| \int_{\pi/2}^{\theta_s}\right|-\eta_o \left| \int_{\pi/2}^{\theta_o}\right|. \label{53}
\end{align}
Where,
\begin{align}
    \eta_s=sign(p^{\theta}_{s})sign(\cos\theta_{s})=(-1)^m sign(p^{\theta}_{o})sign(\cos \theta_s),
    \eta_o=sign(p^{\theta}_{o})sign(\cos\theta_{o})=(-1)^m sign(p^{\theta}_{s})sign(\cos \theta_o), \label{54}
\end{align}
and $m$ is the number of turning points in the latitudinal direction. We have already obtained all the general solutions thus we will directly apply our solutions to \cref{53}.

In \ac{kds} spacetime we obtain, 
\begin{align}
    \mathbb{I}_\theta= \dfrac{2 m}{\sqrt{-u_{-}\Xi}} K\left( k\right) + \dfrac{ (-1)^m sign(p^{\theta}_{o})}{\sqrt{-u_{-}\Xi}} F\left( x_{s}| k\right)-\dfrac{sign(p^{\theta}_{o})}{\sqrt{-u_{-}\Xi}} F\left( x_{o}| k\right), \label{pr10}\\
    \mathbb{I}_\phi=\dfrac{2 m}{\sqrt{-u_{-}\Xi}}\Pi \left( -u_{+} \mathcal{Y};|k\right)+\dfrac{(-1)^m sign(p^{\theta}_{o})}{\sqrt{-u_{-}\Xi}}\Pi \left( -u_{+} \mathcal{Y};x_s|k\right)-\dfrac{ sign(p^{\theta}_{o})}{\sqrt{-u_{-}\Xi}}\Pi \left( -u_{+} \mathcal{Y};x_o|k\right), \label{pr11}\\
    \mathbb{I}_t= \frac{2 m u_{+} J(-u_+ \mathcal{Y}|k)}{\sqrt{-u_{-}\Xi}}+\frac{(-1)^m sign(p^{\theta}_{o}) u_{+} J(-u_+ \mathcal{Y},\arcsin t_s|k)}{\sqrt{-u_{-}\Xi}}-\frac{ sign(p^{\theta}_{o}) u_{+} J(-u_+ \mathcal{Y},\arcsin t_o|k)}{\sqrt{-u_{-}\Xi}}, \label{pr12}
\end{align}
\begin{multline}
     \Bar{\mathbb{I}}_\phi=\dfrac{2 m}{\sqrt{-u_{-}\Xi}} \dfrac{1}{1+\mathcal{Y}}[\Pi(u_{+}|k)+\mathcal{Y} \Pi(-u_{+}\mathcal{Y}|k) ]+\dfrac{(-1)^m sign(p^{\theta}_{o})}{\sqrt{-u_{-}\Xi}} \dfrac{1}{1+\mathcal{Y}}[\Pi(u_{+};\arcsin t_s|k)+\mathcal{Y} \Pi(-u_{+}\mathcal{Y};\arcsin t_s|k) ] \\-\dfrac{ sign(p^{\theta}_{o})}{\sqrt{-u_{-}\Xi}} \dfrac{1}{1+\mathcal{Y}}[\Pi(u_{+};\arcsin t_o|k)+\mathcal{Y} \Pi(-u_{+}\mathcal{Y};\arcsin t_o|k) ],\label{pr13}
\end{multline}
while in \ac{rkds} spacetime we obtain,
\begin{align}
    \hat{\mathbb{I}}_\theta=\dfrac{2 m}{\sqrt{-\hat{u}_{-}a^2}} K\left( \hat{k}\right) + \dfrac{ (-1)^m sign(p^{\theta}_{o})}{\sqrt{-\hat{u}_{-}a^2}} F\left(  \hat{x}_{s}|  \hat{k}\right)-\dfrac{sign(p^{\theta}_{o})}{\sqrt{-\hat{u}_{-}a^2}} F\left(  \hat{x}_{o}|  \hat{k}\right), \label{d5}\\
    \hat{\mathbb{I}}_\phi=\dfrac{2m}{\sqrt{-\hat{u}a^{2}}} \Pi(\hat{u}_{+} |\hat{k} )+\dfrac{(-1)^m sign(p^{\theta}_{o})}{\sqrt{-\hat{u}a^{2}}} \Pi[\hat{u}_{+};\hat{x}_{s} |\hat{k}]-\dfrac{sign(p^{\theta}_{o})}{\sqrt{-\hat{u}a^{2}}} \Pi[\hat{u}_{+};\hat{x}_{o} |\hat{k}], \label{d6}
\end{align}
\begin{multline}
     \hat{\mathbb{I}}_t=\dfrac{2m}{\sqrt{-\hat{u}a^{2}}} \frac{\hat{u}_{+}(F(\hat{k}) -E(\hat{k}))}{\hat{k}}+\dfrac{(-1)^m sign(p^{\theta}_{o})}{\sqrt{-\hat{u}a^{2}}} \frac{\hat{u}_{+}(F(\hat{x}_{s} |\hat{k}) -E(\hat{x}_{s} |\hat{k}))}{\hat{k}}\\-\dfrac{ sign(p^{\theta}_{o})}{\sqrt{-\hat{u}a^{2}}} \frac{\hat{u}_{+}(F(\hat{x}_{o} |\hat{k}) -E(\hat{x}_{o} |\hat{k}))}{\hat{k}}. \label{d7}
\end{multline}
The $sign(p^{\theta}_{o})$ is equivalent to $sign(y)$. 
The above equations are only valid for observers at latitude $\theta_o \in (0,\pi/2)$. \\
We postpone the case of $\theta_o=0,\pi/2$ to our future work.

\section{Change in $\phi$ and $t$ for subsequent images} \label{apIII}\label{subsimages}
In \cref{lensingandphotonrings}, an analysis of direct images, lensing rings and photon rings has been done. In the same section we have slightly talked about subsequent images, ($(m+1)^{th}$,$m^{th}$), undergoing a rotation due to change in azimuthal angle and a delay in time of detection. In  this part we explicitly derive these quantities. Giving the explicit forms of these quantities is important as we shall see that they are related to the critical parameters controlling the photon ring structure.

To obtain the \ac{kds} change in azimuthal angle between subsequent images $  \Delta \phi_{m+1}-\Delta \phi_{m}$ we proceed as follows,
  \begin{subequations}
    \begin{multline}
        \Delta \phi_{m+1}-\Delta \phi_{m}= \left( \int_{r_{o}}^{r_{s}} \frac{\left(a L^2\right) \left(a (a-\lambda )+\text{r}^2\right) \text{dr}}{\pm_{r} \text{$\Delta_{r} $} \sqrt{R(r)}}  -L^{2}a \mathbb{I}_{\phi}+ L^{2}\lambda \Bar{\mathbb{I}}_{\phi}\right)_{m+1}-\\ \left( \int_{r_{s}}^{r_{o}} \frac{\left(a L^2\right) \left(a (a-\lambda )+\text{r}^2\right) \text{dr}}{\pm_{r} \text{$\Delta_{r} $} \sqrt{R(r)}}  -L^{2}a \mathbb{I}_{\phi}+ L^{2}\lambda \Bar{\mathbb{I}}_{\phi}\right)_{m},\label{D8a}
    \end{multline}
   \begin{multline}
       = \int_{r_{s}^{m+1}}^{r_{s}^{m}} \frac{\left(a L^2\right) \left(a (a-\lambda )+\text{r}^2\right) \text{dr}}{\pm_{r} \text{$\Delta_{r} $} \sqrt{R(r)}} -\dfrac{2 L^{2}a}{\sqrt{-u_{-}\Xi}}\Pi \left( -u_{+} \mathcal{Y};|k\right)+ \dfrac{2 L^{2}\lambda}{(1+\mathcal{Y})\sqrt{-u_{-}\Xi}} [\Pi(u_{+}|k)+\mathcal{Y} \Pi(-u_{+}\mathcal{Y}|k) ], \label{D8b} 
    \end{multline}
    \begin{multline}
       = \frac{\left(a L^2\right) \left(a (a-\lambda )+\text{r}^2\right) }{\text{$\Delta_{r} $}} \Gamma_\theta  -\dfrac{2 L^{2}a}{\sqrt{-u_{-}\Xi}}\Pi \left( -u_{+} \mathcal{Y};|k\right)+ \dfrac{2 L^{2}\lambda}{(1+\mathcal{Y})\sqrt{-u_{-}\Xi}} [\Pi(u_{+}|k)+\mathcal{Y} \Pi(-u_{+}\mathcal{Y}|k) ]. \label{D8c}
    \end{multline}
 \end{subequations}
 In \cref{D8b} we have performed a subtraction alongside inserting \cref{pr11} and \cref{pr13}. Subsequent images are just images of same sources that have undergone extreme lensing. Therefore, for each image, $r_{s}^{m+1}$ is simply the same as $r_{s}^{m}$. Thus on the radial integral, we have kept $r$ to be constant. Additionally, $r_{s}^{m}$ to $r_{s}^{m+1}$ is equivalent to half an orbit. The Mino period for half an orbit is as given in \cref{202}. This is the approach we have used to arrive at \cref{D8c}. 
 
 Performing the same analysis for $\Delta t$ gives us,
 \begin{multline}
      \Delta t_{m+1}-\Delta t_{m}= \dfrac{L^{2}((r^{2}+a^{2})^{2}-a\lambda(a^{2}+r^{2}))}{ \Delta_{r}} \Gamma_\theta +\frac{2 a^2 L^2 u_{+} J(-u_+ \mathcal{Y}|k)}{\sqrt{-u_{-}\Xi}}-\dfrac{2 a L^2(a-\lambda)}{\sqrt{-u_{-}\Xi}}\Pi \left( -u_{+} \mathcal{Y};|k\right). \label{tkds}
    \end{multline}
Likewise in \ac{rkds} spacetime we obtain these parameters as,
\begin{align}
    \hat{\Delta}\phi_{m+1}-\hat{\Delta}\phi_{m} = \left( \frac{a r^{2}+a^{3}-a \hat{\Delta}-a^{2}\hat{\lambda}}{\hat{\Delta}} \right) \hat{\Gamma}_\theta+\dfrac{2 \hat{\lambda}}{\sqrt{-u_{-}a^{2}}} \Pi(\hat{u}_{+} |\hat{k} ), \label{D10} \\
    \hat{\Delta}t_{m+1}-\hat{\Delta} t_{m}= \left( \frac{(r^{2}+a^{2})(r^{2}+a^{2}-a \lambda)}{\hat{\Delta}} +a \hat{\lambda}-a^{2} \right) \hat{\Gamma}_\theta + \dfrac{2 a^{2}}{\sqrt{-u_{-}a^{2}}} \frac{\hat{u}_{+}(E(\hat{k}) -E(\hat{k}))}{\hat{k}}. \label{trkds}
\end{align}

\clearpage
\newpage
\begin{acknowledgments}

The authors thank FAPES/FAPEMIG/CNPq/CAPES for financial support. The authors thank various anonymous referees for valuable comments that improved the quality of the paper.
\end{acknowledgments}
\section*{Data Availability Statement}
Data sharing is not applicable to this article as no datasets were generated or analysed during the current study.
\bibliography{Refs}

\end{document}